\RequirePackage{rotating}
\documentclass[a4paper,fleqn,usenatbib]{mnras}

\usepackage{mathptmx}
\usepackage[T1]{fontenc}
\usepackage{ae,aecompl}
\usepackage{graphicx}	
\usepackage{amsmath}	
\usepackage{amssymb}	
\usepackage{caption}

\usepackage[dvipsnames]{xcolor}
\usepackage[normalem]{ulem}
\usepackage{rotating}

\title[Testing the RPM with LF and kHz QPOs]{Testing the relativistic precession model using low frequency and kHz quasi-periodic oscillations in neutron star low mass X-ray binaries with known spin}

\author[van Doesburgh and van der Klis]{
Marieke van Doesburgh,$^{1}$\thanks{E-mail: m.j.vandoesburgh@uva.nl}
Michiel van der Klis$^{1}$
\\
$^{1}$Anton Pannekoek Institute, University of Amsterdam, Science Park 904, Postbus 94249, 1090 GE Amsterdam, The Netherlands\\
}

\date{Accepted 2016 November 11. Received 2016 October 19; in original form 2016 August 5}

\pubyear{2016}

\begin{document}
\label{firstpage}
\pagerange{\pageref{firstpage}--\pageref{lastpage}}
\maketitle

\begin{abstract}
We analyze all available RXTE data on a sample of 13 low mass X-ray binaries with known neutron star spin that are not persistent pulsars. 
We carefully measure the correlations between the centroid frequencies of the quasi-periodic oscillations (QPOs). We compare these correlations to the prediction of the relativistic precession model (RPM) that, due to frame dragging, a QPO will occur at the Lense-Thirring precession frequency $\nu_{LT}$ of a test particle orbit whose orbital frequency is the upper kHz QPO frequency $\nu_u$.
Contrary to the most prominent previous studies, we find two different oscillations in the range predicted for $\nu_{LT}$ that are simultaneously present over a wide range of $\nu_u$. Additionally, one of the low frequency noise components evolves into a (third) QPO in the $\nu_{LT}$ range when $\nu_u$ exceeds 600 Hz. The frequencies of these QPOs all correlate to $\nu_u$ following power laws with indices between 0.4--3.3, significantly exceeding the predicted value of 2.0 in 80$\%$ of the cases (at 3 to >20$\sigma$). Also, there is no evidence that the neutron star spin frequency affects any of these three QPO frequencies as would be expected for frame dragging. Finally, the observed QPO frequencies tend to be higher than the $\nu_{LT}$ predicted for reasonable neutron star specific moment of inertia. In the light of recent successes of precession models in black holes, we briefly discuss ways in which such precession can occur in neutron stars at frequencies different from test particle values and consistent with those observed. A precessing torus geometry and other torques than frame dragging may allow precession to produce the observed frequency correlations, but can only explain one of the three QPOs in the $\nu_{LT}$ range. 
\end{abstract}

\begin{keywords}
X-rays: binaries -- accretion, accretion disks -- stars: neutron -- binaries: close
\end{keywords}



\section{Introduction}
\label{sec:intro}
General Relativity predicts that frame dragging causes nodal precession of misaligned orbits around spinning objects.
Quasi periodic oscillations (QPOs) in the Fourier power spectra of neutron star low mass X-ray binaries (NS-LMXBs, see \citealt{Klis:2006book} for a review) have been linked to orbital motion in the inner accretion disk.  Closest to the neutron star Keplerian orbital frequencies of $\sim$1 kHz are expected, and QPO pairs of this frequency are widely found. The behaviour of kHz QPOs is well studied as they are thought to provide an excellent opportunity to test GR in the strong field regime. Their frequencies vary with time, and correlate with those of QPOs at much lower frequency (<80 Hz) \citep{Ford:1998, Psaltis:1999}.  
The relativistic precession model (RPM) of \cite{Stella:1998} explains the QPOs at frequencies of a few tens of Hz as being due to nodal precession of the orbits at the inner edge of the disk whose orbital frequency is identified with the frequency of the upper (highest frequency) kHz QPO. The predicted QPO frequency is once or twice (due to twofold symmetry existing in the tilted accretion disk geometry) the Lense-Thirring precession frequency of a test particle \begin{align}
 \nu_{LT}
    &= \frac{ 8\pi^{2}\emph{I}\nu_{K}^{2}\nu_{s}}{\emph{c}^{2}\emph{M}} = 13.2  \emph{I}_{45}\emph{m}^{-1}\nu_{\emph{K,3}}^{2} \nu_{\emph{s,2.5}}\text{ Hz }, 
    \label{eq:LT}
\end{align} 

where $M=m$$\cdot$$M_{\odot}$ and $I=10^{45}I_{45}$ g cm$^{2}$ are the neutron star mass and moment of inertia, respectively,  $\nu_{\emph{K}}=10^3\nu_{K,3}$ Hz is the Keplerian orbital frequency, and $\nu_{\emph{s}}=300\nu_{s,2.5}$ Hz the neutron star spin frequency: the frequency of the low frequency QPO is predicted to be proportional to the spin frequency and quadratically related to the upper kHz QPO frequency. For realistic equations of state $I_{45}/m$ ranges from 0.5 to 2 \citep{Friedman:1986}. \\
 In a study of the QPOs in three LMXBs with neutron stars of uncertain spin this quadratic dependence was confirmed by \cite{vanStraaten:2003} to remarkable precision. The authors reported a best-fit power law index of 2.01$\pm$0.02. \\
In the past decade many LMXB neutron star spin frequencies have been measured using pulsations and burst oscillations (see \citealt{Patruno:2012} for a review). It was found that the three sources studied by \cite{vanStraaten:2003} have quite different spin frequencies, raising the issue as to why their QPO frequency correlations coincide \citep{Klis:2006book}.

Black hole LMXBs show similar low-frequency (<80 Hz) QPOs to NS-LMXBs \citep{Wijnands:1999, Klein:2008} and they may follow similar frequency correlations \citep{Psaltis:1999}.
Also, a set of three simultaneous QPO frequencies showing a remarkable match to the RPM prediction for the orbital, periastron and Lense-Thirring frequencies was reported in the BH binary GRO\ J1655--40 by \cite{Motta:2014}.
Currently, the model best explaining the X-ray spectral variations with black hole Type C QPO frequency  involves solid-body Lense-Thirring precession of a hot inner flow \citep{Ingram:2009}.
In BH-LMXBs, the spin is a free parameter. As in NS-LMXBs we can measure the spin frequency directly, and contrary to high-frequency QPOs in BH, NS kHz QPOs are observed to vary over a wide range of frequencies, we can further constrain the precession model by studying the QPOs in these systems.

With this motivation, we undertook to test the relativistic precession model in the entire sample of neutron star sources in the (now complete) RXTE archive for which the spin frequency is known and that show the relevant QPOs. 
 These conditions are met by burst oscillation sources and by accreting millisecond X-ray pulsars (AMXPs).
 
For the AMXPs, \cite{vanStraaten:2005} found that the correlations between QPO frequencies are offset from one another and from those observed in non-pulsating LMXBs by varying factors. They found that a shift in kHz QPO frequency was the simplest explanation for the offsets, but a clear physical origin could not be identified. Recently, \cite{Bult:2015} showed that the pulse amplitude of pulsar SAXJ1808.4--3658 differs markedly depending on whether the kHz QPO frequency is higher or lower than the spin frequency of the neutron star; a strong indication that the accretion flow is affected by the magnetic field. Additionally, \cite{Altamirano:2012} show for  the 11 Hz pulsar IGR\ J17480--2446 that QPOs at $\sim$35 Hz cannot arise due to Lense-Thirring precession if the kHz QPOs are identified as $\nu_K$. For an 11 Hz spin frequency, $\nu_{LT}$ should be  $\lesssim$0.8 Hz.
 
 Clearly additional complications are present in the frequency correlations of AMXPs. In this paper we therefore concentrate on the non-pulsating sources. 
 We do include two sources that each have been seen pulsating once for a brief interval, as their aperiodic timing behaviour strongly resembles that of the other sources in our sample (Aquila X-1, \citealt{ALtamirano:Aquila} and 4U\ 1636--53, \citealt{Strohmayer:2002}). Our sample includes the three sources analyzed by \cite{vanStraaten:2003}, but each with a much larger data set, as well as 10 other neutron star LMXBs.

In this paper we present a timing analysis of all RXTE archival data on 13 neutron star LMXBs. We  determine the correlations of the highest frequency kHz QPO (upper kHz QPO) with features at lower frequency using a new, statistically particularly careful method, and report these findings in Section \ref{sec:results}. Our results differ from those of \cite{vanStraaten:2003}, as our larger data set allows a better identification of the LF QPOs. 
In section \ref{section:Models}, we discuss our result in the context of the relativistic precession model and more sophisticated precession models. 

\section{Observations and data analysis}

We analyze all archival RXTE data on the 13 sources that meet our requirements of known spin, no or only one brief interval of pulsations, and the presence of kHz QPOs. We list these objects along with their respective spin frequencies and observational statistics in Table ~\ref{sources}.
\subsection{Spectral analysis}
  We obtain Crab-normalized hard and soft colours for each source from the Standard 2 data following the procedure described in \cite{vanStraaten:2002}.  We remove all type I X-ray bursts prior to analysis. The soft colour is defined as the ratio between the counts in the 3.5--6.0 keV and 2.0--3.5 keV energy bands, and the hard colour as the ratio between the 9.7--16.0 keV and 6.0--9.7 keV bands. We use the colours to assess the accretion state of the source \citep{Hasinger:1989}. 
\subsection{Timing analysis}
To calculate the power spectra we use Event, Single Bit and Good Xenon data with a time resolution of 1/8192 s ($\sim$122 $\mu$s) or better. We take all available energy channels into account (optimizing the energy range depending on QPO type \citep{vanStraaten:2000}, does not significantly improve the $\nu_{LT}$-$\nu_K$ correlation measurement), rebin if necessary to 1/8192 s, and divide the data into segments of 16 seconds. This results in power spectra with a Nyquist frequency of 4096 Hz and a lowest frequency and frequency resolution of 0.0625 Hz. 
We do not perform any background or dead time corrections prior to calculating the power spectra but correct for these effects after averaging the Leahy-normalized power spectra. To do so, we subtract a counting noise model spectrum incorporating dead-time effects \citep{Zhang:1995} following the method of \cite{Klein:2004PhD} and renormalize the power spectra such that the square root of the integrated power in the spectrum equals the fractional root mean square (rms) of the variability in the signal \citep{Klis:1989}. 

\subsection{Selection of power spectra}
\label{sec:selection}
We initially average the power spectra obtained within a single observation (RXTE ObsID; typically containing $\sim$1.5 ks of data).
We then preselect for further analysis those average power spectra that by visual inspection appear to contain QPOs in both the 200--1200 Hz and 0.0625--80 Hz range. For power spectra in which we do not find or significantly fit the QPOs, we attempt to increase signal to noise by averaging power spectra of multiple observations. We only do this for observations consecutive in time to limit the broadening of narrow components. These consecutive observations differ in colour by $<$2$\%$, indicating that no state transition occurs between them. 
The number of observations used for each source are listed in Table \ref{sources}. The presence of simultaneous high and low frequency QPOs depends on source state, and their detection significance on feature strength and width, as well as the observing time and source brightness. The rejection rate is high for weak sources and sources with many short observations (A\ 1744--361, EXO\ 0748--676, XTE\ J1739--285, 4U\ 1636--53, 4U\ 1608--52), as expected. It is also high for sources that mostly populate the high or low hard colour (EXO 0748--676, 4U\ 1608--52, Aquila X-1, KS\ 1731--260, SAX\ J1750.8--2890), as the QPOs are strongest for intermediate hard colour. The sources with the lowest rejection rates are strong (4U\ 1728--34 and 4U\ 0614+09) and/or observed at intermediate hard colour for a significant part of the observing campaign (4U\ 1728--34, 4U\ 0614+09, 4U\ 1702--43, 4U\ 1915--05 and IGR\ J17191--2821).
\subsection{Power spectral fitting}
We fit the power spectra with the sum of several Lorentzians. This phenomenological model enables us to monitor the QPOs and band limited noise as the accretion state of the source changes.
A Lorentzian can be written as  

\begin{equation}
\label{Lor}
P(\nu) = \frac{(\text{rms})^{2}\Delta}{\pi}\frac{1}{(\nu-\nu_{0})^{2}+(\Delta)^{2}},
\end{equation}
where $\Delta$ is the half width at half maximum and $\nu_{0}$ is the centroid frequency ($P(\nu)$ reaches its maximum here).
We fit our model for $\nu_{0}$, the power integrated between 0 and infinity, and the quality factor Q, which is a measure of the coherence of the Lorentzian (Q=$\nu_{0}/$2$\Delta$). In this work we only report $\nu_0$, because models such as the relativistic precession model predict centroid frequencies.   
In order to characterize both narrow QPOs and broad power spectral features with ill-constrained centroid frequencies using the same model, 
 characteristic frequencies ($\nu_{\max}\equiv\sqrt{{\nu_0}^{2}+\Delta^2}$) are  commonly used \citep{Nowak:2000, Belloni:2002}. Lorentzians peak at $\nu_{\max}$ when plotting $\nu$P($\nu$). For narrow features, $\nu_{\max}$ and $\nu_0$ will be similar. For broad features however, $\nu_{\max}$ approaches $\Delta$.

 \subsection{Naming of power spectral components}
\begin{figure}
	\includegraphics[width=\columnwidth]{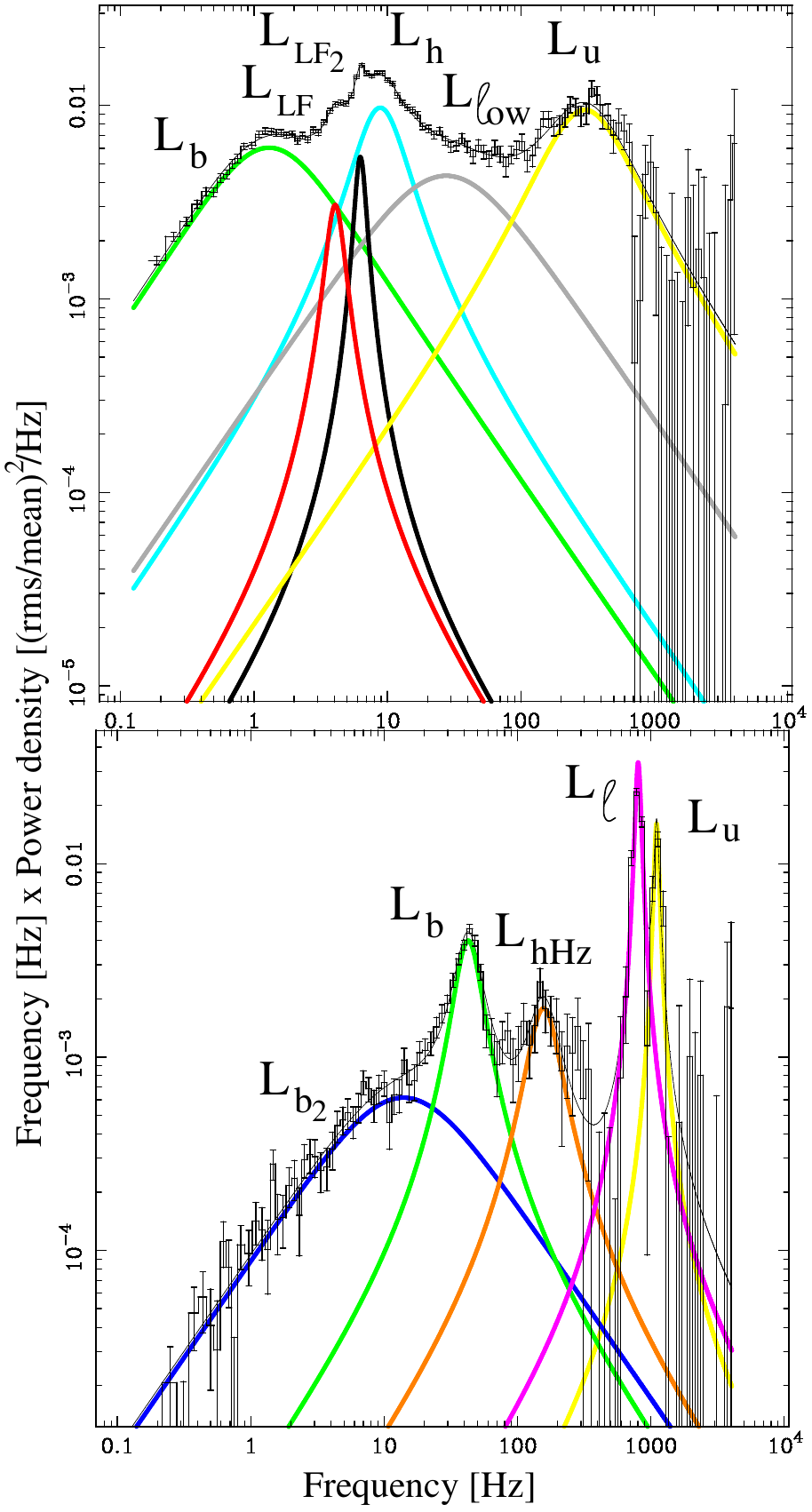}
    \caption{Representative power spectra of two different accretion states of 4U\ 1728--34 illustrating the naming scheme we use. The top and bottom panel are the same as panel A and D, respectively, in Figure \ref{overview}.}
    \label{fig:Alta}
\end{figure}

We identify power spectral components (see Section \ref{ID}) using the identification scheme and naming convention used by \cite{Altamirano:2008} (based on \citealt{vanStraaten:2002}) where features are identified (see Figure \ref{fig:Alta} and Table \ref{scheme}) as break (L$_b$), second break (L$_{b_2}$), low frequency QPO (L$_{LF}$), harmonic of the low frequency QPO (L$_{LF_2}$), hump (L$_{h}$), hectoHz (L$_{hHz}$), low frequency Lorentzian (L$_{{\ell}ow}$), lower kHz QPO (L$_{\ell}$), or upper kHz QPO (L$_u$). This scheme relies on the location of components in the power spectrum,  the correlations between characteristic frequencies ($\nu_{\max}$) of power spectral components and the similarities in the appearance of power spectra between different sources. We write Q$_i$ and rms$_i$ for the quality factor and fractional rms, respectively, of component L$_i$. We discuss the identification of features for all sources together in Section \ref{ID}. A detailed discussion of each source in our sample can be found in Appendix \ref{sources}. 

\begin{table}
\centering
  \tabcolsep=0.1cm
 {
\scalebox{1}{
\small
	\begin{tabular*}{\columnwidth}{@{\extracolsep{\fill}}ll}
   \hline 
\vspace*{0.05cm}
Component name & Symbol (L$_i$)\\
\hline
Break & L$_b$\\
Second break &L$_{b_2}$\\
Low frequency QPO & L$_{LF}$\\
Low frequency QPO harmonic & L$_{LF_2}$\\
Hump & L$_{h}$ \\
HectoHz & L$_{hHz}$\\
Low frequency Lorentzian & L$_{{\ell}ow}$ \\
Lower kHz QPO & L$_{\ell}$ \\
Upper kHz QPO & L$_u$ \\
\hline
 \\

  \end{tabular*}
  }}
  \caption{Identification scheme of power spectral components used in this paper, see Figure \ref{fig:Alta}.}
  \label{scheme}
\end{table}

\subsection{Power law fitting}
In order to compare our measured QPO frequencies to model predictions, we fit power laws (of the form $y=a\cdot x^{b}$) to the sample of ($\nu_{LT}$, $\nu_{K}$) measurements.

Our measurements have asymmetric errors in both coordinates. Rather than relying on an approximate method such as converting to logspace and symmetrizing the error bars to be able to use a 2-dimensional linear regression method, we introduce a new fitting method that is mathematically identical to performing a fit to all power spectra simultaneously with the frequency pair ($\nu_{LT}$, $\nu_K$) tied via a power law relation. This method is described in Appendix \ref{fitting}.

\begin{table}
\centering
  \tabcolsep=0.1cm
 {
\scalebox{1}{
\small
	\begin{tabular*}{\columnwidth}{@{\extracolsep{\fill}}lccc}
   \hline 
\vspace*{0.05cm}
 Source  & Spin (Hz) & Obs. in Archive  & Obs. used  \\
\hline
4U\ 1728--34 & 363 & 423 & 210 \\
4U\ 0614+09 & 415& 494 & 164\\
4U\ 1636--53 & 581& 1555 & 83 \\
4U\ 1702--43 & 329  &255 & 72\\
4U\ 1608--52 & 620  & 1072 & 43 \\
Aquila X-1 &  550  & 583 & 40\\
KS\ 1731--260 & 524 &  86 & 22\\
4U\ 1915--05 & 270  & 56 & 21\\
IGR\ J17191--2821 & 294 &  19 & 11\\
SAX\ J1750.8--2900 & 601  &  129 & 7\\
XTE\ J1739--285 & 1122&  9 & -\\
A\ 1744--361 & 530& 53 & -\\
EXO\ 0748--676 & 552 &  749 & -\\
    \hline 
  \end{tabular*}
  }}
  \caption{The sources included in our sample. The neutron star spin frequency was inferred from burst oscillations (or from intermittent pulsations in 4U\ 1636--53 and Aquila X-1). \citep{Watts:2012, Ritter:2003}. }
  \label{sources}
\end{table}

\section{Results}
\label{sec:results}
We present our best-fit power laws in Table \ref{fit}. 
We find that A\ 1744--361,  EXO\ 0748--676  and XTE\ J1739--285 are unsuitable for further analysis due to low signal to noise.
With our data selection and averaging criteria (see Section \ref{sec:selection}) we fail to detect the upper kHz QPO in these sources above a significance level of 2$\sigma$.

\subsection{Identification of power spectral features}
\label{ID}

In Figure \ref{fig:freq_all} we plot all measured  frequencies for all sources vs. the upper kHz QPO frequency. Since we use centroid frequency  ($\nu_{0}$), the correlations traced out differ from those based on $\nu_{\max}$. As a result, the  $\nu_{\max}$-based classification of (especially the broad) features does not necessarily agree with one that would have likely been used based on the centroid frequency correlations. To illustrate this we plot the same results converted to $\nu_{\max}$ in Figure \ref{fig:freq_numax}. Notable differences include the L$_{b}$ vs. L$_{b_2}$ and L$_{h}$ vs. L$_{hHz}$ identifications. 
Also note the more pronounced flattening of the centroid frequency relations below $\nu_u\sim$350 Hz as compared to $\nu_{\max}$. 
We comment below on how these issues could affect the $\nu_{LT}$-$\nu_K$ relations that are the subject of this paper. 

In order to fit power laws to the frequency correlations we divide the data into groups, as indicated by the ellipses in Figure \ref{fig:freq_all}. The precise frequency ranges defining these groups for each source can be found in Table \ref{tab:gr_All}. Group 1 is  composed of power spectra with a broad upper kHz QPO (see Section \ref{LFLF2H}), Group 3 comprises power spectra in which we do not simultaneously fit L$_h$ and L$_{hHz}$, or with $\nu_u$>1000 (see Sections \ref{hHz} and \ref{high} for more detail).
In Figures \ref{overview} and \ref{overview_2} we show the representative power spectra for 4U\ 1728--34, 4U\ 0614+09, 4U\ 1608--52, 4U\ 1636--53, 4U\ 1702--43 and Aquila X-1. The upper kHz QPO frequency ($\nu_u$) increases from $\sim$250 Hz in row A, via $\sim$500 Hz in row B and $\sim$700 Hz in row C, to $\sim$1000 Hz in row D. Clearly, the power spectra in each row are very similar. The colours of the best-fit Lorentzian components plotted correspond to those used for plotting their respective frequencies in Figures \ref{fig:freq_all} and \ref{fig:freq_numax}.\\
The centroid frequency of broad Lorentzians (L$_b$, L$_{b_2}$, L$_{\ell ow}$ and L$_{hHz}$) can be very small and even slightly negative; they are therefore not always present in our logarithmic centroid frequency-frequency plots. When a negative centroid frequency occurs in a fit, we fix it to 0.

\subsubsection{The break and second break Lorentzians}
 We identify L$_b$ and L$_{b_2}$ by their appearance in the power spectrum. At $\nu_u$<700 Hz (rows A-C in Figure \ref{overview}), L$_b$ (\textit{green}) is the broad component with the lowest frequency. For $\nu_u$>700 Hz (row D in Figure \ref{overview}), Q$_b$ increases and a separate broad low frequency component is needed to obtain a satisfactory power spectral fit, this is L$_{b_2}$ (\textit{dark blue}).  
So, the identification of L$_b$ and L$_{b_2}$ is straightforward and confirmed by the $\nu_{u,\max}$ vs. $\nu_{b,\max}$, $\nu_{u,\max}$ vs. $\nu_{b_2,\max}$ frequency correlations, see Figure \ref{fig:freq_numax}. In order to keep the link to earlier works, we maintain the $\nu_{\max}$-based identifications for these low frequency features, in spite of the different behaviour of $\nu_b$ above and below $\nu_u\sim$700 Hz when plotting centroid frequencies (see Figure \ref{fig:freq_all}). Since Q$_b$ increases for $\nu_u$>700 Hz, we regard L$_b$ as a candidate for precession and fit power laws to the $\nu_b$-$\nu_u$ frequency pairs when Q$_b$>0.5. The second break Lorentzian, L$_{b_2}$, is a broad feature that is often best characterized by a zero-centered Lorentzian. 


\subsubsection{The LF, LF$_2$ QPOs and hump Lorentzians}
\label{LFLF2H}
\begin{figure*}

	\includegraphics[width=5in]{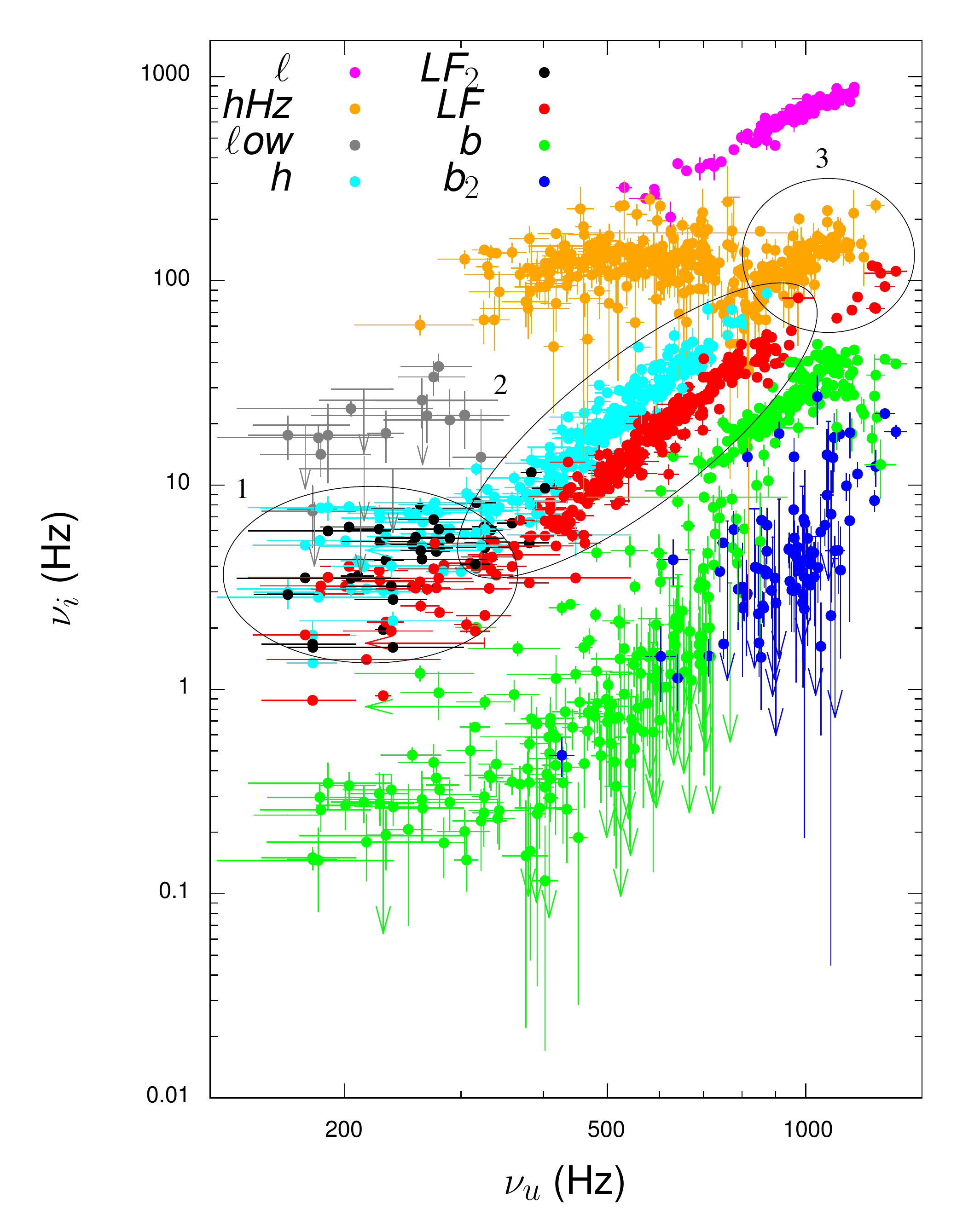}
    \caption{Frequencies of all sources plotted against the upper kHz QPO frequency. We fit power laws to distinct groups that appear to be present in both the LF QPO and hump feature. For clarity, we indicate the 98$\%$ upper limit to frequencies of broad $\ell ow$, hHz, b and b$_2$ Lorentzians with arrows.}
    \label{fig:freq_all}
\end{figure*}

\begin{figure*}
	\includegraphics[width=5in]{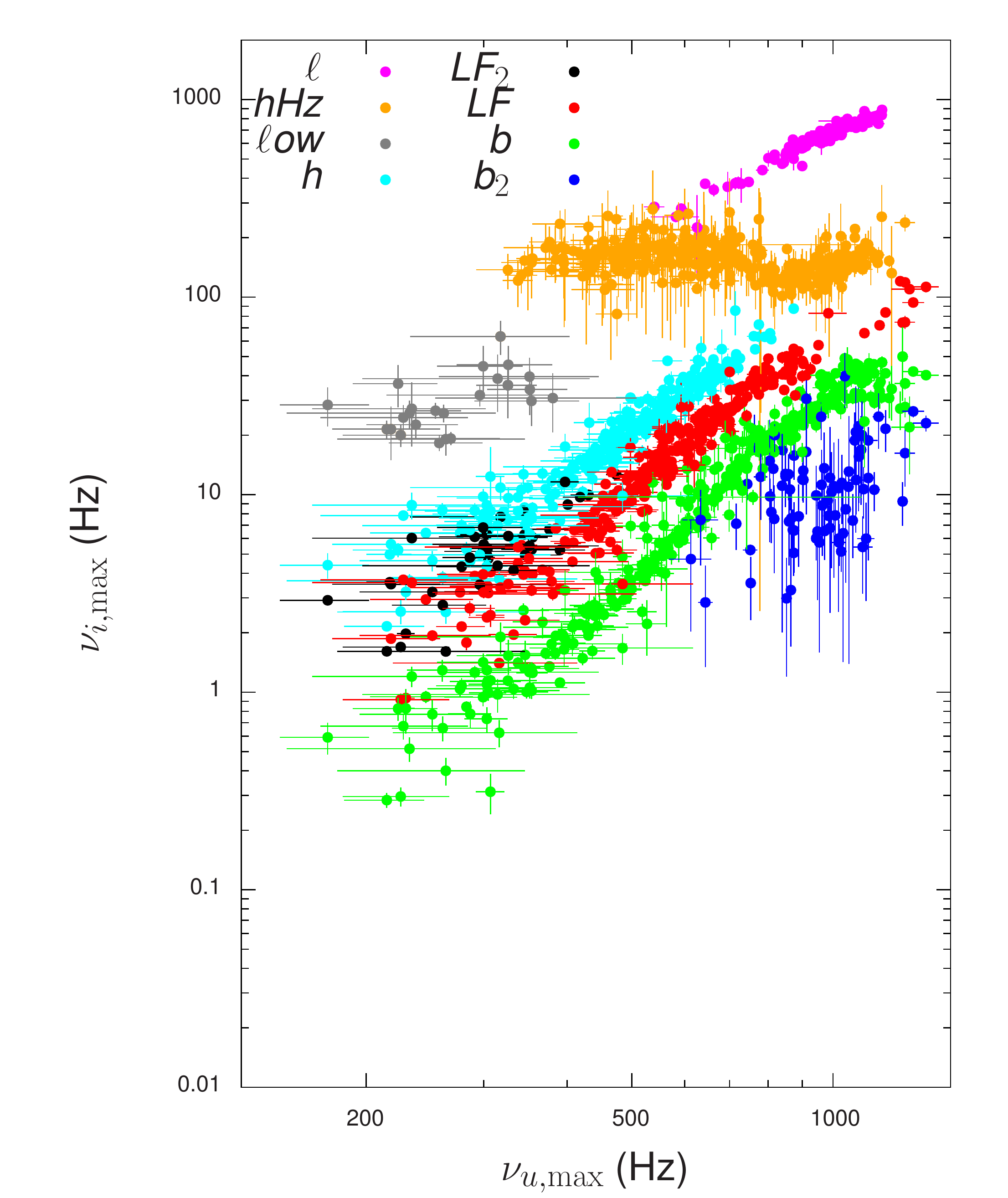}
    \caption{As in Figure \ref{fig:freq_all}, but with $\nu_{\max}$.}
    \label{fig:freq_numax}
\end{figure*}

In rows B and C ($\nu_u\sim$500 Hz), Group 2 in Figure \ref{fig:freq_all}, the identification of L$_{LF}$ (\textit{red}) and L$_h$ (\textit{cyan}) again is straightforward. Our power spectra closely resemble those previously reported in the literature  (see Figure \ref{fig:Alta}) and $\nu_h$ and $\nu_{LF}$ fall on different correlations (with always $\nu_h$>$\nu_{LF}$) both in $\nu_{\max}$ and $\nu_0$ \citep{vanStraaten:2002, Altamirano:2008}.
 We find that for $\nu_u$> 400 Hz, Q$_h$ increases, Q$_{LF}$ stays roughly constant,  rms$_h$ decreases, and  rms$_{LF}$ reaches a maximum at $\nu_u\sim$600 Hz and then decreases as $\nu_u$ increases (see Figures \ref{fig:rms_All} and \ref{fig:Q_All}). 


\begin{figure}
	\includegraphics[width=0.9\columnwidth]{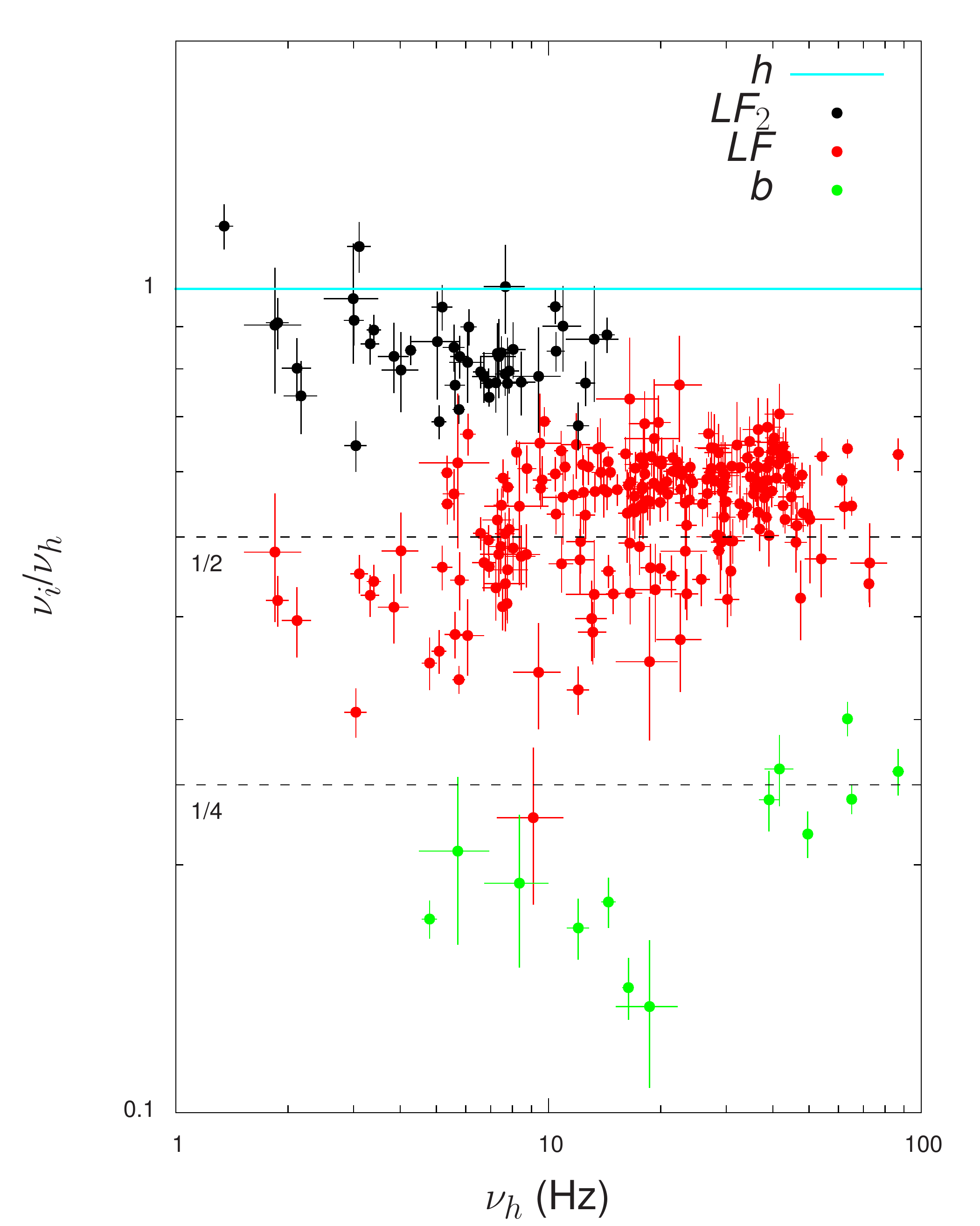}
    \caption{Best-fit $\nu_{LF_2}$, $\nu_{LF}$ and $\nu_{b}$ of all sources divided by $\nu_{h}$, plotted vs. $\nu_h$.} 
    \label{fig:LF_H}
\end{figure}

\begin{figure}
	\includegraphics[width=0.9\columnwidth]{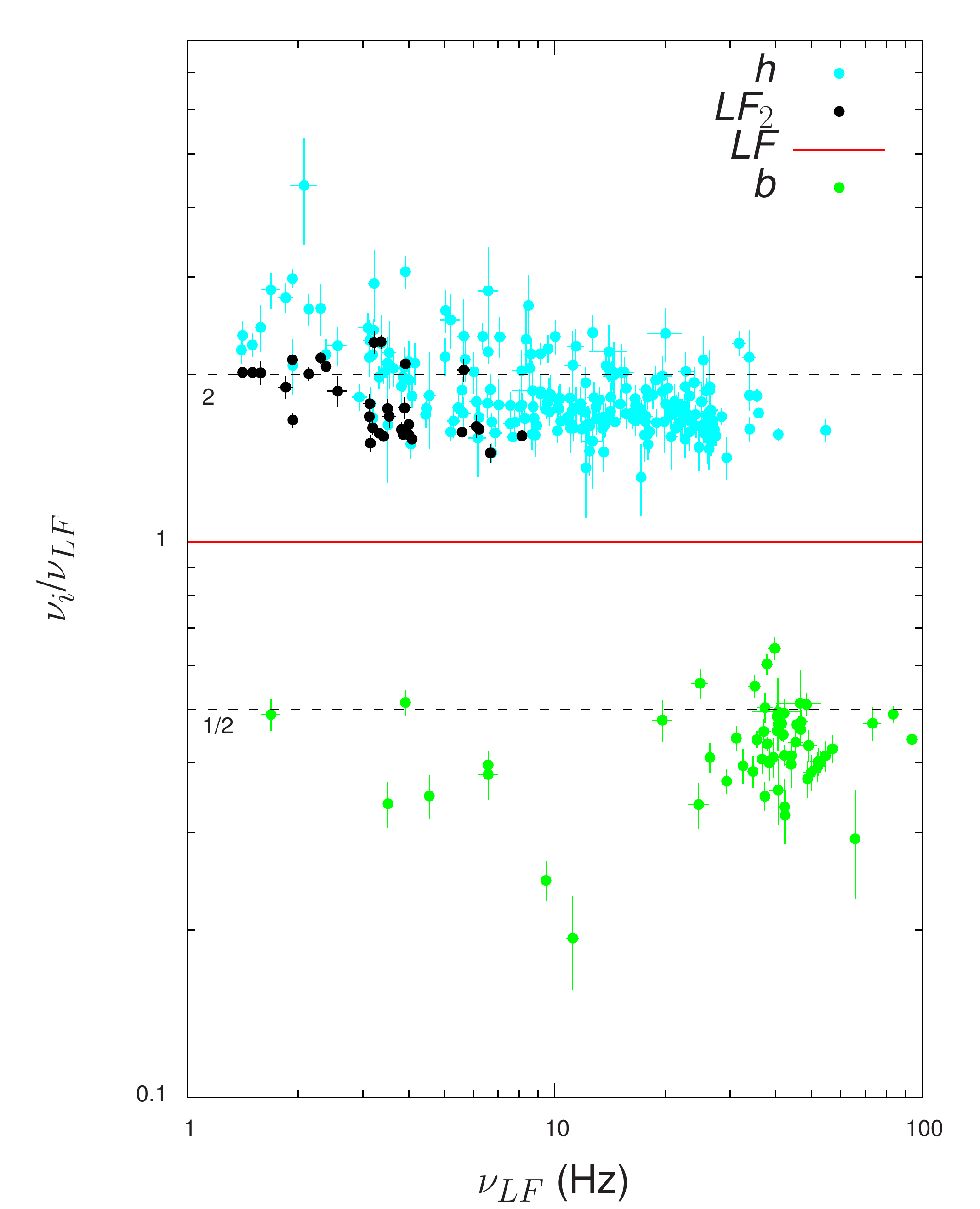}
    \caption{Best-fit $\nu_{LF_2}$, $\nu_{h}$ and $\nu_{b}$ of all sources divided by $\nu_{LF}$, plotted vs. $\nu_{LF}$.} 
    \label{fig:LF_LF}
\end{figure}

 The correlations traced out by $\nu_h$ and $\nu_{LF}$ flatten in some sources when $\nu_{u}$<400 Hz (For instance in 4U\ 1728--34, Group 1, see Figure \ref{fig:freq_1728}). This effect is less pronounced in the $\nu_{\max}$ representation. It can be explained by L$_u$ becoming broader at low (<400 Hz) frequency (compare L$_u$ (\textit{yellow}) in row A to L$_u$ in row B in Figures \ref{overview} and \ref{overview_2}).\\ 
 When we fit two narrow QPOs and a broad component at low frequency we identify them, in order of increasing centroid frequency, as $L_{LF}$ (\textit{red}), L$_{LF_2}$ (\textit{black}), and L$_h$ (\textit{cyan}).  
 As can be seen in Figures \ref{overview} and \ref{overview_2} (for instance, in row A, compare 4U\ 1728--34 and Aquila X-1), L$_{LF}$ and L$_{LF_2}$ are not always simultaneously present. In Figure  \ref{fig:Q_All} we see that the Q-values of L$_{LF}$ and L$_{LF_2}$ are similar for $\nu_u$<400 Hz. Identification based solely on the appearance of these features in the power spectrum is therefore not straightforward. To resolve the ambiguity in this identification, we use the respective correlations of $\nu_{LF_2}$ and $\nu_{LF}$ with $\nu_h$ as an additional tool to correctly differentiate between these low frequency QPOs (see Figure \ref{fig:LF_H}). 
We note that centroid frequencies of L$_b$, L$_{LF}$, L$_{LF_2}$, and L$_h$ roughly follow $\nu_{LF}$/$\nu_{h}$=0.4-0.7, $\nu_{LF_2}$$\approx$$\nu_h$ and $\nu_{b}$/$\nu_h$=0.15-0.3. As can be seen from Figures \ref{fig:LF_H} and \ref{fig:LF_LF}, there is no strong evidence for harmonic relations between any two components. Furthermore, for L$_{LF}$ and L$_h$, rms and Q behave differently when plotted vs. $\nu_u$ (see Figures \ref{fig:rms_All} and \ref{fig:Q_All}) suggesting a different physical origin for these features. 
We discuss the identification of L$_{LF}$ in Group 3 in section \ref{high}.

 \subsubsection{HectoHz, low frequency Lorentzians and kHz QPOs}
 \label{hHz}
\begin{figure*}

\includegraphics[width=6.5in]{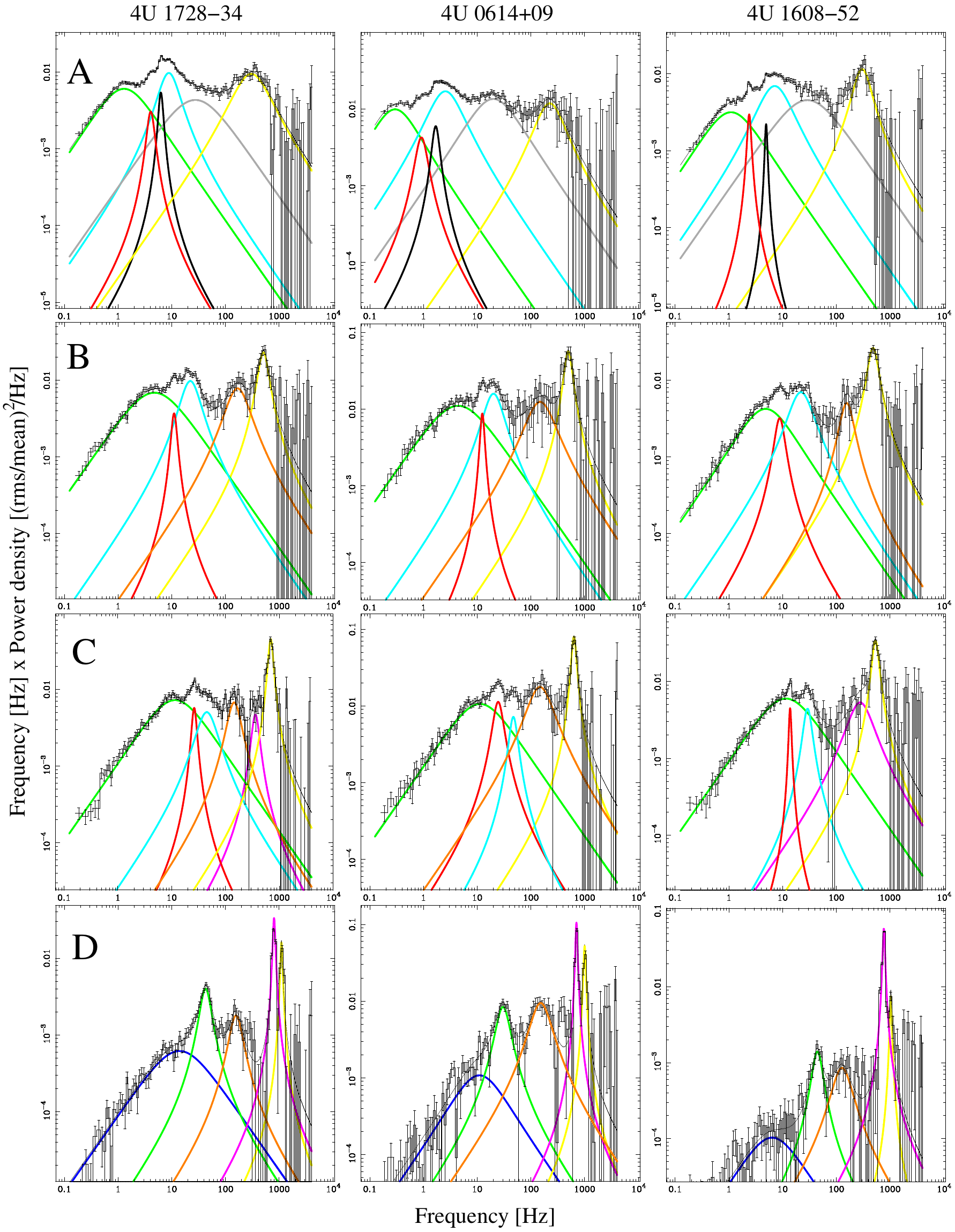}
\caption{Representative power spectra of 4U\ 1728--34, 4U\ 0614+09 and 4U\ 1608--52 in different accretion states. Colours correspond with Figure \ref{fig:freq_all}. The frequency of the upper kHz QPO (\textit{yellow}) increases from $\sim$250 Hz in row A, via $\sim$500 Hz in row B and $\sim$700 Hz in row C, to $\sim$1000 Hz in row D. The low frequency QPO (L$_{LF}$, \textit{red}) and the hump component (L$_h$, \textit{cyan}) are present in rows A-C. The low frequency QPO is accompanied by its 'harmonic' (L$_{LF_2}$, \textit{black}) in row A. The lower kHz QPO (L$_{\ell}$, \textit{magenta}) shows up in row D, as well as an extra low frequency noise component (L$_{b_2}$, \textit{dark blue}). The break and hHz noise components L$_{b}$ (\textit{green}) and L$_{hHz}$ (\textit{orange}) are present in all power spectra shown here, L$_{\ell ow}$ (\textit{gray}) is only present in row A. The observations used in this figure are listed in Table \ref{tab:fig_obs}.}
\label{overview}
\end{figure*}

\begin{figure*}
\includegraphics[width=6.5in]{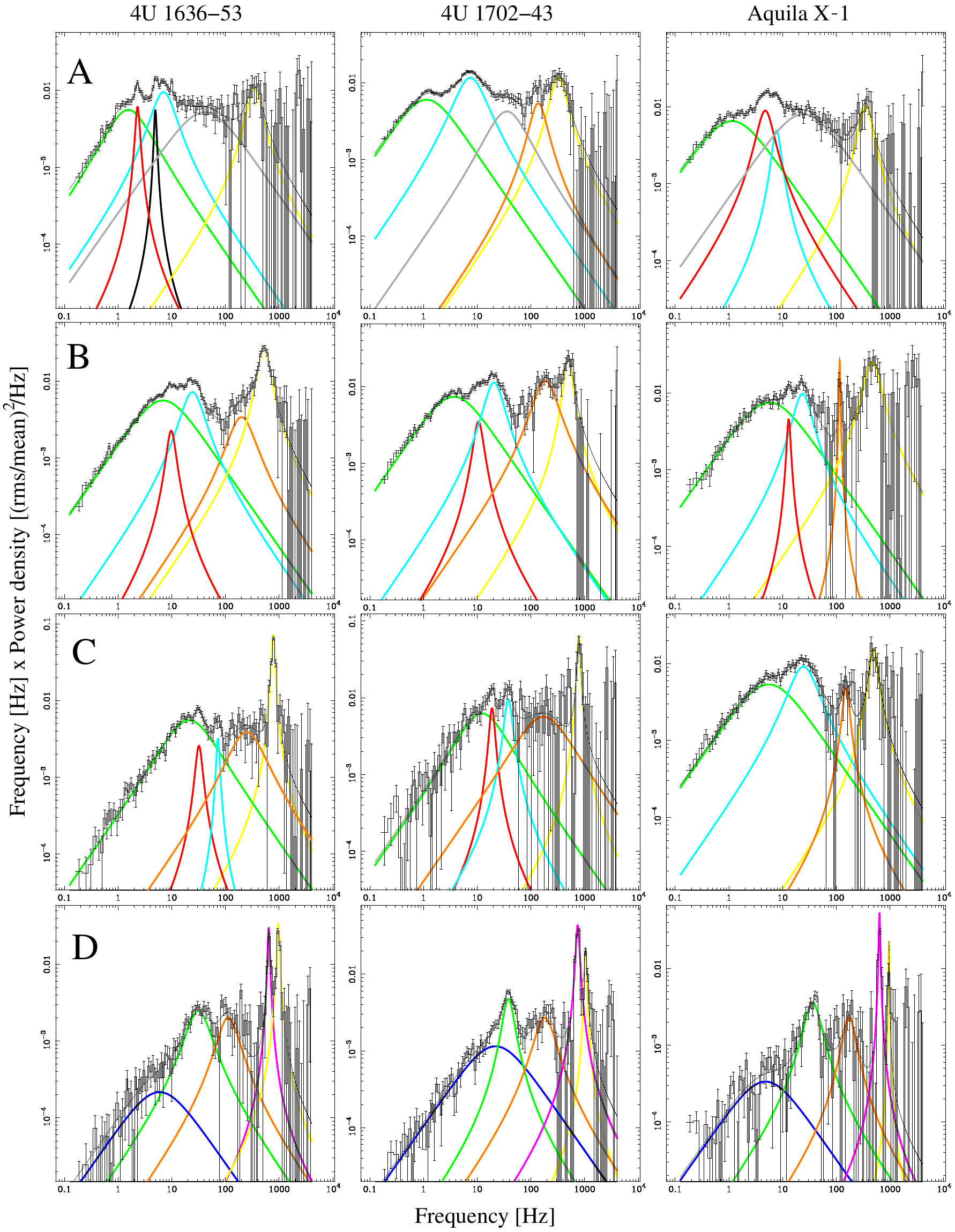}
\caption{As in Figure \ref{overview}, but for 4U\ 1636--53, 4U\ 1702--43 and Aquila X-1. }
\label{overview_2}
\end{figure*}

 We identify the broad component with $\nu\sim$100 Hz in rows B-C of Figures \ref{overview} and \ref{overview_2} as L$_{hHz}$ (\textit{orange}). 
 We find that for $\nu_u>$700 Hz, i.e. in Group 3, Q$_{hHz}$ increases while $\nu_{hHz}$ falls on the extrapolation of the correlation traced out by $\nu_h$-$\nu_u$. In the $\nu_{\max}$ representation this is not obvious. We do not detect a separate L$_h$ (with an expected frequency around $\sim$100 Hz) in power spectra with $\nu_u>$700 Hz;  we therefore suspect  that we are fitting a blend of L$_h$ and L$_{hHz}$ there.
 The low frequency Lorentzian (L$_{\ell ow}$, \textit{gray}) is a broad feature with $\nu_{hHz}>\nu_{\ell ow}$>$\nu_h$ and Q$_{\ell ow}$>Q$_h$ which is occasionally needed to fit the broad band noise between L$_h$ and L$_u$ in power spectra such as illustrated in row A of Figures \ref{overview} and \ref{overview_2}.
 Finally, of the kHz QPOs, L$_{u}$ (\textit{yellow}) is always the component with the highest frequency in the power spectrum. The lower kHz QPO L$_{\ell}$ (\textit{magenta}) appears in rows C and D in Figures \ref{overview} and \ref{overview_2}. Always, $\nu_u$>$\nu_{\ell}$>$\nu_{hHz}$ and also Q$_{\ell}$>Q$_{hHz}$.

 \subsubsection{Power spectra with high $\nu_u$}
 \label{high}
\begin{figure*}

\includegraphics[width=6in]{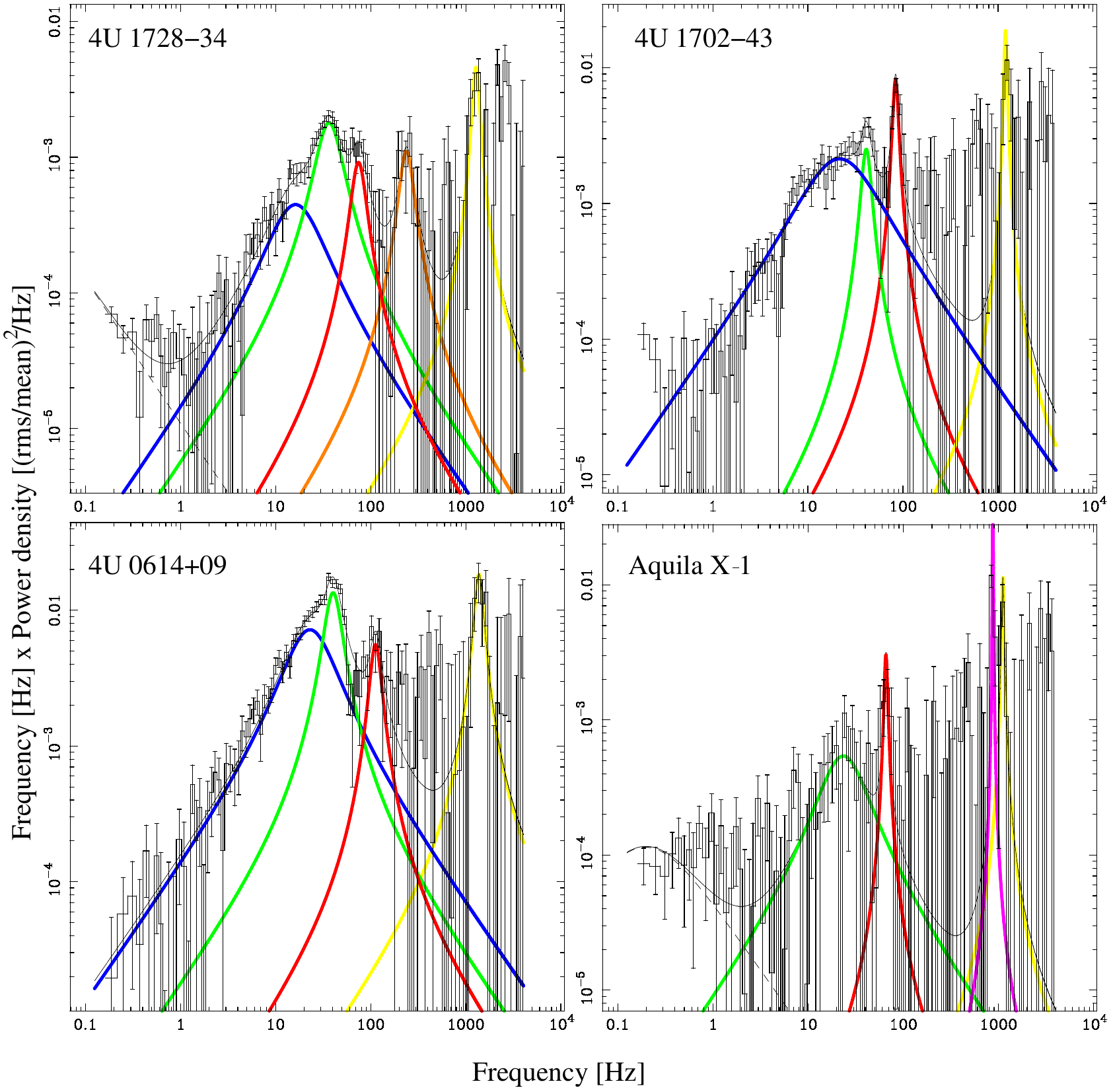}
\caption{Examples of power spectra of 4U\ 1728--34, 4U\ 0614+09, 4U\ 1702--43 and Aquila X-1 with the highest $\nu_u$, all found at high luminosity. The frequencies fall in Group 3, shown in Figure \ref{fig:freq_all}. The power spectra have similar upper kHz QPOs (\textit{yellow}) with $\nu_{u}>1000$ Hz, and are accompanied by QPOs with 70$<\nu<$100 Hz. We tentatively identify the latter as $\nu_{LF}$ (\textit{red}), see text. The observations used in this figure are listed in Table \ref{tab:fig_obs}, in the column "High $\nu_u$". }
\label{overview_high}
\end{figure*}

 We find characteristic power spectra with $\nu_u$ in the 1000--1400 Hz range (in Group 3 of Figure \ref{fig:freq_all}) for 4U\ 1728--34 (3 cases), 4U\ 0614+09 (5 cases), 4U1702--43 (1 case) and Aquila X-1 (1 case; see Figure \ref{overview_high} for examples), at relatively high luminosity and low hard color. They are similar to the power spectra illustrated in row D of Figures \ref{overview} and \ref{overview_2}, but need to be fitted with an extra Lorentzian below 100 Hz.
 As data are sparse for these high values of $\nu_u$ we have to extrapolate the frequency and rms trends for the feature identification. 
The centroid frequency of the extra Lorentzian falls between $\nu_{hHz}$ (\textit{orange}) and $\nu_b$ (\textit{green}), and roughly on the extrapolation of the $\nu_{LF}$-$\nu_u$ correlation when plotted vs. $\nu_u$.
We tentatively identify this component as $\nu_{LF}$ (\textit{red}).
In 4U\ 1728--34 we identify the Lorentzian with a centroid frequency of 234.5 Hz as L$_{hHz}$ (\textit{orange}, top panel of Figure \ref{overview_high}). Its frequency falls on the extrapolation of the $\nu_{h}$--$\nu_u$ correlation (Figure \ref{fig:freq_all}), however when regarding the Lorentzian as L$_{h}$, $\nu_{LF}$ does not behave as expected from Figure \ref{fig:LF_H}. We consider the identifications of L$_{LF}$ and L$_{hHz}$ in this case as ambiguous.

We use the extrapolation of the $\nu_b$-$\nu_u$ and $\nu_{b_2}$-$\nu_u$ correlations at lower frequencies for the identification of L$_b$ (\textit{green}) and L$_{b_2}$ (\textit{dark blue}) in the power spectra with $\nu_u$>1000 Hz, see Figure \ref{fig:freq_all}.


\subsubsection{Best-fit power laws}
\label{sec:bestfit}
\begin{figure*}
\includegraphics[scale=1]{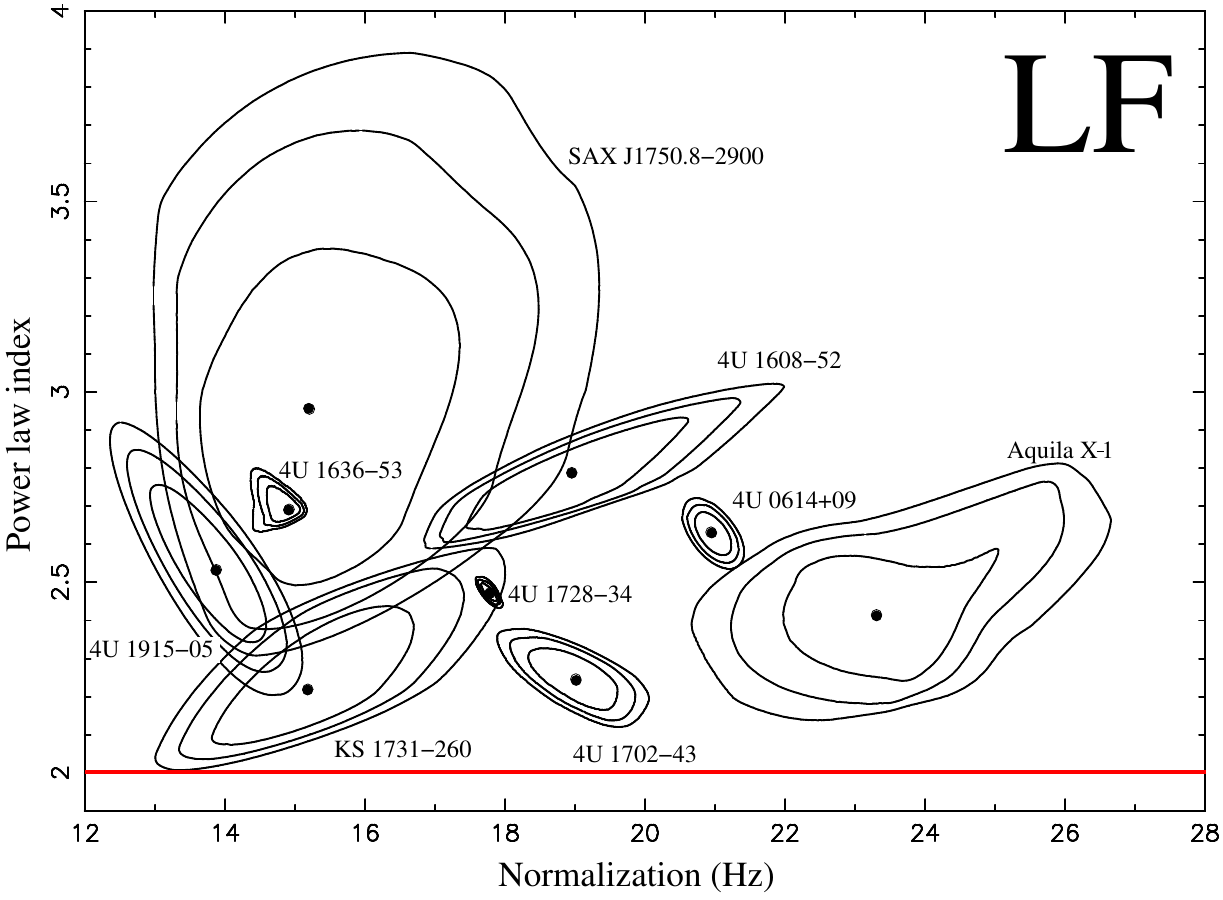}
\caption{Confidence contours for the best-fit power law index and normalization at $\nu_u$=600 Hz of the $\nu_{LF}$-$\nu_u$ correlations in each source, with $\nu_{LF}$ in Group 2 (see Table \ref{fit}). We plot the 75$\%$ (inner), 95$\%$ (middle) and 99$\%$ (outer) two-parameter confidence limits (corresponding to $\Delta\chi^2$=3.13,6.17 and 9.21, respectively). The best-fit values are indicated by the dots, the red line corresponds a power law index of 2. }
\label{fig:contLF}
\end{figure*}

\begin{figure*}
\includegraphics[scale=1]{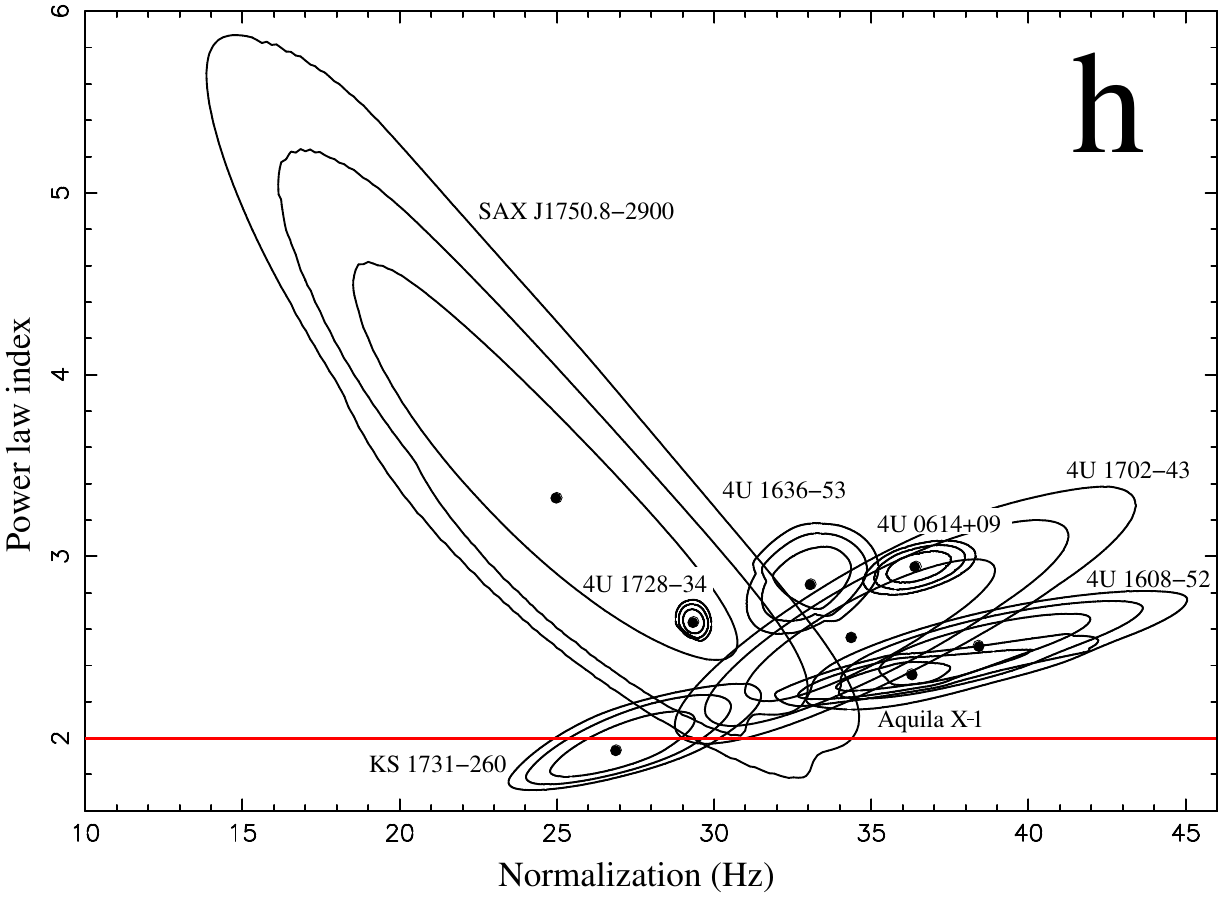}
\caption{As in Figure \ref{fig:contLF}, but for the $\nu_{h}$-$\nu_u$ correlations (with $\nu_h$ in Group 2). }
\label{fig:conth}
\end{figure*}

\begin{figure*}
\includegraphics[scale=1]{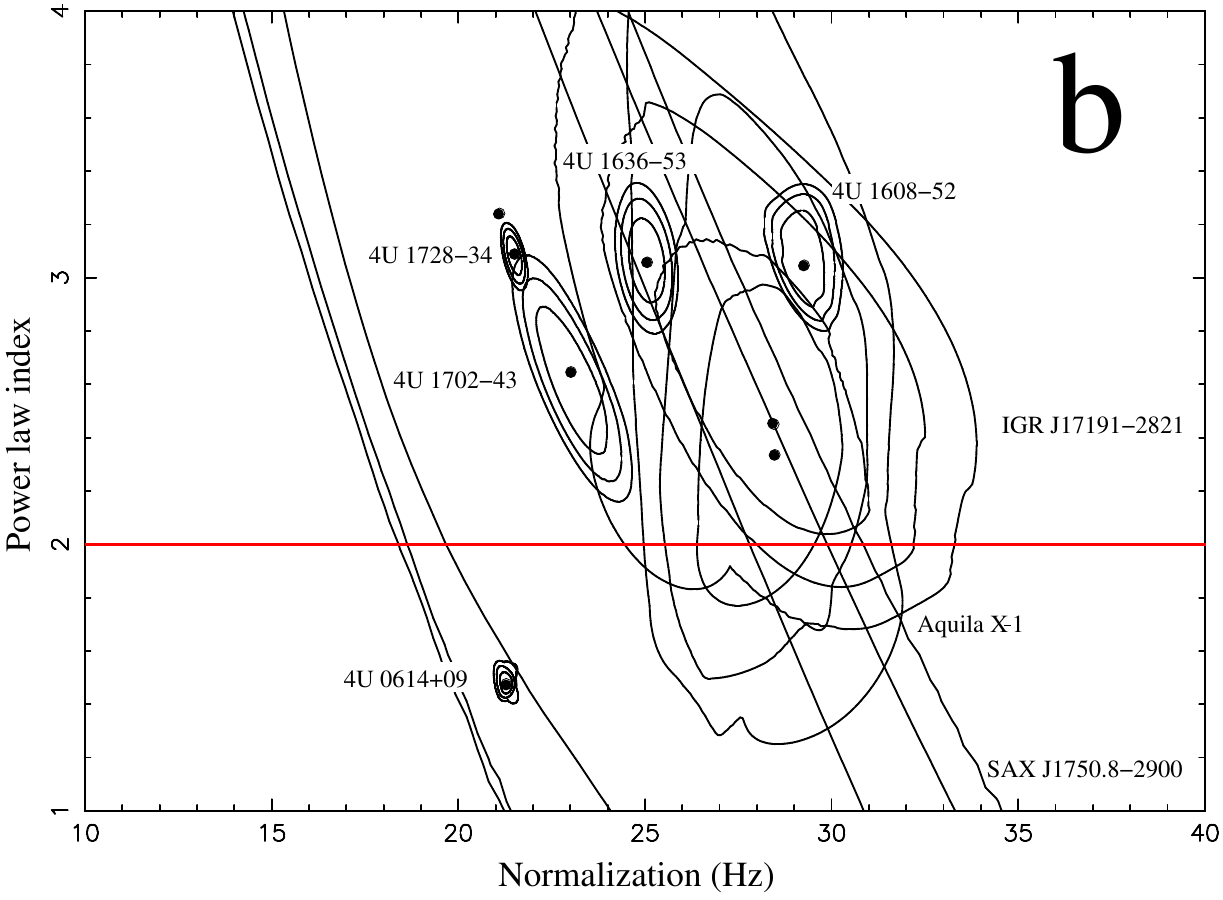}
\caption{As in Figure \ref{fig:contLF}, but with the normalization at 900 Hz, for the $\nu_b$-$\nu_u$ correlations including $\nu_b$ only when Q$_b$>0.5. }
\label{fig:contb}
\end{figure*}

In Table \ref{fit} we present the parameters of the best-fit power laws to our frequency groups specified in Table \ref{tab:gr_All}. 
In the last column we quote the upper and lower limits on power law indices and associated confidence levels to illustrate the deviation of the best-fit power law index from the RPM-prediction of 2.0. When we assess this significance, we cap the fraction of extrapolated $\chi^2$ maps used for our power law fits (see Appendix \ref{fitting}) to <10$\%$, which leads to different values for the limits quoted.
Overall, the best-fit power law index exceeds 2.0 in 80$\%$ of the cases at >3$\sigma$. In 4U 1728--34 and 4U\ 0614+09 the best-fit power law index to $\nu_{LF}$ and $\nu_h$ in Group 2 is in excess of 2.2 at >10$\sigma$.
 \\\\
The best-fit power laws to frequencies in Group 2 significantly differ from one another, see Figures \ref{fig:contLF}, \ref{fig:conth}, \ref{fig:contb} where we show the confidence contours for the best-fit power law index and normalization.\\\\

Interestingly, power laws fitted to $\nu_{LF}$ in Groups 2 and 3 combined in 4U\ 1728--34, 4U\ 0614+09, 4U\ 1702--43 and Aquila X-1 all have indices that are somewhat lower than the power laws fitted to $\nu_{LF}$ in Group 2 alone (for $\nu_h$ this is not the case). This might indicate a change in power law index as $\nu_u$ increases. However, the differences in index are not very significant.
To further test this possibility, and also because the identification of $\nu_{LF}$ in Group 3 is somewhat ambiguous (as explained in Section \ref{high}), we fit power laws to frequencies below (Group 2a) and above (Group 2b) $\nu_u$=600 Hz in Group 2 in 4U 1728--34 and 4U 0614+09. We use these two sources as they have the largest data sets. In 4U\ 1728--34, we indeed find that the power law index is lower for $\nu_{LF}$ in Group 2b than 2a, see Table \ref{fit} (at 2.28$\pm$0.03 it is still in excess of 2.1 at 7$\sigma$). For 4U\ 0614+09 no significant flattening towards higher frequency is detected by this method. \\

Although we obtain a reduced $\chi^2$ of $\sim$1, the corresponding formal probabilities are typically low. This is the result of a combination of small deviations of the power spectral model from the data and fitting many power spectra simultaneously. The fits to individual power spectra are all acceptable, see Table \ref{tab:pars} for the $\chi^2$/$dof$.


\begin{figure}
	\includegraphics[width=0.9\columnwidth]{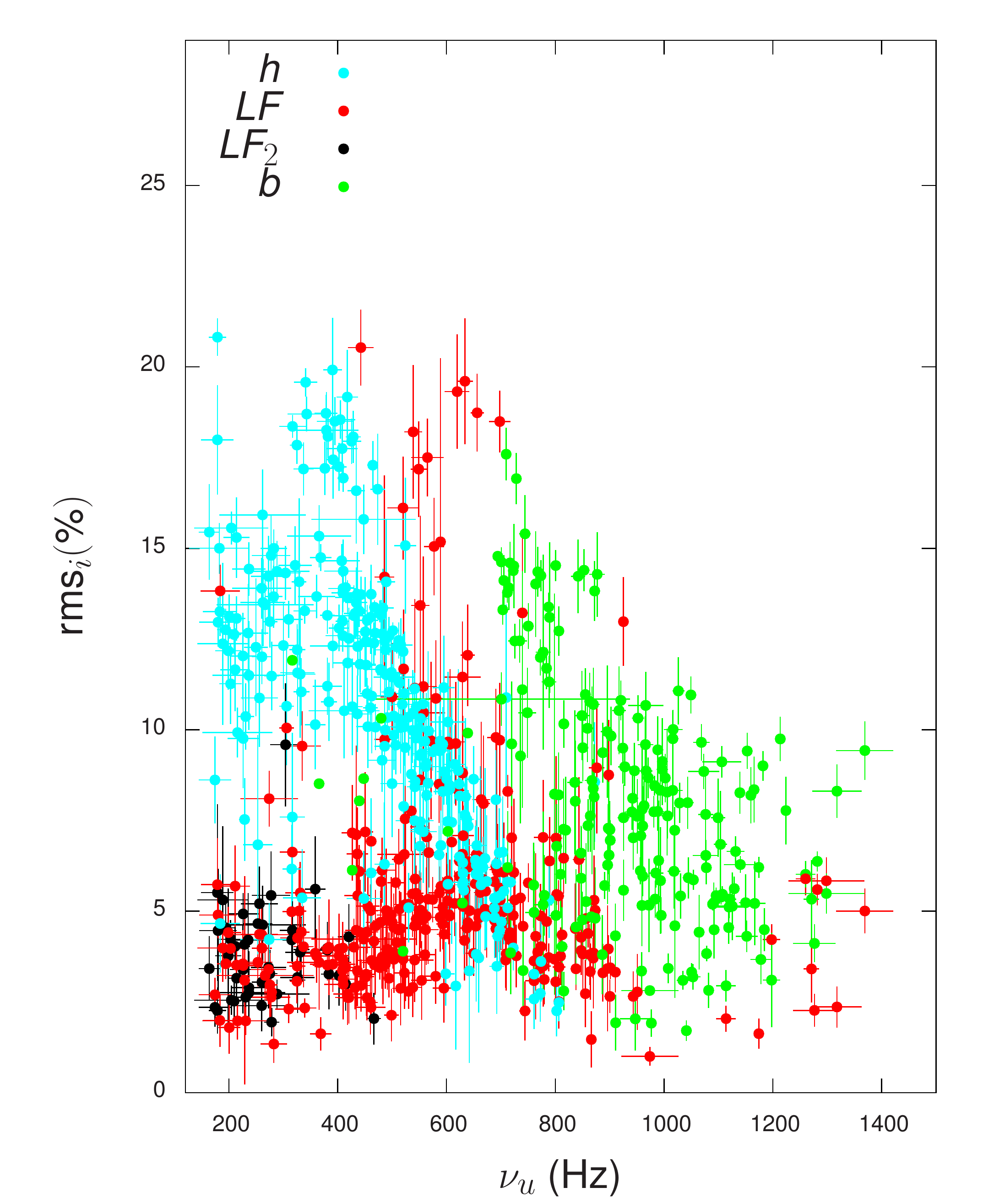}
    \caption{Fractional rms levels of L$_u$, L$_h$, L$_{LF}$, L$_{LF_2}$, and L$_{b}$ (L$_{b}$ for Q$_b$>0.5 only) vs. upper kHz QPO frequency in all sources.} 
    \label{fig:rms_All}
\end{figure}

\begin{figure}
	\includegraphics[width=0.9\columnwidth]{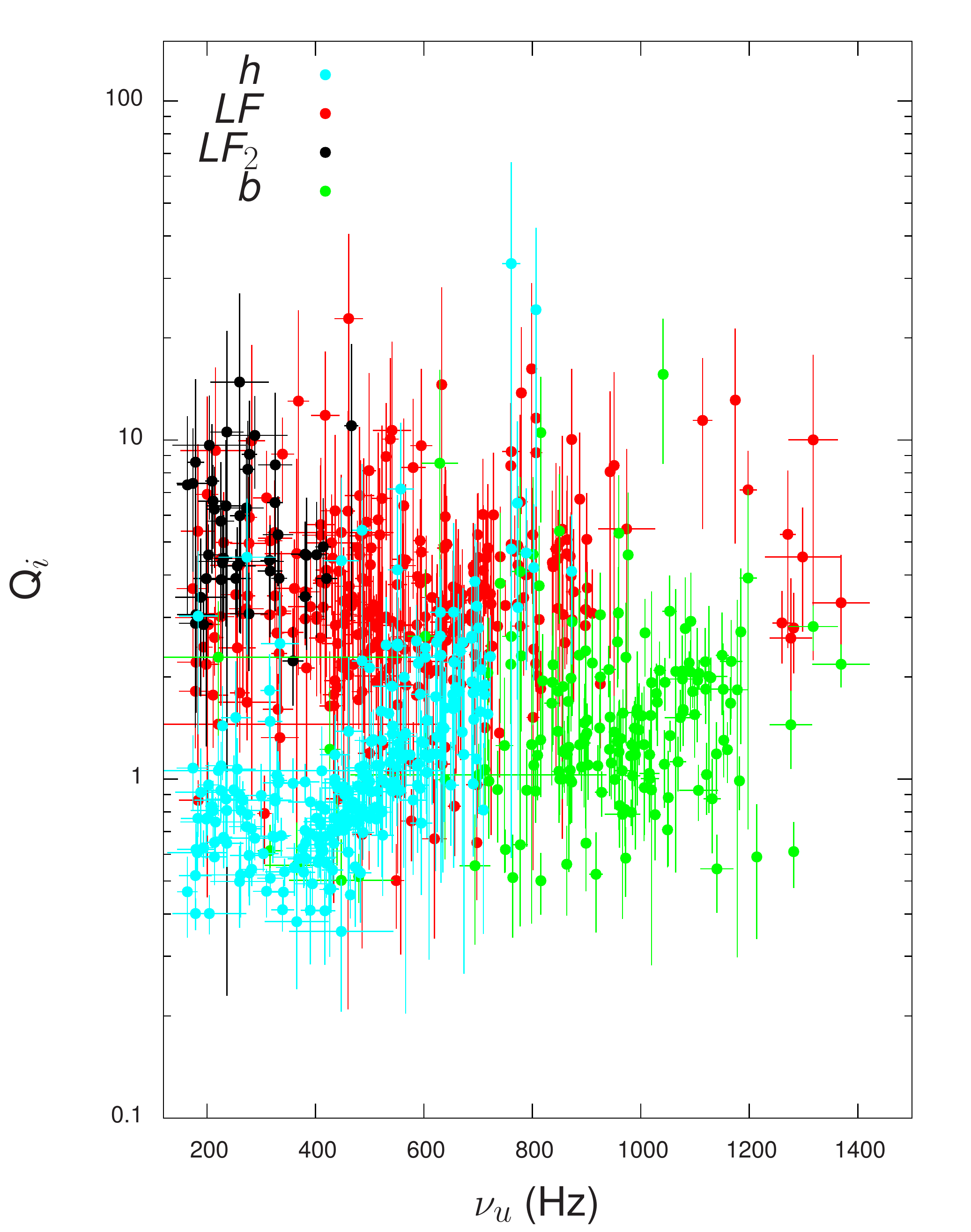}
    \caption{Q-factors of L$_u$, L$_h$, L$_{LF}$, L$_{LF_2}$, and L$_{b}$ (L$_{b}$ for Q$_b$>0.5 only) vs. upper kHz QPO frequency in all sources.} 
    \label{fig:Q_All}
\end{figure}

\begin{table*}
\centering
  \tabcolsep=0.1cm
 {
\scalebox{0.8}{
\small
  \begin{tabular}{l c c c c c c c }

   \hline 
\vspace*{0.05cm}
 Source  & L$_{i}$ & Group & No. of freq. & Norm. (Hz) & Index  & Reduced & Limit on index (significance) \\
  & & &  pairs ($\nu_i$, $\nu_u$) & at $\nu_u$=600 Hz & &  $\chi^2$ ($dof$)  & \\
\hline
4U\ 1728--34 & LF & 1 & 12 & 5.1$\pm{0.2}$  & 0.43$^{+0.04}_{-0.06}$ & 1.10 (3864) & <0.8 (3.3$\sigma$) \\[0.1cm]
\vspace*{0.1cm}
 &           h  & 1 & 11 & 11.2$\pm{1.1}$ & 0.66$\pm$0.13 & 1.08 (3541) & <2.0 (5.0$\sigma$)\\
\vspace*{0.05cm}
  &          LF   & 2 & 133 & 17.79$^{+0.04}_{-0.07}$ & 2.47$\pm$0.01& 1.05 (43389) & >2.2 (23.9$\sigma$)\\
\vspace*{0.05cm}
 &          h    & 2 & 103 & 29.4$\pm{0.2}$ & 2.64$^{+0.04}_{-0.03}$ & 1.05 (33626) & >2.2 (13.4$\sigma$)\\
\vspace*{0.05cm}
 &          LF   & 2+3 & 136 & 17.8$\pm{0.1}$ & 2.45$\pm$0.01 & 1.05 (44509) & >2.2 (21.2$\sigma$ )\\
 \vspace*{0.05cm}
 &          LF   & 2a & 56 & 5.9$^{+0.1}_{-0.2}$* & 2.76$^{+0.11}_{-0.08}$ & 1.08 (18258) & >2.5 (10.0$\sigma$) \\
  \vspace*{0.05cm}
 &          LF   & 2b & 77 & 36.0$\pm{0.1}$** & 2.28$\pm{0.03}$ & 1.03 (25133) &  >2.1 (7.0$\sigma$) \\
\vspace*{0.05cm}
 &          h   & 2+3 & 161 & 29.5$\pm{0.2}$ & 2.75$\pm$0.02 & 1.06 (53366) & >2.4 (17.0$\sigma$) \\
\vspace*{0.05cm}
 &          b   &  -   & 62 & 6.1$\pm{0.1}$ &3.09$^{+0.03}_{-0.04}$ & 1.07 (21573) & >2.6 (12.6$\sigma$) \\
     \hline
 4U\ 0614+09  &    LF  & 2  & 70 & 20.9$^{+0.2}_{-0.1}$  &   2.63$^{+0.03}_{-0.04}$ & 1.04 (22708) & >2.2 (13.4$\sigma$ )\\ [0.1cm]
\vspace*{0.05cm}
  &  h  &  2  & 55 &  36.4$^{+0.6}_{-0.5}$ &   2.95$\pm$0.05  & 1.06 (17785) & >2.4 (11.4$\sigma$) \\
\vspace*{0.05cm}
 &  LF  & 2+3&  75 & 21.0$\pm$0.1  & 2.58$^{+0.03}_{-0.02}$ & 1.05 (24772) & >2.2 (14.1$\sigma$) \\
\vspace*{0.05cm}
 &          LF   & 2a & 32 & 7.1$^{+0.2}_{-0.1}$* & 2.69$\pm{0.07}$ & 1.06 (10497) &  >2.3 (5.5$\sigma$)\\
  \vspace*{0.05cm}
 &          LF   & 2b & 38 & 44.4$\pm{0.4}$** & 2.57$\pm{0.06}$ & 1.02 (12213) & >2.2 (6.3$\sigma$) \\
\vspace*{0.05cm}
 &  h   &  2+3 &   98 &  34.8$^{+0.3}_{-0.4}$   &    2.70$^{+0.03}_{-0.02}$    & 1.05 (32346) & >2.5 (9.5$\sigma$)\\
\vspace*{0.05cm}
 &          b  &  -   & 58  & 11.7$\pm{0.1}$ &1.48$\pm$0.02 & 1.04 (19926) & <2.0 (14.1$\sigma$) \\
    \hline
 4U\ 1608--52 & LF  & 2   & 14  & 19.0$\pm{0.9}$ &  2.78$^{+0.09}_{-0.11}$ & 1.06 (4388) & >2.5 (6.3$\sigma$)\\ [0.1cm]
\vspace*{0.05cm}
 &  h &  2  & 14 &  38.4$^{+1.9}_{-1.8}$ &   2.51$\pm$0.10  & 1.05 (4388) & >2.1 (3.5$\sigma$) \\
\vspace*{0.05cm}
   &  h &  2+3  & 23 &   41.2$\pm{0.8}$ &  2.56$\pm$0.03  & 1.40 (8136) & >2.2 (7.7$\sigma$)  \\
\vspace*{0.05cm}
     &          b  &  -   & 15  & 8.5$^{+0.3}_{-0.4}$ &3.05$^{+0.09}_{-0.07}$ & 1.58 (5722) & >2.6 (5.7$\sigma$)\\
\hline
 4U\ 1702--43  & LF   & 2  & 15  &  19.4$\pm$0.4 & 2.51$\pm$0.07 & 1.07 (4882) &  >2.2 (3.9$\sigma$) \\[0.1cm]
\vspace*{0.05cm}
  &  h & 2  &  12  &     34.4$^{+2.5}_{-1.9}$    &   2.56$^{+0.23}_{-0.20}$ & 1.06 (3902) & >2.1 (3.6$\sigma$) \\
\vspace*{0.05cm}
    & LF   & 2+3  & 16   &  19.0$^{+0.3}_{-0.4}$ & 2.25$\pm$0.05 & 1.08 (5210) & >2.0 (3.2$\sigma$) \\
\vspace*{0.05cm}
  &  h & 2+3  &  20  &     34.8$\pm{0.9}$  &  2.59$\pm$0.07         & 1.08 (6561) & >2.2 (5.4$\sigma$)\\
\vspace*{0.05cm}
  &          b   &  -   & 10  & 7.9$^{+0.8}_{-0.6}$ &2.64$^{+0.13}_{-0.18}$ & 1.09 (3314) & >2.0 (4.5$\sigma$)\\
\hline
  4U\ 1636--53  & LF    & 2  & 24   & 14.9$^{+0.1}_{-0.2}$  & 2.69$^{+0.03}_{-0.02}$ & 1.01 (7734) & >2.6 (4.5$\sigma$)\\[0.1cm] 
\vspace*{0.05cm}
     &  h  & 2  &  22  &    33.1$^{+0.7}_{-0.8}$ &   2.85$^{+0.12}_{-0.09}$& 1.01 (7154) & >2.5 (3.7$\sigma$)\\
\vspace*{0.05cm}
    &  h & 2+3  &  35  & 33.4$^{+0.5}_{-0.6}$  &   2.91$^{+0.04}_{-0.03}$ & 1.06 (11381) & >2.5 (10$\sigma$) \\
\vspace*{0.05cm}
    &          b    &  -   & 17  & 7.3$\pm{0.3}$ &3.05$^{+0.01}_{-0.08}$ & 1.15 (5464) & >2.8 (3.2$\sigma$)\\
\hline
  4U\ 1915--05 & LF    & 2  & 15   &13.9$^{+0.4}_{-0.6}$  & 2.53$^{+0.13}_{-0.11}$ & 1.05 (4947) & >2.1 (4.5$\sigma$)\\[0.1cm]
\hline
    Aquila X-1 & LF    & 2  & 15   & 23.3$\pm{0.7}$ & 2.42$^{+0.08}_{-0.10}$ & 1.04 (5067) & >2.1 (3.6$\sigma$) \\[0.1cm]
\vspace*{0.05cm}
     &  h  & 2  &  28   & 36.3$^{+0.6}_{-1.0}$ &2.35$\pm$0.04 & 1.05 (9316) & >2.2 (3$\sigma$)\\
  \vspace*{0.05cm}      
  & LF   & 2+3  & 16    & 21.5$^{+0.3}_{-0.2}$ & 2.13$^{+0.02}_{-0.03}$ & 1.04 (5393) & >2.0 (4.8$\sigma$)\\
\vspace*{0.05cm}
      &  h & 2+3 &  29   & 36.3$^{+0.6}_{-0.8}$ &2.35$^{+0.04}_{-0.03}$  & 1.06 (9701) & >2.2 (3.8$\sigma$)\\ 
  \vspace*{0.05cm}      
 &          b   &  -  & 3  &10.9$^{+1.5}_{-1.4} $ & 2.38$^{+0.28}_{-0.34}$ & 1.04 (1036) & >2.0 (1.0$\sigma$)  \\ 
\hline
    KS\ 1731--260 & LF    & 2  & 10   & 15.2$\pm{0.8}$ & 2.22$^{+0.11}_{-0.10}$  & 1.07 (3237) & >2.0 (3.2$\sigma$)\\[0.1cm]
      \vspace*{0.05cm}  
       &  h  & 2  &  11  & 26.9$^{+1.4}_{-1.2}$ &1.93$^{+0.12}_{-0.10}$ & 1.07 (3493) & <2.0 (0.6$\sigma$) \\
\hline
   SAX\ J1750.8--2900  & LF    & 2  & 3    & 15.2$^{+1.1}_{-9.6}$ &2.95$^{+0.21}_{-0.22}$ & 0.98 (981) & >2.4 (2.2$\sigma$) \\[0.1cm]
\vspace*{0.05cm}
     &  h  & 2  &  4 & 25.0$^{+3.2}_{-3.8}$ &3.32$^{+0.70}_{-0.52}$  & 0.97 (1312) & >2.0 (3.0$\sigma$) \\
\vspace*{0.05cm}
    &  b  & -  &  2 & 5.7$^{+10.5}_{-4.6}$ &3.24$^{+3.46}_{-1.97}$  & 0.89 (662) & >2.0 (0.7$\sigma$)\\
\hline
 IGR\ J17191--2821   & b    & -  & 4    & 10.6$^{+1.4}_{-1.8} $ & 2.45$^{+0.32}_{-0.21}$ & 0.97 (1439) & >2.0 (1.7$\sigma$) \\[0.1cm]
\hline
 \end{tabular}
  }}
  \caption{Power law fit parameters for $\nu_i$ vs. $\nu_u$ correlations. See the main text for explanation of groups. When we fit  $\nu_h$ in Group 2+3, we include the $\nu_{hHz}$ measured in Group 3 as $\nu_h$. We use $\nu_b$ for the power law fits to the $\nu_b$-$\nu_u$ correlation only when Q$_b$>0.5. Errors quoted here use $\Delta\chi^2$=1.(* normalization at  $\nu_u$ = 400 Hz,  ** normalization at  $\nu_u$ = 800 Hz.)
  }
  \label{fit}
\end{table*}



\subsection{Comparison to earlier work}

The power law indices we find are similar to earlier results from \cite{vanStraaten:2000} on 4U\ 0614+09 in which a power law index of 2.46$\pm$0.07 was reported for the $\nu_{LF}$-$\nu_u$ correlation. In that work the averaging of power spectra was limited due to a small data set. In later work  on 4U\ 1728--34 \citep{Ford:1998, vanStraaten:2002}, 4U\ 0614+09 \citep{vanStraaten:2002} and 4U\ 1608--52 \citep{vanStraaten:2003}, the authors found power laws with indices of around 2.0. In these studies many power spectra with similar colours were averaged which led to considerable broadening of, in particular, L$_{LF}$. The resulting blend of L$_{LF}$ and L$_h$ was then fitted with a single Lorentzian.

\label{sec:compare}
\label{C1728}

\begin{figure}

	\includegraphics[width=\columnwidth]{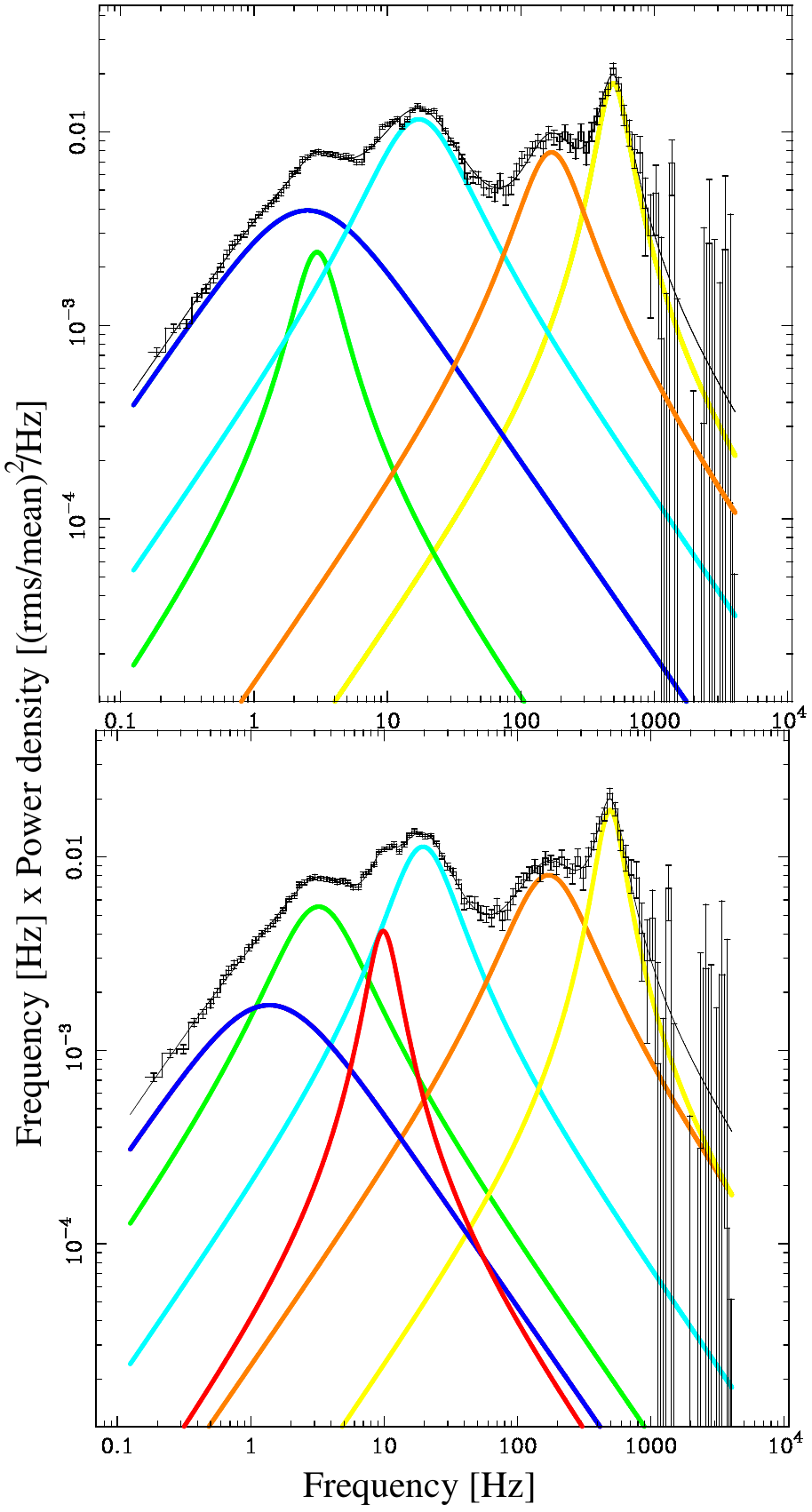}
    \caption{Upper panel: fit to the power spectrum of an observation of 4U\ 1728--34 (10073-01-07-000)  with 5 Lorentzians as in \protect\cite{DiSalvo:2001, vanStraaten:2002} ($\chi^2$/dof=425/327). Lower panel: fit with 6 Lorentzians ($\chi^2$/dof=343/324).}
    \label{fig:vS}
\end{figure}

With our strict data selection and averaging rules and careful identification we are able to detect and separately fit both L$_{LF}$ and L$_h$ over a large range of $\nu_u$. In Figure \ref{fig:vS} we illustrate this using an observation of 4U\ 1728--34 used in \cite{vanStraaten:2002} and in our analysis, fitted with 5 (upper panel, as in the literature) and 6 (lower panel, as in our work) components. We find the fit is significantly better when fitting L$_{LF}$ and L$_h$ as two components (F-statistic probability P=0.53$\times$10$^{-14}$).

\begin{figure}
	\includegraphics[width=\columnwidth]{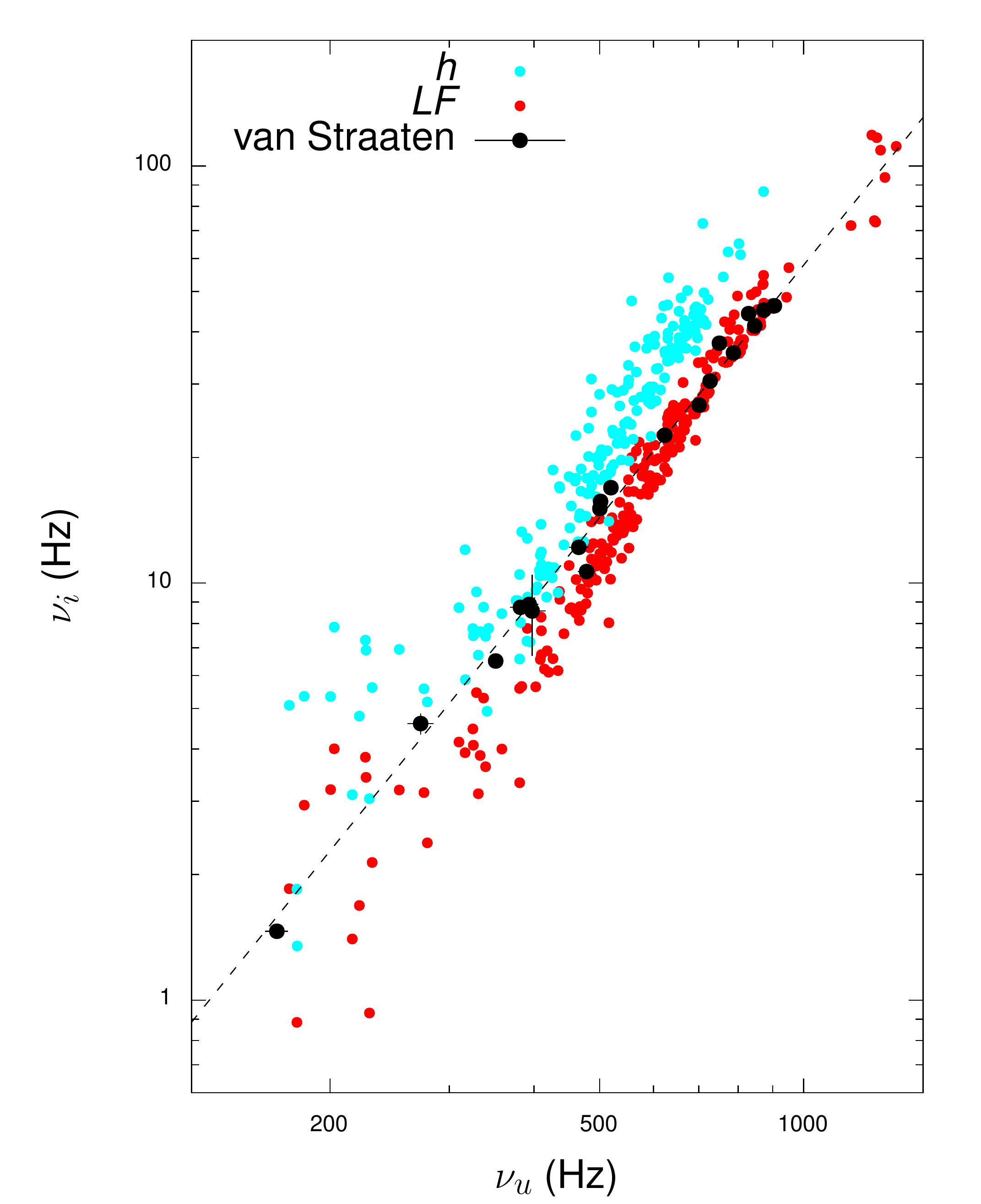}
    \caption{Frequencies previously reported by \protect\cite{vanStraaten:2003} in the range predicted for the Lense-Thirring precession frequency for 4U\ 1728--34, 4U\ 0614+09 and 4U\ 1608--52 plotted together with our results for these sources. The dashed line is the best-fit power law with index 2.01 reported in \protect\cite{vanStraaten:2003}.}
    \label{fig:freq_vS}
\end{figure}

We find that as $\nu_u$ increases, L$_{LF}$ becomes stronger compared to L$_h$ (rms$_{h}$/rms$_{LF}$ drops,  Q$_h$/Q$_{LF}$ rises but is always <1, see Figures \ref{fig:rms_All} and \ref{fig:Q_All}) and L$_h$ blends with L$_{hHz}$. 
Due to these effects, a blend of L$_{LF}$ and L$_{h}$ is fit with a centroid frequency close to $\nu_h$ for low $\nu_u$, and close to $\nu_{LF}$ for high $\nu_u$. 
This results in a shallower power law when fitting the frequency correlation with $\nu_u$. We plot the frequencies and best-fit power law with index 2.01 reported in \cite{vanStraaten:2002} ($\textit{black}$) together with our results ($\textit{blue}$ and $\textit{red}$) for 4U\ 1728--34, 4U\ 0614+09 and 4U\ 1608--52 in Figure \ref{fig:freq_vS}. 
Our best-fit power laws are characterized by indices significantly higher than 2, as reported in Table \ref{fit}.

\clearpage

\section{Discussion}
We inspected power spectra of a sample of 13 NS-LMXBs and in a careful analysis measured the centroid frequency correlations of three different peaks with $\nu$<80 Hz; L$_h$, L$_{LF}$ and L$_b$ with always $\nu_h$>$\nu_{LF}$>$\nu_b$. We found that more indiscriminate averaging of power spectra than performed by us led previous authors (e.g., \citealt{vanStraaten:2003}) to fit L$_h$ and L$_{LF}$ as a single blended feature which has a correlation with $\nu_u$ that is less steep than those obtained for the individual components. Our results are in correspondence with previous studies using small data sets or those in which power spectra were averaged over a limited time span \citep{vanStraaten:2000, Altamirano:2008}. We note that while the frequencies of L$_h$, L$_{LF}$ and L$_b$ depend similarly on $\nu_u$, their rms dependencies on $\nu_u$ markedly differ (see Figure \ref{fig:rms_All}), suggesting differences in their formation physics.

For the correlations of $\nu_{LF}$, $\nu_{h}$ and $\nu_b$ vs. $\nu_u$ we find best-fit power law indices that are significantly higher than 2 for all well constrained sources in our sample; see Table \ref{fit}. 
We find that the frequencies of power spectral components behave similarly in 4U\ 1728--34, 4U\ 0614+09, 4U\ 1608--52, 4U\ 1702--43 and 4U\ 1636--53 when plotted against $\nu_u$. The correlations of $\nu_{LF}$, $\nu_h$ and $\nu_b$ with $\nu_u$ we fit in these sources are very similar, but not identical, as the joint probability distributions of the power law indices and normalizations  differ significantly (see Figures \ref{fig:contLF}, \ref{fig:conth} and \ref{fig:contb}). 

Aquila X-1, KS\ 1731--260, SAX\ J1750--2900, IGR\ J17191--2821, and 4U\ 1915--05 have either low signal to noise or a small data set. The QPOs we find in these sources appear similar to the first group but their identification is less secure. However, none of the power laws we fit in these sources contradict the high power law indices found in sources with higher signal to noise.

\subsection{Precession due to frame dragging}

\subsubsection{Test particle}
\label{section:Models}
Frequencies compatible with Lense-Thirring (LT) precession due to frame dragging of a test particle orbit as proposed by \cite{Stella:1998} are not in correspondence to our findings, as the correlations to $\nu_u$ traced out by the candidate $\nu_{LT}$ frequencies ($\nu_h$, $\nu_{LF}$ and $\nu_b$) mostly have power law indices significantly higher than 2.
For this reason, the relativistic precession model in its original form is incompatible with our data.\\

As noted previously \citep{Stella:1998, vanStraaten:2003} the observed frequencies also tend to be higher than predicted for acceptable values of $I/M$.
\begin{figure*}
	\includegraphics[scale=0.5]{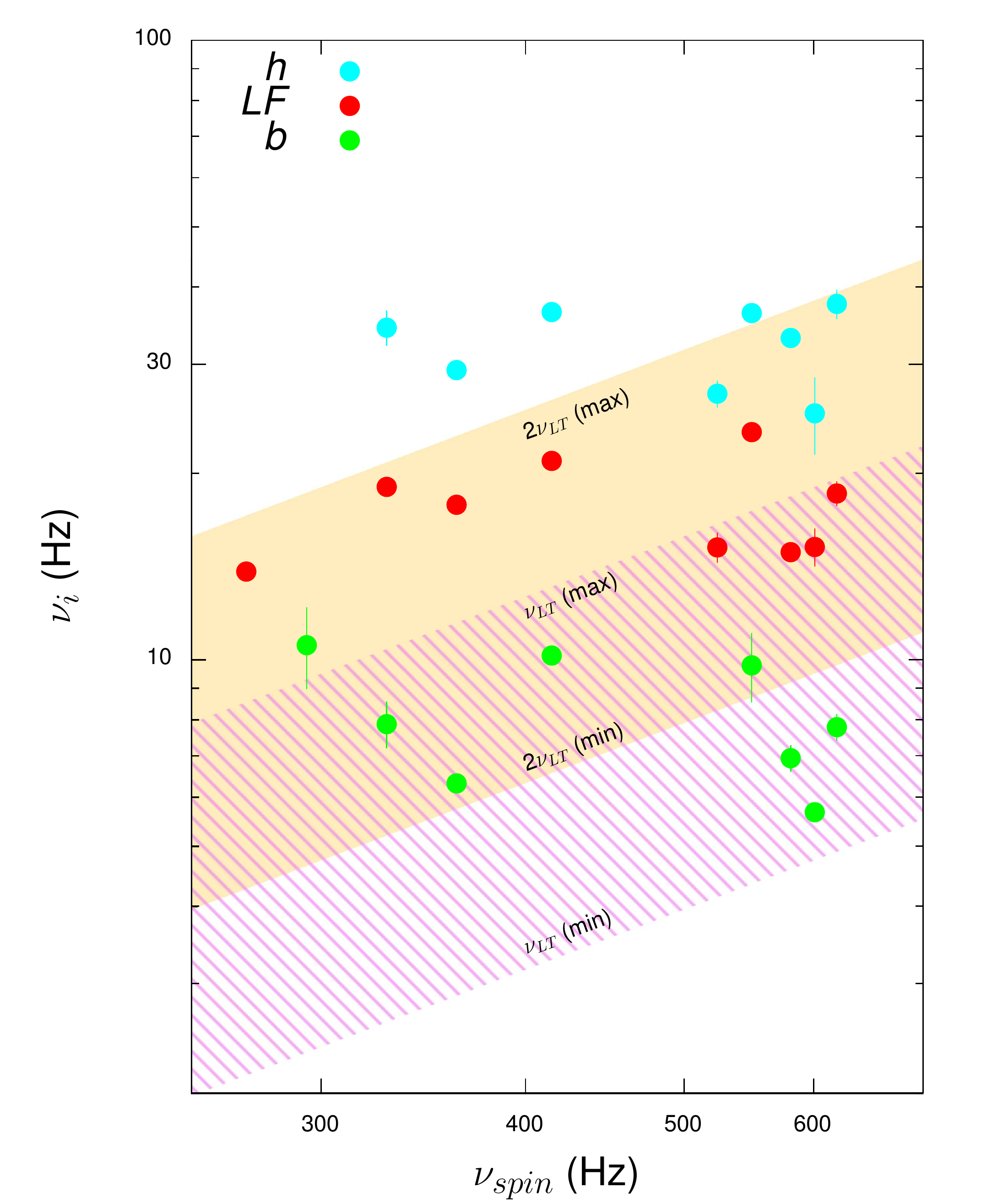}
    \caption{Frequencies $\nu_b$, $\nu_{LF}$ and $\nu_h$ for each source corresponding to $\nu_u$=600 Hz vs. spin frequency, plotted with $\nu_{LT}$ model predictions, see text.}

    \label{fig:spin}
\end{figure*}
Figure \ref{fig:spin} shows the range of possible relativistic precession frequencies $\nu_{LT}$ vs. spin frequency  together with our measurements of $\nu_b$, $\nu_{LF}$ and $\nu_h$ at $\nu_u$=600 Hz.  Realistic equations of state limit the value of $I_{45}/m$ to a maximum of 2 and a minimum of 0.5 (\textit{hatched region} in the Figure; \citealt{Stella:1998, Friedman:1986, Cook:1994}) which is still well below the $I_{45}/m$ required by our measured  $\nu_h$, and also below $\nu_{LF}$ for five out of nine sources. 
 Remarkably, all our measured $\nu_b$ fit this prediction.
 If we assume the observed frequencies are twice the precession frequency (\textit{yellow region}), which is not unrealistic \citep{Stella:1998},  the model predictions are in the observed $\nu_{LF}$ range, but still below $\nu_h$ in four out of eight sources. A dependence of the precession frequency on spin is not evident in our data.\\
 A latitude dependent radiation field as proposed by \cite{Miller:1999} can increase the test particle precession frequency significantly, even in low luminosity sources.  
 Radiation forces could possibly affect the index of a power law correlation between the precession frequency and $\nu_K$ as they are strongest for small radii, and have a much smaller effect on orbital frequencies. We note that \cite{Miller:1999} argues that narrow QPOs are hard to form via precession when including radiation forces. Asymmetry in the radiation field for instance would increase the FWHM of the QPO significantly. 

\subsubsection{Torus}

The model of \cite{Ingram:2009} describes the inner accretion flow as a torus characterized by an inner and outer radius that precesses as a solid body due to frame dragging. Outside the outer radius of the torus is the thin disk. 
In \cite{Ingram:2010} the authors propose that the frequencies of broad components L$_b$ and L$_{hHz}$ in NS-LMXBs represent the viscous timescales at the outer and inner radius, respectively. 
The emergence of L$_{b_2}$ for $\nu_u$>$\sim$700 Hz is explained by the disk moving inward, penetrating the hot flow. L$_{b_2}$ then tracks the viscous timescale in the overlap region. 
By parametrizing the viscous timescale, the inner and outer radius of the torus can be obtained from the measured frequencies. 
The nodal precession frequency of the torus ($\nu_{LT,t}$) is a mass-weighted average of Lense-Thirring precession frequencies at different radii in the torus, and is proposed to be associated with $\nu_{LF}$. What mechanism sets $\nu_h$ has not been specified in this model.\\
As the disk moves in, it is possible that the torus narrows, so its mass can effectively become increasingly concentrated towards the inner radius, favoring progressively higher precession frequencies. In this way, identifying $\nu_u$ with the orbital frequency $\nu_K$ at the inner disk edge outside the torus, a $\nu_{LT,t}$-$\nu_{K}$ correlation with a power law index higher than 2 might be obtained. As the precession frequency originates closer to the neutron star than $\nu_u$, lower values of $I/M$ are required to match the data, so that $I/M$ is closer to the range predicted for realistic equations of state. We note that for high $\nu_u$, when we presume the torus to be narrower, fewer radii contribute to the precession frequency, and the test particle case is approximately recovered. So probably this interpretation still requires $\nu_{LF}$ = 2$\nu_{LT,t}$ for the predicted $\nu_{LT,t}$ to match the data at high $\nu_u$ for realistic $I/M$. 
This could also explain a possible decrease of the power law index toward higher $\nu_u$ which, however, only in 4U 1728--34 we can detect significantly, see Section \ref{sec:bestfit} and Table \ref{fit}, as well as the fact that the index significantly exceeds 2.0 even at the highest frequencies.

\subsection{Classical and magnetic precession}
\cite{Altamirano:2012} found that in the 11 Hz pulsar IGR\ J17480--2446, the LF-QPO (35--50 Hz) cannot be caused by frame dragging, which is prograde with respect to the spin. They suggest that the LF-QPO might be mainly attributable to retrograde classical and magnetic precession. These additional torques are expected to operate on the disk around a neutron star due to magnetic stresses and stellar oblateness. If the magnetic field axis does not coincide with the spin axis of the neutron star, which is the case for pulsars, the magnetic torque causes misalignment of the disk angular momentum with the spin axis of the neutron star and drives magnetic precession (with frequency $\nu_m$). Stellar oblateness causes classical precession with frequency $\nu_{cl}$ \citep{Morsink:1999, Shirakawa:2002}. The three precessional effects depend differently on radius and hence on Keplerian orbital frequency; $\nu_{LT}\propto\nu_K^2$, $\nu_{m}\propto$ -- ($\nu_K^{14/3}$), $\nu_{cl}\propto$ -- ($\nu_K^{7/3}$).  Depending on the coupling mechanism between different radii in the accretion disk and the amount of warping, a net overall precession frequency can arise from the combination of these three torques.
Depending on the system parameters, either the prograde (frame dragging) or the retrograde (magnetic and classical) precession will dominate.  
 Adopting a realistic parameter set for a weakly magnetized (B$\sim$10$^8$ G) neutron star, taking into account all three precession effects as well as warping,  \cite{Shirakawa:2002} predict a prograde precession frequency that correlates with the Keplerian orbital frequency according to a power law with an index below 2 (no exact number is given). Increasing the warping or the magnetic dipole moment lowers the index of the correlation even further.  So, the high power law indices of the correlations we measure cannot be explained by adding classical and magnetic precession effects to (dominant) Lense-Thirring precession. 

If instead the system parameters would be such that (retrograde) magnetic precession dominates, a much steeper power law index may be obtained as $\nu_{m}\propto$ -- ($\nu_K^{14/3}$). 
The observational result of \cite{Bult:2015b}, where the 410 Hz QPO in the 401 Hz AMXP SAX\ J1808.4--3658 is explained as a beat of the spin frequency with a retrograde precession frequency would support this scenario.\\

Our observational results are incompatible with frame-dragging induced precession taking place at test particle frequencies. However, they may be explained in the scenario where the entire inner flow precesses due to frame dragging, in which case multiple radii (with variable weighting) are expected to contribute to the precession frequency \citep{Ingram:2009}. Since a variety of torques are expected around neutron stars, if a QPO is produced by precession, frame dragging is likely only one of the torques contributing to that precession. In any case, unless differential precession \citep{Jakob:2016} affects different harmonics unequally, precession can only explain the occurrence of one frequency, while we observe three QPOs in the $\nu_{LT}$ range with no strong evidence for integer frequency ratios.\\
In black holes as well, then, LF QPO frequencies might be expected to differ from test-particle Lense-Thirring values. This is in accordance with recent findings on the iron line modulation with QPO phase in the black hole H\ 1743--322 which suggest that a precessing torus produces the LF-QPO in that system \citep{Ingram:2016}.

\section{Acknowledgements} 
This research has made use of data obtained through the High Energy Astrophysics Science Archive Research Center Online Service, provided by the NASA/Goddard Space Flight Center.
This work is (partly) financed by the Netherlands Organisation for Scientific Research (NWO).  
M. van Doesburgh thanks Peter Bult, Diego Altamirano and Adam Ingram for many useful discussions. We thank the referee for the constructive feedback and suggestions that improved the quality of the paper.




\bibliographystyle{mnras}
\bibliography{Bibliography} 

\begin{thebibliography}{}
\makeatletter
\relax
\def\mn@urlcharsother{\let\do\@makeother \do\$\do\&\do\#\do\^\do\_\do\%\do\~}
\def\mn@doi{\begingroup\mn@urlcharsother \@ifnextchar [ {\mn@doi@}
  {\mn@doi@[]}}
\def\mn@doi@[#1]#2{\def\@tempa{#1}\ifx\@tempa\@empty \href
  {http://dx.doi.org/#2} {doi:#2}\else \href {http://dx.doi.org/#2} {#1}\fi
  \endgroup}
\def\mn@eprint#1#2{\mn@eprint@#1:#2::\@nil}
\def\mn@eprint@arXiv#1{\href {http://arxiv.org/abs/#1} {{\tt arXiv:#1}}}
\def\mn@eprint@dblp#1{\href {http://dblp.uni-trier.de/rec/bibtex/#1.xml}
  {dblp:#1}}
\def\mn@eprint@#1:#2:#3:#4\@nil{\def\@tempa {#1}\def\@tempb {#2}\def\@tempc
  {#3}\ifx \@tempc \@empty \let \@tempc \@tempb \let \@tempb \@tempa \fi \ifx
  \@tempb \@empty \def\@tempb {arXiv}\fi \@ifundefined
  {mn@eprint@\@tempb}{\@tempb:\@tempc}{\expandafter \expandafter \csname
  mn@eprint@\@tempb\endcsname \expandafter{\@tempc}}}

\bibitem[\protect\citeauthoryear{{Altamirano}, {van der Klis}, {M{\'e}ndez},
  {Jonker}, {Klein-Wolt}  \& {Lewin}}{{Altamirano}
  et~al.}{2008}]{Altamirano:2008}
{Altamirano} D.,  {van der Klis} M.,  {M{\'e}ndez} M.,  {Jonker} P.~G.,
  {Klein-Wolt} M.,   {Lewin} W.~H.~G.,  2008, \mn@doi [\apj] {10.1086/590897},
  \href {http://adsabs.harvard.edu/abs/2008ApJ...685..436A} {685, 436}

\bibitem[\protect\citeauthoryear{{Altamirano}, {Ingram}, {van der Klis},
  {Wijnands}, {Linares}  \& {Homan}}{{Altamirano}
  et~al.}{2012}]{Altamirano:2012}
{Altamirano} D.,  {Ingram} A.,  {van der Klis} M.,  {Wijnands} R.,  {Linares}
  M.,   {Homan} J.,  2012, \mn@doi [\apjl] {10.1088/2041-8205/759/1/L20}, \href
  {http://adsabs.harvard.edu/abs/2012ApJ...759L..20A} {759, L20}

\bibitem[\protect\citeauthoryear{{Belloni}, {Psaltis}  \& {van der
  Klis}}{{Belloni} et~al.}{2002}]{Belloni:2002}
{Belloni} T.,  {Psaltis} D.,   {van der Klis} M.,  2002, \mn@doi [\apj]
  {10.1086/340290}, \href {http://adsabs.harvard.edu/abs/2002ApJ...572..392B}
  {572, 392}

\bibitem[\protect\citeauthoryear{{Bult} \& {van der Klis}}{{Bult} \& {van der
  Klis}}{2015a}]{Bult:2015}
{Bult} P.,  {van der Klis} M.,  2015a, \mn@doi [\apjl]
  {10.1088/2041-8205/798/2/L29}, \href
  {http://adsabs.harvard.edu/abs/2015ApJ...798L..29B} {798, L29}

\bibitem[\protect\citeauthoryear{{Bult} \& {van der Klis}}{{Bult} \& {van der
  Klis}}{2015b}]{Bult:2015b}
{Bult} P.,  {van der Klis} M.,  2015b, \mn@doi [\apj]
  {10.1088/0004-637X/806/1/90}, \href
  {http://adsabs.harvard.edu/abs/2015ApJ...806...90B} {806, 90}

\bibitem[\protect\citeauthoryear{{Casella}, {Altamirano}, {Patruno}, {Wijnands}
   \& {van der Klis}}{{Casella} et~al.}{2008}]{ALtamirano:Aquila}
{Casella} P.,  {Altamirano} D.,  {Patruno} A.,  {Wijnands} R.,   {van der Klis}
  M.,  2008, \mn@doi [\apjl] {10.1086/528982}, \href
  {http://adsabs.harvard.edu/abs/2008ApJ...674L..41C} {674, L41}

\bibitem[\protect\citeauthoryear{{Cook}, {Shapiro}  \& {Teukolsky}}{{Cook}
  et~al.}{1994}]{Cook:1994}
{Cook} G.~B.,  {Shapiro} S.~L.,   {Teukolsky} S.~A.,  1994, \mn@doi [\apj]
  {10.1086/173721}, \href {http://adsabs.harvard.edu/abs/1994ApJ...422..227C}
  {422, 227}

\bibitem[\protect\citeauthoryear{{Di Salvo}, {M{\'e}ndez}, {van der Klis},
  {Ford}  \& {Robba}}{{Di Salvo} et~al.}{2001}]{DiSalvo:2001}
{Di Salvo} T.,  {M{\'e}ndez} M.,  {van der Klis} M.,  {Ford} E.,   {Robba}
  N.~R.,  2001, \mn@doi [\apj] {10.1086/318278}, \href
  {http://adsabs.harvard.edu/abs/2001ApJ...546.1107D} {546, 1107}

\bibitem[\protect\citeauthoryear{{Ford} \& {van der Klis}}{{Ford} \& {van der
  Klis}}{1998}]{Ford:1998}
{Ford} E.~C.,  {van der Klis} M.,  1998, \mn@doi [\apjl] {10.1086/311638},
  \href {http://adsabs.harvard.edu/abs/1998ApJ...506L..39F} {506, L39}

\bibitem[\protect\citeauthoryear{{Friedman}, {Ipser}  \& {Parker}}{{Friedman}
  et~al.}{1986}]{Friedman:1986}
{Friedman} J.~L.,  {Ipser} J.~R.,   {Parker} L.,  1986, \mn@doi [\apj]
  {10.1086/164149}, \href {http://adsabs.harvard.edu/abs/1986ApJ...304..115F}
  {304, 115}

\bibitem[\protect\citeauthoryear{{Hasinger} \& {van der Klis}}{{Hasinger} \&
  {van der Klis}}{1989}]{Hasinger:1989}
{Hasinger} G.,  {van der Klis} M.,  1989, \aap, \href
  {http://adsabs.harvard.edu/abs/1989A26A...225...79H} {225, 79}

\bibitem[\protect\citeauthoryear{{Ingram} \& {Done}}{{Ingram} \&
  {Done}}{2010}]{Ingram:2010}
{Ingram} A.,  {Done} C.,  2010, \mn@doi [\mnras]
  {10.1111/j.1365-2966.2010.16614.x}, \href
  {http://adsabs.harvard.edu/abs/2010MNRAS.405.2447I} {405, 2447}

\bibitem[\protect\citeauthoryear{{Ingram}, {Done}  \& {Fragile}}{{Ingram}
  et~al.}{2009}]{Ingram:2009}
{Ingram} A.,  {Done} C.,   {Fragile} P.~C.,  2009, \mn@doi [\mnras]
  {10.1111/j.1745-3933.2009.00693.x}, \href
  {http://adsabs.harvard.edu/abs/2009MNRAS.397L.101I} {397, L101}

\bibitem[\protect\citeauthoryear{{Ingram}, {van der Klis}, {Middleton}, {Done},
  {Altamirano}, {Heil}, {Uttley}  \& {Axelsson}}{{Ingram}
  et~al.}{2016}]{Ingram:2016}
{Ingram} A.,  {van der Klis} M.,  {Middleton} M.,  {Done} C.,  {Altamirano} D.,
   {Heil} L.,  {Uttley} P.,   {Axelsson} M.,  2016, \mn@doi [\mnras]
  {10.1093/mnras/stw1245}, \href
  {http://adsabs.harvard.edu/abs/2016MNRAS.461.1967I} {461, 1967}

\bibitem[\protect\citeauthoryear{{Klein Wolt}}{{Klein
  Wolt}}{2004}]{Klein:2004PhD}
{Klein Wolt} M.,  2004, PhD thesis, University of Amsterdam

\bibitem[\protect\citeauthoryear{{Klein-Wolt} \& {van der Klis}}{{Klein-Wolt}
  \& {van der Klis}}{2008}]{Klein:2008}
{Klein-Wolt} M.,  {van der Klis} M.,  2008, \mn@doi [\apj] {10.1086/525843},
  \href {http://adsabs.harvard.edu/abs/2008ApJ...675.1407K} {675, 1407}

\bibitem[\protect\citeauthoryear{{Miller}}{{Miller}}{1999}]{Miller:1999}
{Miller} M.~C.,  1999, \mn@doi [\apj] {10.1086/307420}, \href
  {http://adsabs.harvard.edu/abs/1999ApJ...520..256M} {520, 256}

\bibitem[\protect\citeauthoryear{{Morsink} \& {Stella}}{{Morsink} \&
  {Stella}}{1999}]{Morsink:1999}
{Morsink} S.~M.,  {Stella} L.,  1999, \mn@doi [\apj] {10.1086/306876}, \href
  {http://adsabs.harvard.edu/abs/1999ApJ...513..827M} {513, 827}

\bibitem[\protect\citeauthoryear{{Motta}, {Belloni}, {Stella},
  {Mu{\~n}oz-Darias}  \& {Fender}}{{Motta} et~al.}{2014}]{Motta:2014}
{Motta} S.~E.,  {Belloni} T.~M.,  {Stella} L.,  {Mu{\~n}oz-Darias} T.,
  {Fender} R.,  2014, \mn@doi [\mnras] {10.1093/mnras/stt2068}, \href
  {http://adsabs.harvard.edu/abs/2014MNRAS.437.2554M} {437, 2554}

\bibitem[\protect\citeauthoryear{{Nowak}}{{Nowak}}{2000}]{Nowak:2000}
{Nowak} M.~A.,  2000, \mn@doi [\mnras] {10.1046/j.1365-8711.2000.03668.x},
  \href {http://adsabs.harvard.edu/abs/2000MNRAS.318..361N} {318, 361}

\bibitem[\protect\citeauthoryear{{Patruno} \& {Watts}}{{Patruno} \&
  {Watts}}{2012}]{Patruno:2012}
{Patruno} A.,  {Watts} A.~L.,  2012, preprint, \href
  {http://adsabs.harvard.edu/abs/2012arXiv1206.2727P} {} (\mn@eprint {arXiv}
  {1206.2727})

\bibitem[\protect\citeauthoryear{{Press}, {Teukolsky}, {Vetterling}  \&
  {Flannery}}{{Press} et~al.}{1992}]{Press:1992}
{Press} W.~H.,  {Teukolsky} S.~A.,  {Vetterling} W.~T.,   {Flannery} B.~P.,
  1992, {Numerical recipes in FORTRAN. The art of scientific computing}

\bibitem[\protect\citeauthoryear{{Psaltis}, {Belloni}  \& {van der
  Klis}}{{Psaltis} et~al.}{1999}]{Psaltis:1999}
{Psaltis} D.,  {Belloni} T.,   {van der Klis} M.,  1999, \mn@doi [\apj]
  {10.1086/307436}, \href {http://adsabs.harvard.edu/abs/1999ApJ...520..262P}
  {520, 262}

\bibitem[\protect\citeauthoryear{{Ritter} \& {Kolb}}{{Ritter} \&
  {Kolb}}{2003}]{Ritter:2003}
{Ritter} H.,  {Kolb} U.,  2003, \mn@doi [\aap] {10.1051/0004-6361:20030330},
  \href {http://adsabs.harvard.edu/abs/2003A26A...404..301R} {404, 301}

\bibitem[\protect\citeauthoryear{{Shirakawa} \& {Lai}}{{Shirakawa} \&
  {Lai}}{2002}]{Shirakawa:2002}
{Shirakawa} A.,  {Lai} D.,  2002, \mn@doi [\apj] {10.1086/324548}, \href
  {http://adsabs.harvard.edu/abs/2002ApJ...565.1134S} {565, 1134}

\bibitem[\protect\citeauthoryear{{Stella} \& {Vietri}}{{Stella} \&
  {Vietri}}{1998}]{Stella:1998}
{Stella} L.,  {Vietri} M.,  1998, \mn@doi [\apjl] {10.1086/311075}, \href
  {http://adsabs.harvard.edu/abs/1998ApJ...492L..59S} {492, L59}

\bibitem[\protect\citeauthoryear{{Strohmayer} \& {Markwardt}}{{Strohmayer} \&
  {Markwardt}}{2002}]{Strohmayer:2002}
{Strohmayer} T.~E.,  {Markwardt} C.~B.,  2002, \mn@doi [\apj] {10.1086/342152},
  \href {http://adsabs.harvard.edu/abs/2002ApJ...577..337S} {577, 337}

\bibitem[\protect\citeauthoryear{{Watts}}{{Watts}}{2012}]{Watts:2012}
{Watts} A.~L.,  2012, \mn@doi [\araa] {10.1146/annurev-astro-040312-132617},
  \href {http://adsabs.harvard.edu/abs/2012ARA26A..50..609W} {50, 609}

\bibitem[\protect\citeauthoryear{{Wijnands} \& {van der Klis}}{{Wijnands} \&
  {van der Klis}}{1999}]{Wijnands:1999}
{Wijnands} R.,  {van der Klis} M.,  1999, \mn@doi [\apj] {10.1086/306993},
  \href {http://adsabs.harvard.edu/abs/1999ApJ...514..939W} {514, 939}

\bibitem[\protect\citeauthoryear{{Zhang}, {Jahoda}, {Swank}, {Morgan}  \&
  {Giles}}{{Zhang} et~al.}{1995}]{Zhang:1995}
{Zhang} W.,  {Jahoda} K.,  {Swank} J.~H.,  {Morgan} E.~H.,   {Giles} A.~B.,
  1995, \mn@doi [\apj] {10.1086/176111}, \href
  {http://adsabs.harvard.edu/abs/1995ApJ...449..930Z} {449, 930}

\bibitem[\protect\citeauthoryear{{van Straaten}, {Ford}, {van der Klis},
  {M{\'e}ndez}  \& {Kaaret}}{{van Straaten} et~al.}{2000}]{vanStraaten:2000}
{van Straaten} S.,  {Ford} E.~C.,  {van der Klis} M.,  {M{\'e}ndez} M.,
  {Kaaret} P.,  2000, \mn@doi [\apj] {10.1086/309351}, \href
  {http://adsabs.harvard.edu/abs/2000ApJ...540.1049V} {540, 1049}

\bibitem[\protect\citeauthoryear{{van Straaten}, {van der Klis}, {di Salvo}  \&
  {Belloni}}{{van Straaten} et~al.}{2002}]{vanStraaten:2002}
{van Straaten} S.,  {van der Klis} M.,  {di Salvo} T.,   {Belloni} T.,  2002,
  \mn@doi [\apj] {10.1086/338948}, \href
  {http://adsabs.harvard.edu/abs/2002ApJ...568..912V} {568, 912}

\bibitem[\protect\citeauthoryear{{van Straaten}, {van der Klis}  \&
  {M{\'e}ndez}}{{van Straaten} et~al.}{2003}]{vanStraaten:2003}
{van Straaten} S.,  {van der Klis} M.,   {M{\'e}ndez} M.,  2003, \mn@doi [\apj]
  {10.1086/378155}, \href {http://adsabs.harvard.edu/abs/2003ApJ...596.1155V}
  {596, 1155}

\bibitem[\protect\citeauthoryear{{van Straaten}, {van der Klis}  \&
  {Wijnands}}{{van Straaten} et~al.}{2005}]{vanStraaten:2005}
{van Straaten} S.,  {van der Klis} M.,   {Wijnands} R.,  2005, \mn@doi [\apj]
  {10.1086/426183}, \href {http://adsabs.harvard.edu/abs/2005ApJ...619..455V}
  {619, 455}

\bibitem[\protect\citeauthoryear{{van den Eijnden}, {Ingram}  \& {Uttley}}{{van
  den Eijnden} et~al.}{2016}]{Jakob:2016}
{van den Eijnden} J.,  {Ingram} A.,   {Uttley} P.,  2016, \mn@doi [\mnras]
  {10.1093/mnras/stw610}, \href
  {http://adsabs.harvard.edu/abs/2016MNRAS.458.3655V} {458, 3655}

\bibitem[\protect\citeauthoryear{{van der Klis}}{{van der
  Klis}}{1989}]{Klis:1989}
{van der Klis} M.,  1989, in {{\"O}gelman} H.,  {van den Heuvel} E.~P.~J.,
  eds,  NATO Advanced Science Institutes (ASI) Series C Vol. 262, NATO Advanced
  Science Institutes (ASI) Series C. p.~27

\bibitem[\protect\citeauthoryear{{van der Klis}}{{van der
  Klis}}{2006}]{Klis:2006book}
{van der Klis} M.,  2006, {Rapid X-ray Variability}.
pp 39--112

\makeatother
\end{thebibliography}


\clearpage

\appendix
\section{Map Fitting}
\label{fitting}

Standard $\chi^2$ methods for performing the power-law fits to our frequency correlations run into a number of problems.

First, the data have error bars in both coordinates ($\nu_u$ and $\nu_x$, where $x$ is $b$, $LF$ or $h$) which means that for arbitrary fit functions a well-behaved $\chi^2$ statistic taking into account both errors must be defined.  

Second, the frequency measurements are obtained by performing multi-Lorentzian model fits to power spectra with typically 15-24 free parameters.  Such fits typically result in asymmetric error bars, the interpretation of which, in terms of the contribution to the $\chi^2$ statistic made by a data point depending on whether the fit function passes it on the side of the larger or the smaller error bar, is uncertain, as the information about the actual probability distribution of the parameter is not preserved in the 'error-bar' description.

Third, although in our case this was usually not the case, such fits can result in strongly correlated errors, biasing any true correlation between the two frequencies fitted.

We resolved these issues by employing a method that is mathematically identical to performing a joint fit to all power spectra simultaneously, with all multi-Lorentzian parameters free except that the two frequencies of interest are tied together by the power law relation we desire to fit.  The power law parameters act as two more fit parameters added to the total set of fit parameters.  For a typical power-law fit to 100 frequency pairs this entails a simultaneous fit to 100 power spectra with typically 2000 free parameters and 40000 $\textit{dof}$, which would be unwieldy to perform directly. 

We therefore performed the fit in two steps. In the first step, we fit each individual power spectrum with the multi-Lorentzian model, and perform a scan of the $\chi^2$  values in the $\nu_u$, $\nu_x$ plane around the best fit, leaving all other parameters free to minimize $\chi^2$  in the usual way (e.g., \citealt{Press:1992}).  The resulting $\chi^2$  maps, one for each power spectrum, form the input to the second step, the power law fit.

\begin{figure*}
	\includegraphics[width=5in]{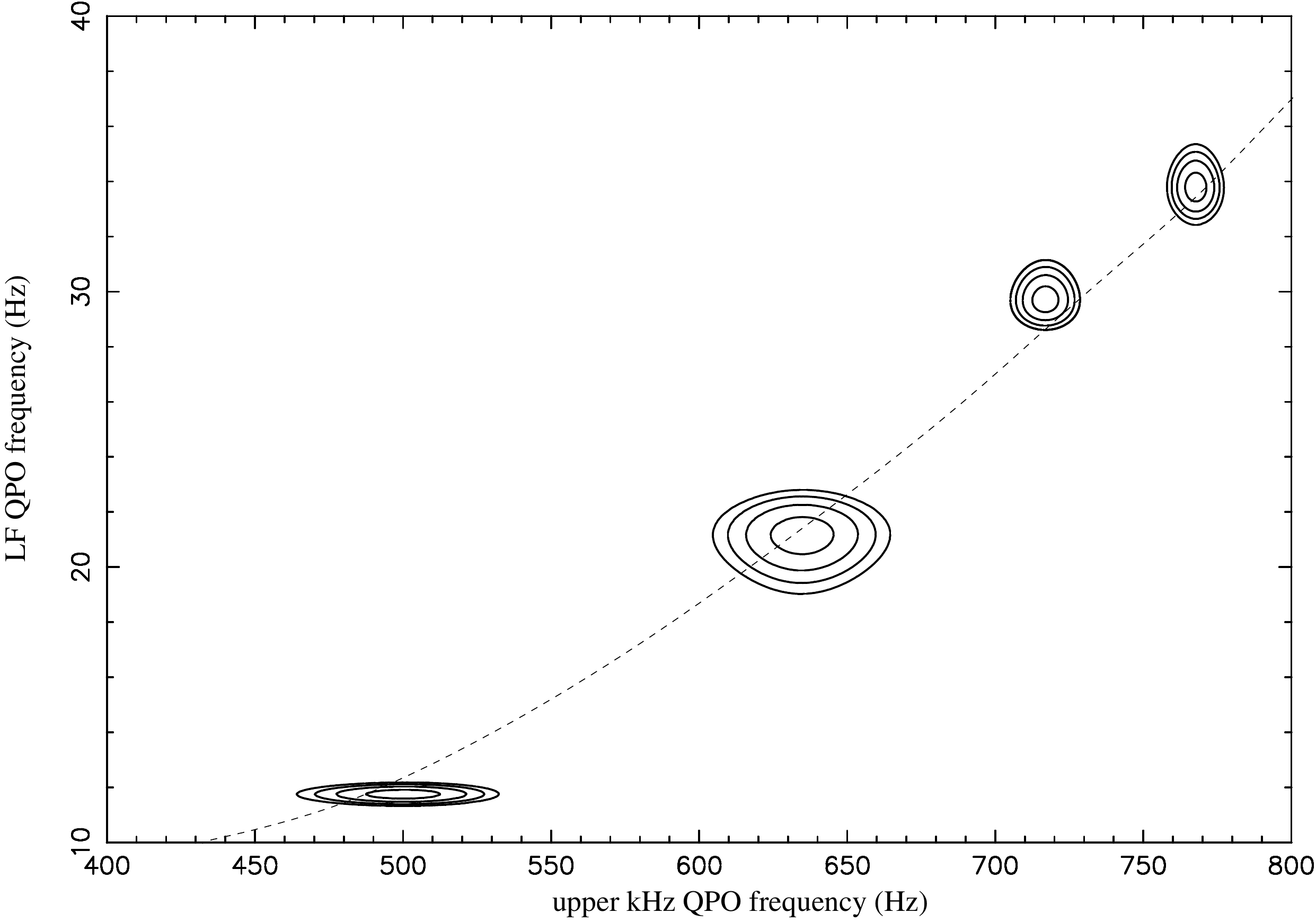}
    \caption{Example of a power law fit to $\chi^2$  maps of $\nu_{LF}$-$\nu_u$ frequency pairs. Shown are the $\Delta\chi^2$=1,3,5 and 7 contours of a selection of measured frequency pairs from Group 2 of 4U\ 1728--34 as well as the best-fit power law ($\textit{dashed line}$).} 
    \label{fig:err}
\end{figure*}

In the second step, we fit a power law to all the $\chi^2$  maps in the $\nu_u$, $\nu_x$ diagram by varying the two power law parameters and minimizing the total $\chi^2$ of the power law fit. This total $\chi^2$  is defined as the sum of the $\chi^2$  contributions of each map, where each map's contribution is just the lowest $\chi^2$ value in the map corresponding to a point on the power law (see Figure \ref{fig:err}). 

A difficulty that had to be overcome in applying this procedure is that sometimes the trial power law runs so far from a $\chi^2$ map center, that the implied $\nu_u$, $\nu_x$ pairs can not fit the power spectrum, as one or both frequencies come too close to other features in the power spectrum than the ones identified as $L_u$ and $L_x$, which results in an altogether different multi-Lorentzian fit where the fitted frequencies no longer represent the intended power spectral components.  In terms of the $\chi^2$  maps, this translates in a secondary minimum in $\chi^2$.  To overcome this problem, we truncated our map scans before the turn-over to the secondary minimum, and when necessary extrapolated the maps using an elliptical paraboloid extrapolation scheme.  So, in cases where the power law runs far from a map center we conservatively assign a larger $\chi^2$ contribution to that map than that corresponding to the statistically 'best' (but physically incorrect) fit, a contribution that is representative of the $\chi^2$  that the map would have contributed if the $L_u$ and $L_x$ components in the fit function would have been forced to continue to describe their intended counterparts in the observed power spectra. In practice, for all our final power law fits, $<$20$\%$ of maps were extrapolated (8, 12, and 20$\%$ for L$_{LF}$, L$_h$ and L$_b$, respectively).\\
We offer a comparison of this method to the result we obtain when minimizing $\chi^2$ with a standard fitting routine in which only vertical errors are taken into account. We use the $\nu_{LF}$ of Group 2 in 4U\ 1728--34 with 133 frequency measurements here. With the standard method we obtain a best-fit with
normalization 1.37($\pm$0.06)$\times$10$^{-6}$ and power law index 2.56$\pm$0.01, with $\chi^2$/$dof$: 1689/131. 
The best fit with the map fitting technique gives a normalization of 2.42($\pm$0.20)$\times$10$^{-6}$, a power law index of 2.47($\pm$0.01) and $\chi^2$/$dof$:   45677/43389.


 \section{Detection significance of QPOs}
 
 We report all fitted Lorentzians that were needed to characterize power spectra and include all centroid frequencies in our power law fits except in cases noted below.  As we are able to track components over different observations, we are able to usefully include centroid frequencies of Lorentzians detected at significance levels <3$\sigma$ (down to 2$\sigma$, calculated from the negative error on the integral power of the Lorentzian). We do this only if the component is detected at >3$\sigma$ in similar observations.  
 In Table \ref{tab:sig_All} we quote the expected number of false positives for L$_b$, L$_{LF}$, L$_h$ and L$_u$ resulting from the ensemble of detection significances for each component in each source. Clearly these numbers are very small.

\begin{table}
	\centering

	\begin{tabular*}{\columnwidth}{@{\extracolsep{\fill}}lcccc}
	
			\hline
 Source &  L$_b$ & L$_{LF}$ &        L$_{h}$& L$_u$ \\
  & false/tot. & false/tot. & false/tot. & false/tot. \\
		\hline
4U\ 1728--34 & 1.38/165 & 1.53/148 & 1.10/118 & 0.35/188 \\
4U\ 0614+09 & 0.81/97 & 2.15/76 & 1.05/55 & 0.34/136 \\
4U\ 1608--52 & 0.04/27 & 0.23/16 & 0.21/16 & 0.55/31 \\
4U\ 1636--53 & 0.22/35 & 0.64/27 & 0.27/25 & 0.06/38 \\
4U\ 1702--43 & 0.38/21 & 0.56/16 & 0.02/12 & 0.03/24 \\
Aquila X-1 & 0.05/22 & 0.47/16 & 0.26/29 & 0.08/34 \\
SAX\ J1750--2900 & 0.03/6 & 0.01/3 & 0.01/4 & 0.03/6 \\
4U\ 1915--05 & 0.18/3 & 0.05/16 & - & 0.05/15 \\ 
KS\ 1731--260 & 0.01/7 & 0.54/10 & 0.01/11 & 0.01/11 \\
IGR\ J17191--2821 & 0.20/5 & - &-& 0.03/5 \\
	\hline	
 	  	   \end{tabular*}	
\caption{Expected false positives out of the total number of fitted Lorentzians per source.}
\label{tab:sig_All}
\end{table}


 \section{Detailed results for individual sources}
 In this appendix we present some details on the results obtained for the individual sources in our sample. For each source separately we display the measured frequencies vs. $\nu_u$ as in Figure \ref{fig:freq_all}, with the best-fit power laws (\textit{dashed}) to Group 2 as specified in Table \ref{tab:gr_All}. 
 All power spectral fit parameters can be found in Table \ref{tab:pars}; the full version is available online. The power law fit results can be found in Table \ref{fit}. For selected sources we also display fractional rms and Q values similarly to Figures \ref{fig:rms_All} and \ref{fig:Q_All}.

\begin{table*}
	\centering
	
	  \tabcolsep=0.1cm
 {
\scalebox{1}{
\small
	\begin{tabular}{lccccc}
			\hline
Source  & Row A & Row B & Row C & Row D & High $\nu_u$ \\
		\hline
4U\ 1728--34 & 30042-03-18-00  & 50023-01-25-00  & 50023-01-30-00 & 20083-01-04-00 & 20083-01-02-000/01 \\
4U\ 0614+09 & 50031-01-04-13 & 30053-01-03-01/02 & 20074-01-07-00 & 80037-01-06-00 & 40030-01-06-00 \\
4U\ 1608--52 & 30062-02-02-00/000  & 60052-03-01-00/01 &60052-02-06/07-00  &30062-02-01-000 & - \\
4U\ 1636--53 &92023-01/02-11-00 &60032-05-10-00/000  & 60032-05-07-00/01 & 60032-01-06-00/000 & - \\
4U\ 1702--43 &50030-01-04-00/000 &80033-01-18-01/02/03 & 40025-04-03-01 & X-06-01/02 and X-04-04  & X-10-03, X-15-01, X-12-05, \\
            &                   &                   &                   &                       & X-12-04, X-14-05, X-15-02, \\
            &                   &                   &                   &                       & X-11-00, X-14-06, X-10-01, \\
            &                   &                   &                   &                       &      X-10-02 \\
Aquila X-1 & 91414-01-08-05/06 & 60054-02-02-02 &90017-01-09-02  &50049-02-15-05/06 & 30188-03-01-00 \\
	\hline	
 	  	   \end{tabular}
 	  	   }
\caption{The observations we used to make the power spectra shown in Figures \ref{overview}, \ref{overview_2} and \ref{overview_high}. Observations "20083-01-02-000/01" for instance means we included both 20083-01-02-000 and 20083-01-02-01,   "X" stands for "80033-01".  }
\label{tab:fig_obs}}
\end{table*}


 \begin{sidewaystable*}

   \tabcolsep=0.1cm
 {
\scalebox{0.83}{
\small
\begin{tabular}{c c c c c c c c c c c c c c c c c}

\hline	
Source & ObsID & No. of &No. of  &   & L$_{b}$ &  &  &  L$_{LF}$ &  &  & L$_h$ & &  & L$_u$ &  & $\chi^2/dof$ \\ 
      &       & obs. &power spectra   & $\nu$ (Hz) & FWHM (Hz)  & IP ($\times$10$^{-3}$) & $\nu$ (Hz) & FWHM (Hz) & IP ($\times$10$^{-3}$) &$\nu$ (Hz)& FWHM (Hz) & IP ($\times$10$^{-3}$)  &$\nu$ (Hz)& FWHM (Hz) & IP ($\times$10$^{-3}$)  & \\
       \hline
4U\ 1728--34 &50030-03-05-02&1&629&0.30$\pm$0.05&1.80$^{+0.10}_{-0.09}$&15.5$^{+0.6}_{-0.5}$&2.93$^{+0.13}_{-0.09}$&0.55$^{+0.49}_{-0.31}$&0.40$^{+0.34}_{-0.20}$&5.35$^{+0.18}_{-0.16}$&6.99$^{+0.52}_{-0.65}$&17.6$^{+1.5}_{-2.0}$&183.3$^{+27.0}_{-34.2}$&327.4$^{+66.8}_{-51.7}$&16.9$^{+2.6}_{-2.2}$&341.98/327\\
&30042-03-01-01&2&264&0.27$^{+0.06}_{-0.08}$&2.01$^{+0.13}_{-0.12}$&15.5$\pm$0.6&3.20$^{+0.10}_{-0.08}$&0.46$^{+0.50}_{-0.32}$&0.32$^{+0.31}_{-0.18}$&5.34$\pm$0.18&6.96$^{+0.49}_{-0.54}$&17.2$^{+1.5}_{-1.7}$&200.5$^{+17.4}_{-20.9}$&380.8$^{+45.7}_{-40.2}$&15.9$^{+1.3}_{-1.2}$&312.54/327\\
&30042-03-18-00&1&589&0.34$^{+0.05}_{-0.06}$&2.50$\pm$0.10&17.0$\pm$0.4&4.00$\pm$0.06&1.40$^{+0.27}_{-0.23}$&1.58$^{+0.39}_{-0.38}$&7.84$\pm$0.46&8.16$^{+0.81}_{-0.92}$&12.7$^{+2.4}_{-2.5}$&203.0$^{+18.6}_{-19.8}$&450.4$^{+36.2}_{-33.7}$&19.5$^{+1.1}_{-1.2}$&332.89/324\\
&30042-03-20-00&1&257&0.31$^{+0.08}_{-0.09}$&2.44$^{+0.15}_{-0.14}$&16.5$\pm$0.7&3.82$^{+0.10}_{-0.11}$&1.27$^{+0.49}_{-0.59}$&1.26$^{+0.67}_{-0.63}$&7.29$^{+0.63}_{-0.54}$&7.88$^{+0.80}_{-0.86}$&14.5$^{+2.6}_{-3.1}$&225.6$^{+22.1}_{-24.2}$&367.0$^{+50.7}_{-46.5}$&15.1$^{+1.5}_{-1.6}$&364.60/324\\
&30042-03-04-00&1&410&0.28$^{+0.06}_{-0.07}$&2.20$^{+0.11}_{-0.14}$&16.4$^{+0.7}_{-1.2}$&3.42$\pm$0.04&0.77$^{+0.15}_{-0.12}$&1.23$^{+0.23}_{-0.21}$&6.90$^{+0.28}_{-0.08}$&6.31$^{+0.90}_{-0.92}$&9.5$^{+2.8}_{-2.5}$&226.1$^{+29.6}_{-26.0}$&406.3$^{+47.7}_{-47.3}$&14.5$^{+3.7}_{-9.7}$&336.18/322\\
&30042-03-01-02&2&229&0.48$\pm$0.04&1.63$^{+0.17}_{-0.16}$&12.7$^{+1.6}_{-1.4}$&3.19$^{+0.07}_{-0.06}$&0.91$^{+0.28}_{-0.22}$&1.29$^{+0.53}_{-0.40}$&6.93$^{+0.14}_{-0.53}$&4.57$^{+2.10}_{-1.53}$&4.7$^{+5.4}_{-2.0}$&253.1$^{+24.5}_{-27.3}$&319.4$^{+57.8}_{-54.9}$&11.1$^{+2.4}_{-2.7}$&368.50/322\\
&30042-03-01-04&1&206&0.37$^{+0.05}_{-0.06}$&1.84$^{+0.13}_{-0.12}$&15.5$\pm$0.7&3.15$^{+0.10}_{-0.11}$&0.91$^{+0.48}_{-0.39}$&0.88$^{+0.58}_{-0.44}$&5.58$^{+0.35}_{-0.34}$&7.10$\pm$0.46&16.9$\pm$1.8&275.3$^{+12.2}_{-13.1}$&261.2$^{+37.5}_{-32.8}$&10.7$\pm$1.0&338.86/324\\
&40033-06-01-00&1&589&0.30$\pm$0.06&2.59$\pm$0.10&16.6$\pm$0.5&4.08$\pm$0.06&1.18$^{+0.27}_{-0.25}$&1.22$^{+0.36}_{-0.32}$&7.48$^{+0.33}_{-0.31}$&8.52$\pm$0.57&14.9$^{+1.6}_{-1.7}$&325.5$^{+13.4}_{-13.6}$&229.9$^{+36.1}_{-35.0}$&9.4$^{+1.3}_{-1.6}$&307.31/321\\
&30042-03-03-01&1&276&0.38$^{+0.05}_{-0.06}$&1.89$\pm$0.15&15.1$\pm$0.8&3.13$^{+0.10}_{-0.12}$&1.95$^{+0.51}_{-0.38}$&3.03$^{+1.04}_{-0.80}$&6.72$^{+0.46}_{-0.44}$&6.89$^{+1.07}_{-1.25}$&13.3$^{+3.7}_{-4.3}$&331.1$^{+23.1}_{-30.6}$&249.7$^{+56.2}_{-52.6}$&9.1$^{+3.2}_{-2.9}$&338.31/320\\
&30042-03-06-00&1&540&0.37$\pm$0.05&2.40$^{+0.11}_{-0.10}$&15.5$\pm$0.4&3.86$^{+0.06}_{-0.07}$&1.65$^{+0.29}_{-0.25}$&1.96$^{+0.33}_{-0.35}$&7.65$^{+0.43}_{-0.42}$&7.73$^{+0.60}_{-0.63}$&12.2$\pm$2.0&333.1$^{+13.4}_{-13.6}$&244.7$^{+32.8}_{-34.4}$&9.1$^{+1.3}_{-1.7}$&408.32/321\\
  &	... & ... & ... & ... & ... & ... & ... & ... & ... & ... & ... & ... & ... & ... & ... & ... \\
 	 \hline		

\end{tabular}
}}
\caption{Shortened table with parameters of the multi-Lorentzian fits per source. The number of averaged subsequent observations and power spectra are given in columns 3 and 4, respectively. We quote the centroid frequency ($\nu$), full width at half maximum (FWHM) and rms-normalized integral power (IP) of L$_b$, L$_{LF}$, L$_h$ and L$_u$. The quoted errors use $\Delta\chi^2$=1. The full version, with all parameters for all sources, is available online.}
\label{tab:pars}
\end{sidewaystable*}

 \begin{table}
	\centering

	\begin{tabular}{lcccr} 
		\hline	 
	Source &Group	&   Range of $\nu_u$ (Hz) &  Range of $\nu_u$ (Hz)    \\  
		   &  & L$_{LF}$                   & L$_h$ \\
               \hline
4U\ 1728--34& 		1 & <381 & <381\\
	&	2 & 381-950 & 381-806 \\
	&	3 & >950  &  >806  \\
\hline
4U\ 0614+09 & 		1 & - & - \\
	 & 	2 & <872 & <872 \\
	 &	3 & >872 &  >872  \\		
\hline
4U\ 1608--52 & 		1 & <215 & <215 \\
	 & 	2 & 215-531  & 215-531 \\
	 &	3 & -  &  >531 \\	
\hline
4U\ 1636--53 & 	1 & - & - \\
	 & 	2 & <772  & <910 \\
	 &	3 & >772 &  - \\
	 \hline
4U\ 1702--43 & 	1 & - & - \\
	 & 	2 & <888  & <542 \\
	 &	3 & >888 &  >542 \\
	 \hline
Aquila X-1 & 	1 & - & - \\
	 & 	2 & <973  & <761 \\
	 &	3 & >973 &  >761 \\	 
	 \hline
	\end{tabular}
	\caption{Specifications of frequency ranges defining groups of data points in the frequency-frequency plots used to fit power laws. In Group 3 we take $\nu_{hHz}$ as $\nu_h$ only if L$_{hHz}$ and L$_h$ are not detected in the same power spectrum.}	\label{tab:gr_All}
\end{table}

\subsection{4U\ 1728--34}

\begin{figure*}
	\includegraphics[width=5in]{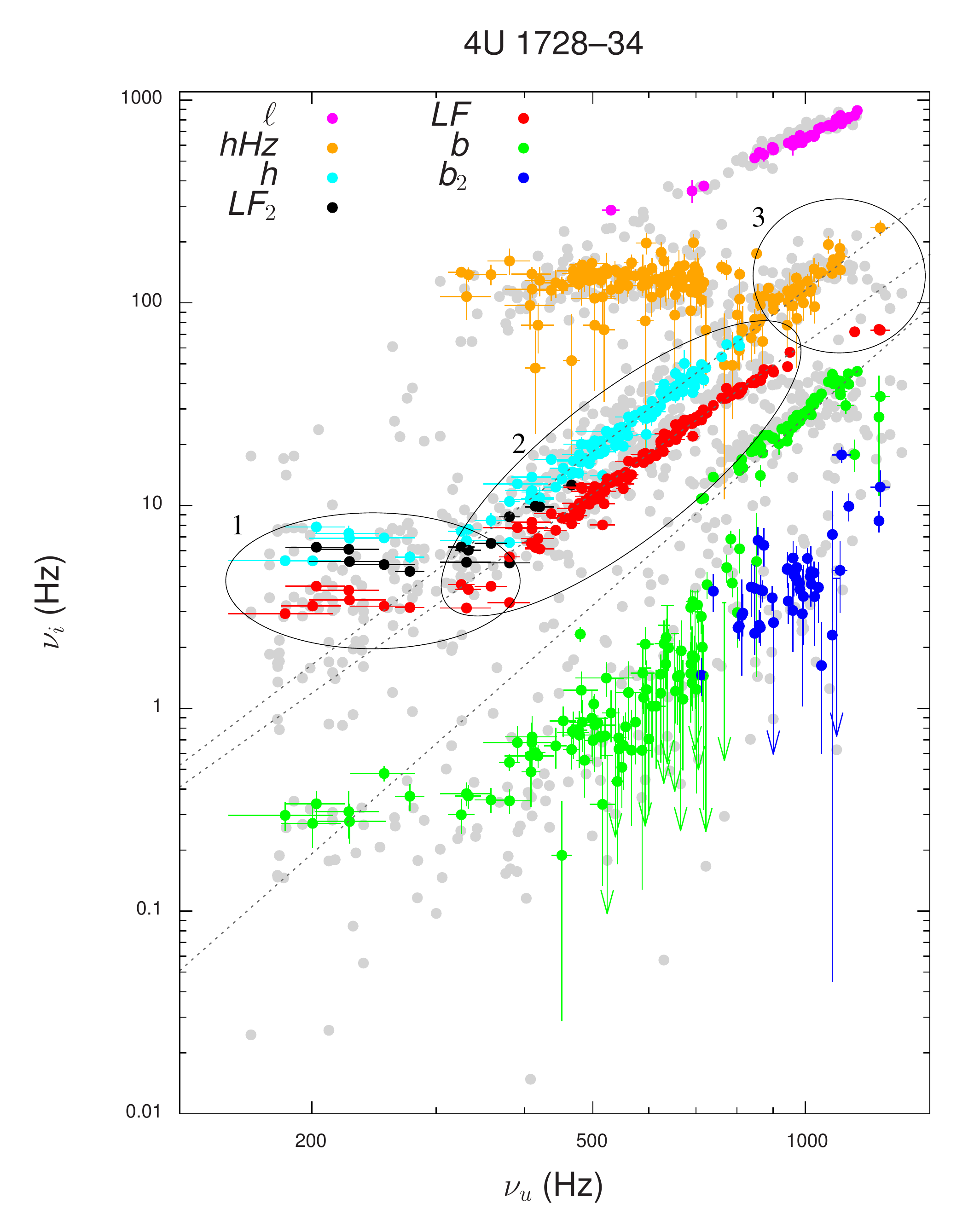}
    \caption{As in Figure \ref{fig:freq_all}, but for 4U\ 1728--34. Best-fit power laws to $\nu_h$ and $\nu_{LF}$ in Group 2 are shown, as well as to $\nu_b$ for data with Q$_b$>0.5. The identification of the $\nu_{LF}$ data in Group 3 (\textit{red}) is tentative. Frequencies measured in all other sources in our sample are plotted in grey. }
   \label{fig:freq_1728}
\end{figure*}

\begin{figure}
	\includegraphics[width=\columnwidth]{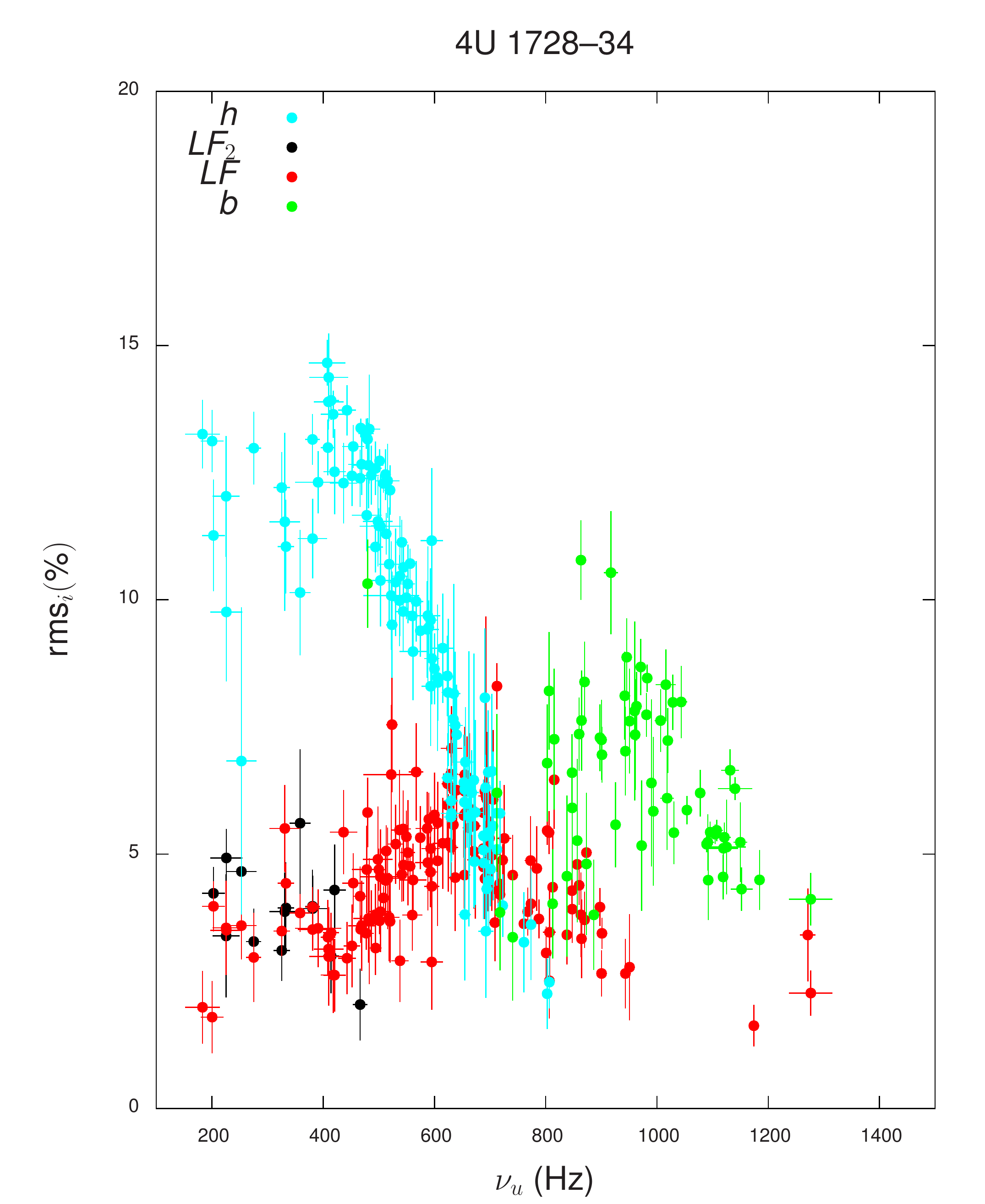}
    \caption{Fractional rms levels of Lorentzians vs. $\nu_u$ in 4U\ 1728--34. We only plot rms$_b$ for Q$_b$>0.5} 
    \label{fig:rms_1728}
\end{figure}

\label{sec:1728}

For $\nu_{u}>$700 Hz, the $\nu_{LF}$-$\nu_u$ correlation in 4U\ 1728--34 slightly flattens (see Figure \ref{fig:freq_1728}), as also reflected in a lower power law index of the best-fit power law to the $\nu_{LF}$-$\nu_u$ frequency pairs when combining Groups 2 and 3, and when fitting $\nu_{LF}$ in Group 2 above $\nu_u$=600 Hz, see Table \ref{fit}. When including $\nu_{LF}$ when $\nu_u$>700 Hz (with $\nu_{LF}$ in Group 2 and 3), we obtain a best-fit power law index of 2.13$^{+0.05}_{-0.04}$, which is still in excess of 2.0 at 3.2$\sigma$.\\
Comparing the low signal to noise power spectra of 91023-01-02-00, 92023-03-41-00, 92023-03-57-00, and 92023-03-82-00 all with $\nu_u\sim$1100 Hz, to high signal to noise power spectra with similar colours, we conclude that the single fitted Lorentzian at low frequency is a blend of L$_b$ and L$_{b_2}$. We designate this blend L$_b$ in Table \ref{tab:pars} and (by plotting colour) in the figures although it is broader than L$_b$ fitted in high signal to noise power spectra and its centroid frequency does not fall on the extrapolation of the $\nu_b$-$\nu_u$ correlation for $\nu_u$<1000 Hz. In 40033-06-03-04, L$_h$ with $\nu_h\sim$41 Hz can only be fitted when fixing the width.

\subsection{4U\ 0614+09}
\label{App_0614}
\begin{figure}
	\includegraphics[width=\columnwidth]{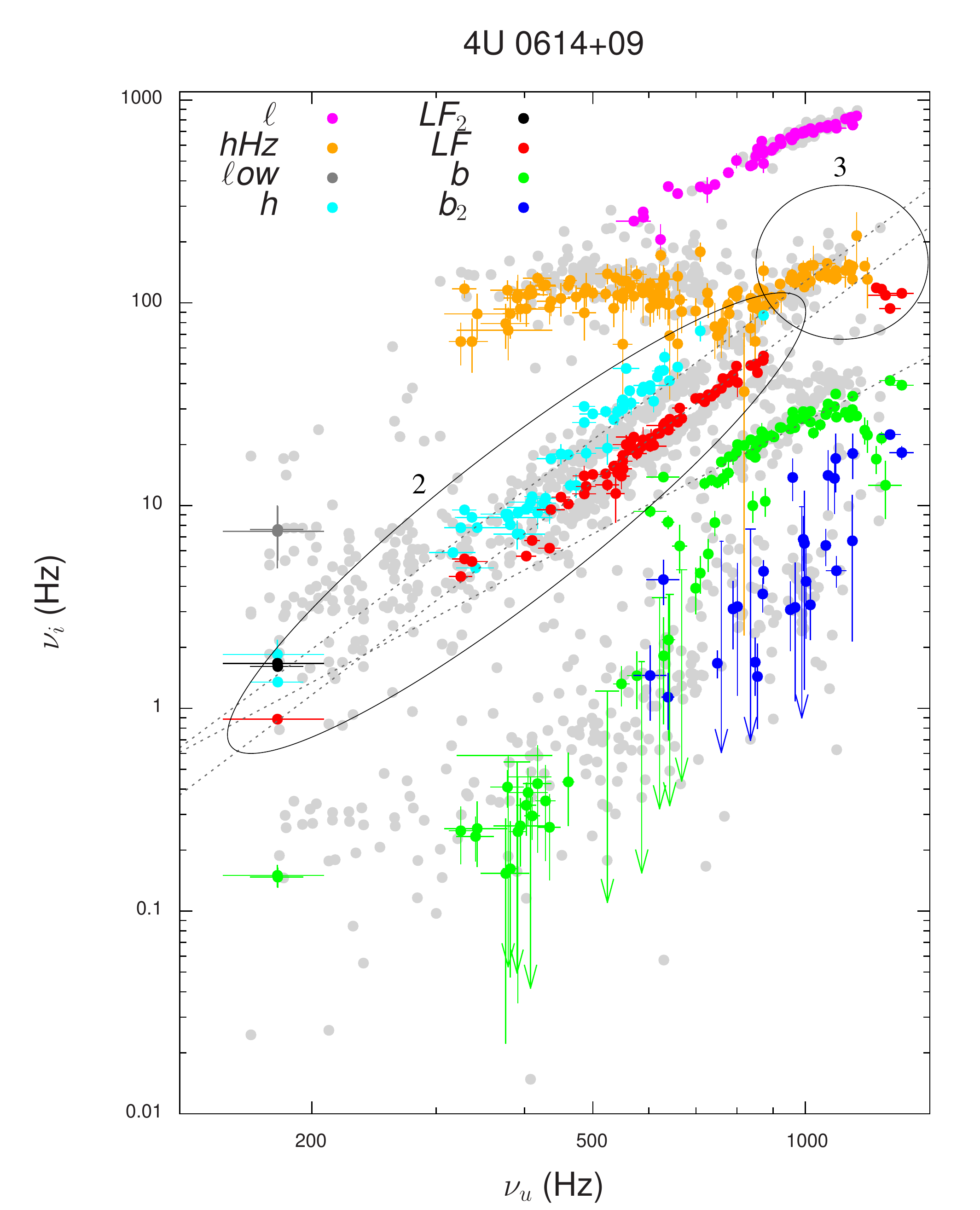}
    \caption{Same as Figure \ref{fig:freq_1728}; but for 4U\ 0614+09.}
    \label{fig:freq_0614}
\end{figure}

\begin{figure}
	\includegraphics[width=\columnwidth]{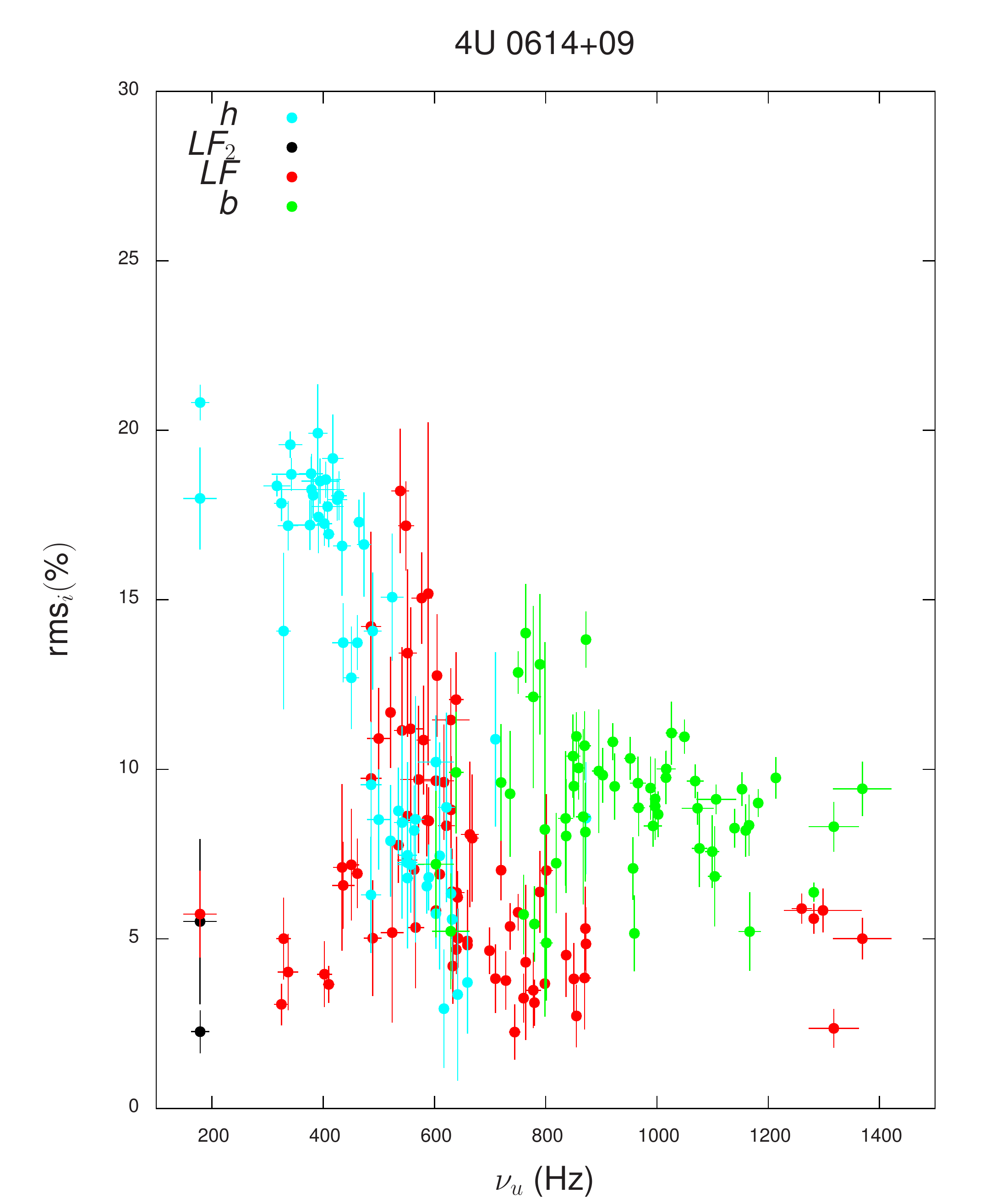}
    \caption{Same as Figure \ref{fig:rms_1728}; but for 4U\ 0614+09.
    }
    \label{fig:rms_0614}
\end{figure}

The timing behaviour of 4U\ 0614+09 strongly resembles that of 4U\ 1728--34 both in terms of frequencies and rms (compare Figures \ref{fig:freq_0614} and \ref{fig:freq_1728}, and Figures \ref{fig:rms_0614} and \ref{fig:rms_1728}, respectively), but the flattening of the $\nu_{h}$-$\nu_u$ and $\nu_{LF}$-$\nu_u$ correlations at $\nu_u$<400 Hz in 4U\ 1728--34, is not evident in 4U\ 0614+09. 
We test whether the power law index changes as $\nu_u$ increases (see Table \ref{fit} for $\nu_{LF}$ within Group 2). When $\nu_u$>700 Hz, including $\nu_{LF}$ in Group 2 and 3, the best-fit power law has index 2.18$\pm$0.05, which is still significantly in excess of 2.0 at 3.8$\sigma$.
The data point with $\nu_{\ell}$ at 205 Hz when $\nu_u\sim$ 600 Hz that falls below the $\nu_{\ell}$-$\nu_u$ correlation corresponds to a 2.5$\sigma$ detection in 91425-01-03-00. By an F-test this component is required at the 3.5$\sigma$ level. We detect a 4$\sigma$ 110 Hz feature in the same power spectrum, which supports our identification of the 205 Hz feature as $\nu_{\ell}$ and not $\nu_{hHz}$. 

At $\nu_u\sim$ 400 Hz, the $\nu_h$ data fall below the correlation traced out for $\nu_u$>500 Hz.  We fail to significantly detect L$_{LF_2}$ in these power spectra, but do detect a power excess at $\sim$2$\nu_{LF}$, suggesting that the lower $\nu_h$ might be a result of blending of L$_h$ and L$_{LF_2}$ (this refers to observations: 50031-01-01-01/02/03/07/08/09, 50031-01-02-01/09, 50031-01-03-01/03/07/08, 50031-01-04-13).

In 10073-01-10-01 (with $\nu_u\sim$380 Hz) Q$_u$ is high (>1.5) which is atypical for Group 1, we therefore add $\nu_{LF}$ and $\nu_h$ to Group 2.


\subsection{4U\ 1608--53}
\begin{figure}
	\includegraphics[width=\columnwidth]{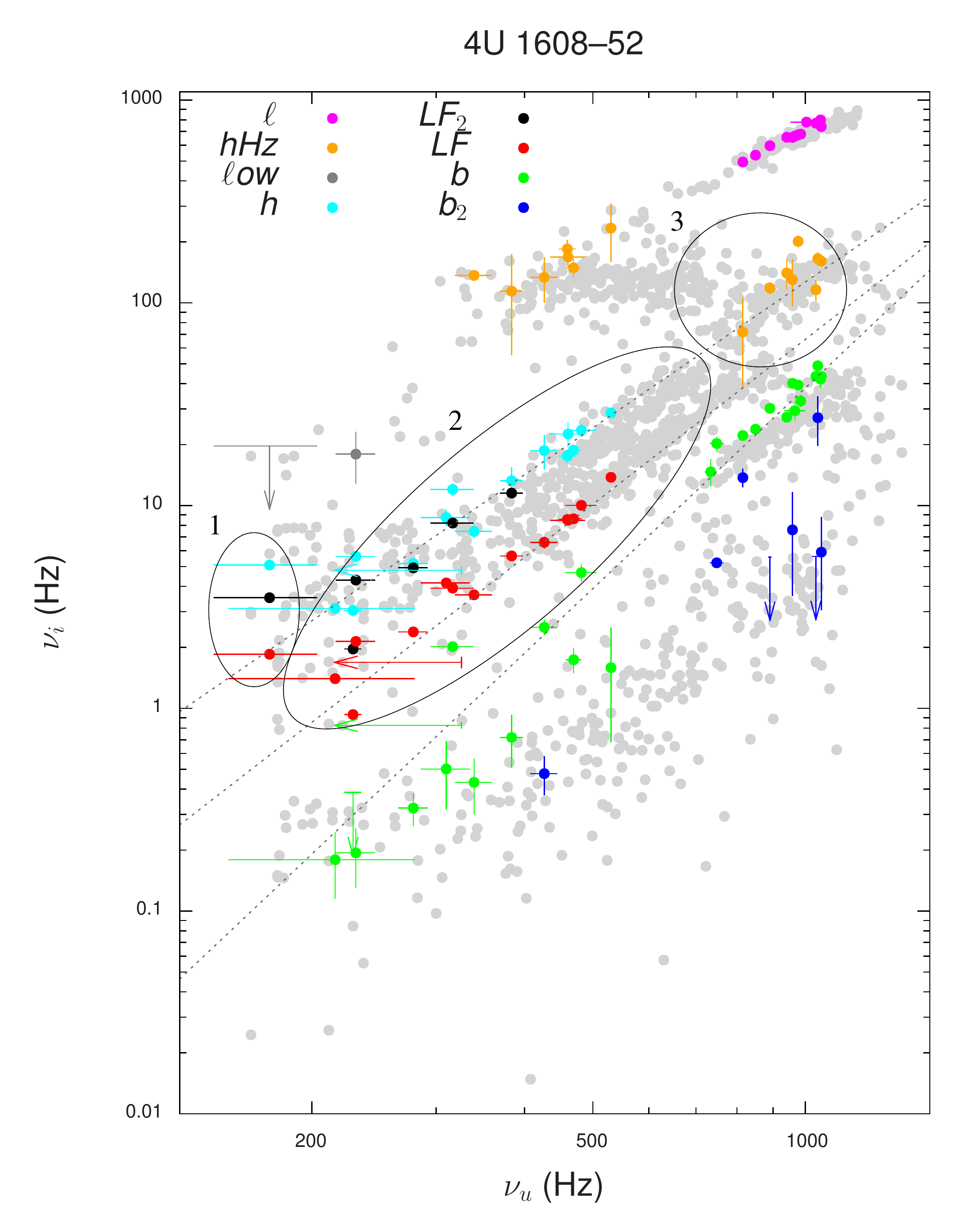}
    \caption{Same as Figure \ref{fig:freq_1728}; but for 4U\ 1608--52. }
    \label{fig:freq_1608}
\end{figure}

\begin{figure}
	\includegraphics[width=\columnwidth]{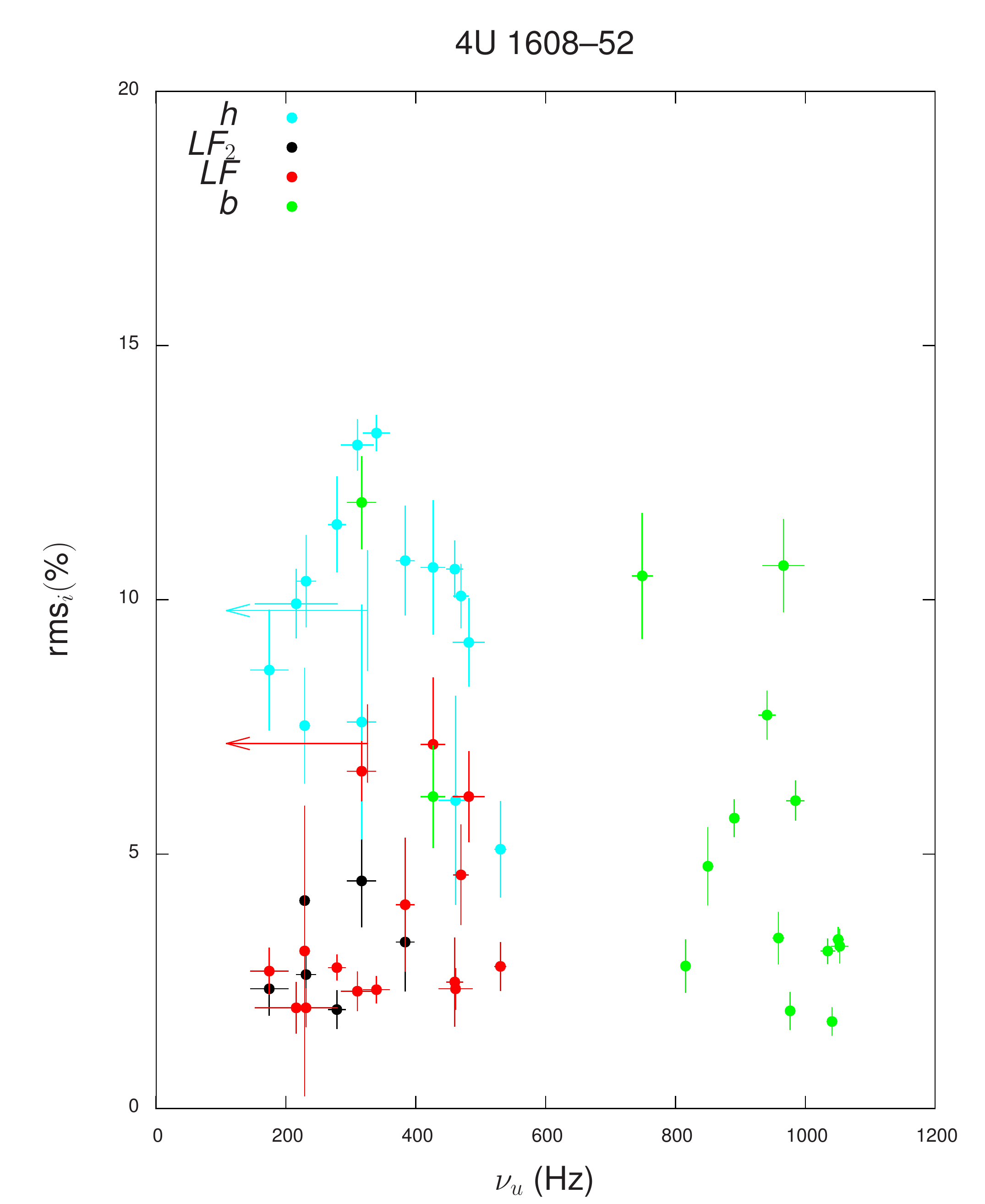}
    \caption{Same as Figure \ref{fig:rms_1728}; but for 4U\ 1608--52. }
    \label{fig:rms_1608}
\end{figure}

Although with fewer data points, the timing behaviour of 4U\ 1608--53 is similar to that of 4U 1728--34, although we do not detect L$_{LF}$ for $\nu_u$>500 Hz (see Figures \ref{fig:freq_1728}, \ref{fig:freq_1608} and \ref{fig:rms_1608}). 
\label{App_1608}

\subsection{4U\ 1636--52}
\label{App_1636}
\begin{figure}
	\includegraphics[width=\columnwidth]{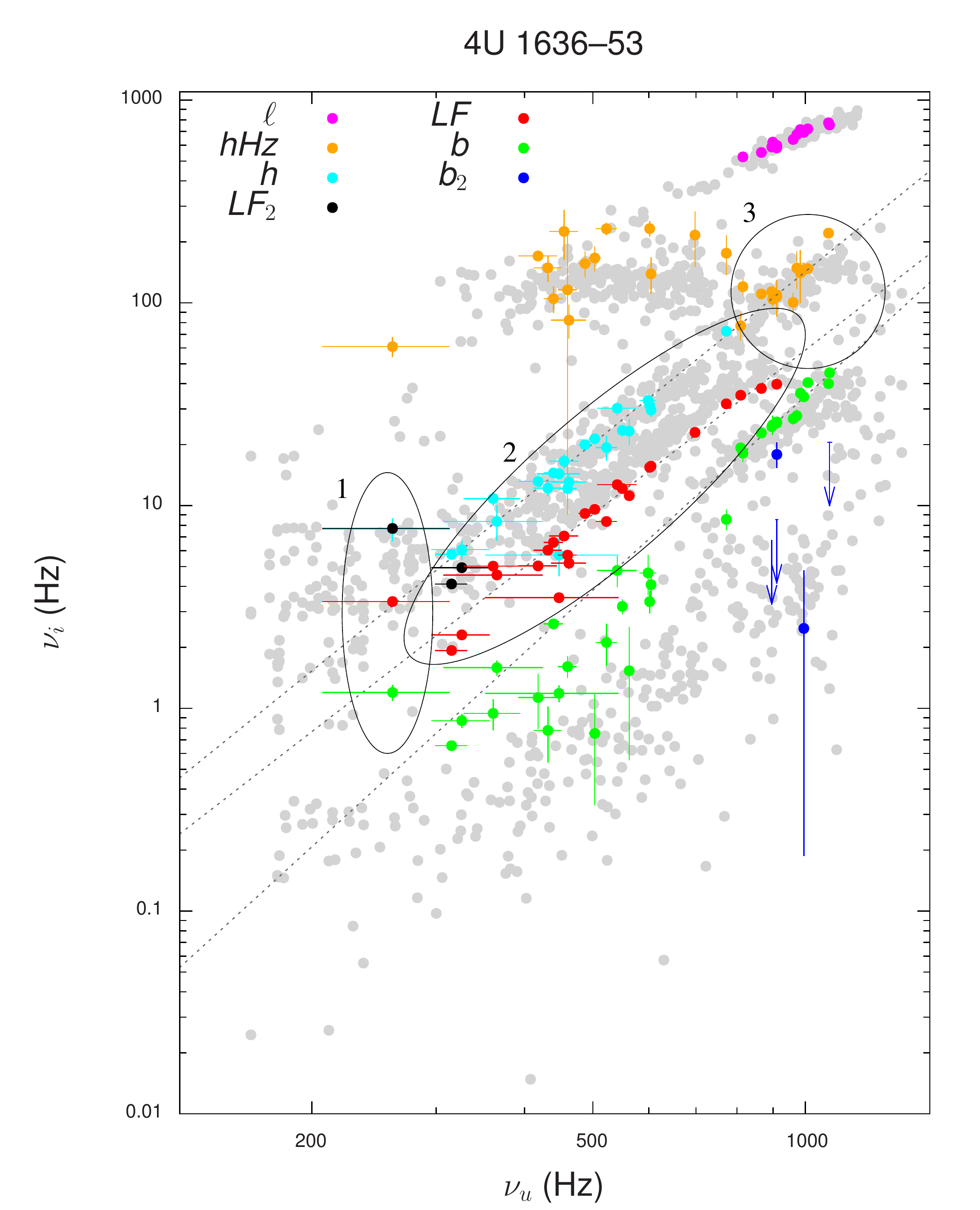}
    \caption{Same as Figure \ref{fig:freq_1728}; but for 4U\ 1636--53.}
    \label{fig:freq_1636}
\end{figure}

The data of 4U\ 1636--52 contain many short observations, and as the source is not bright we can by our selection criteria (which do not allow to combine data over wide time spans) only use a small fraction of the available data. As previously noted by \cite{Altamirano:2008}, this source behaves similarly to 4U\ 1728--34, 4U\ 0614+09 and 4U\ 1608--53 (for instance, compare Figures \ref{fig:freq_1636} and \ref{fig:freq_1728}).
In 95087-01-50-10 we fit a broad zero-centered Lorentzian where, based on other data with similar colours (e.g. 91024-01-75-00), one would expect a kHz QPO with a frequency around 366 Hz. As this may indicate that the features moved during the observation, we discarded this power spectrum from our analysis.
In 92023-01-02-10, L$_h$ and L$_{LF}$ are characterized by unusually low centroid frequencies ($\nu_u\sim$450 Hz). Apart from the low $\nu_h$ and $\nu_{LF}$ this power spectrum is very similar to that of, e.g., 80425-01-04-02 supporting our identification of these features. We therefore include these frequencies in our frequency-frequency correlation fitting.  \\

\subsection{4U\ 1702--43}
\begin{figure}
	\includegraphics[width=\columnwidth]{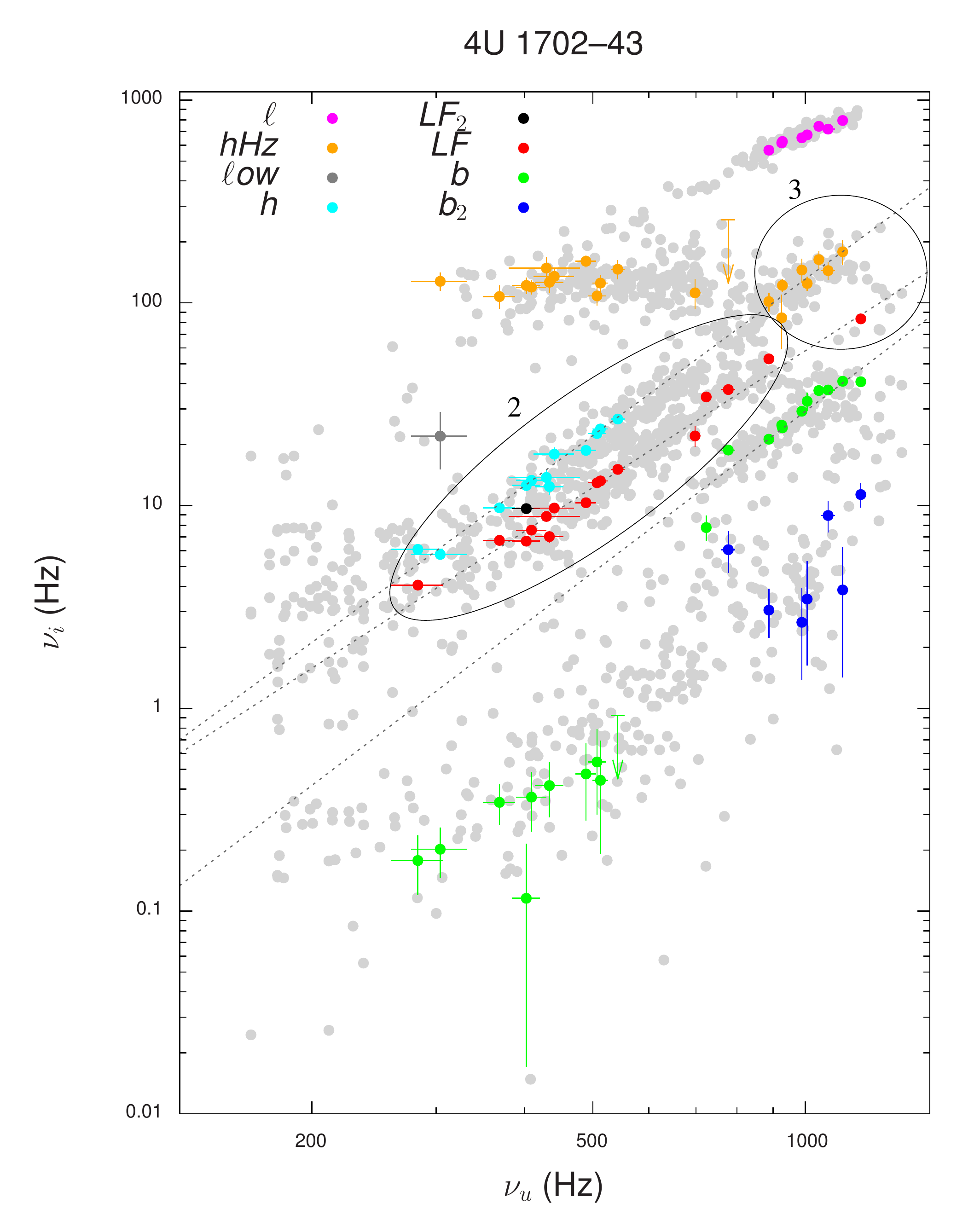}
    \caption{Same as Figure \ref{fig:freq_1728}; but for 4U\ 1702--43.}
    \label{fig:freq_1702}
\end{figure}

In 4U\ 1702-43, we do not see a flattening of the frequency correlations at low $\nu_u$ (see Figure \ref{fig:freq_1702}). We therefore do not define a frequency Group 1.

\label{App_1702}

\subsection{Aquila X-1}
\label{App_aql}

\begin{figure}
	\includegraphics[width=\columnwidth]{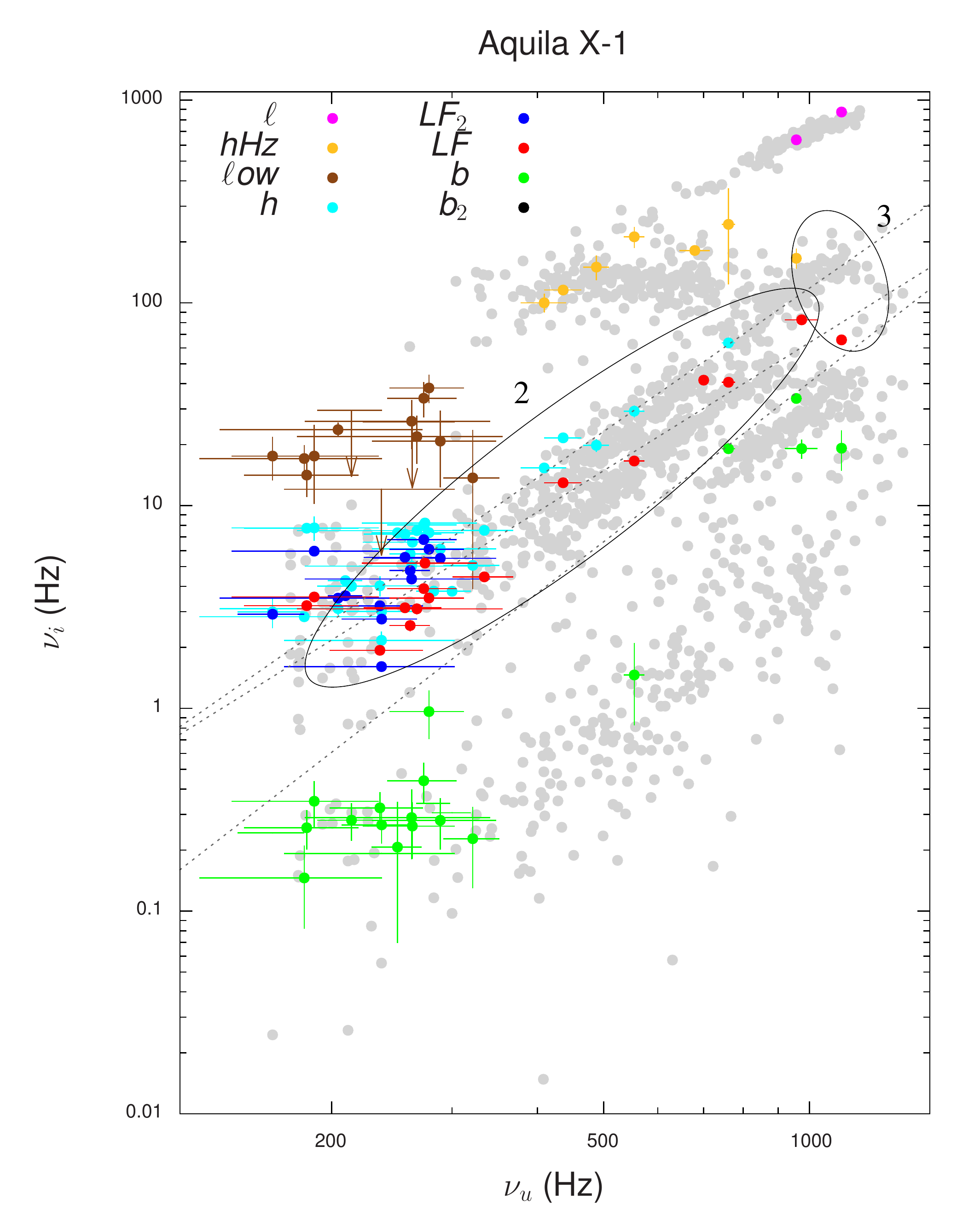}
    \caption{Same as Figure \ref{fig:freq_1728}; but for Aquila X-1.}
    \label{fig:freq_Aql}
\end{figure}

\begin{figure}
	\includegraphics[width=\columnwidth]{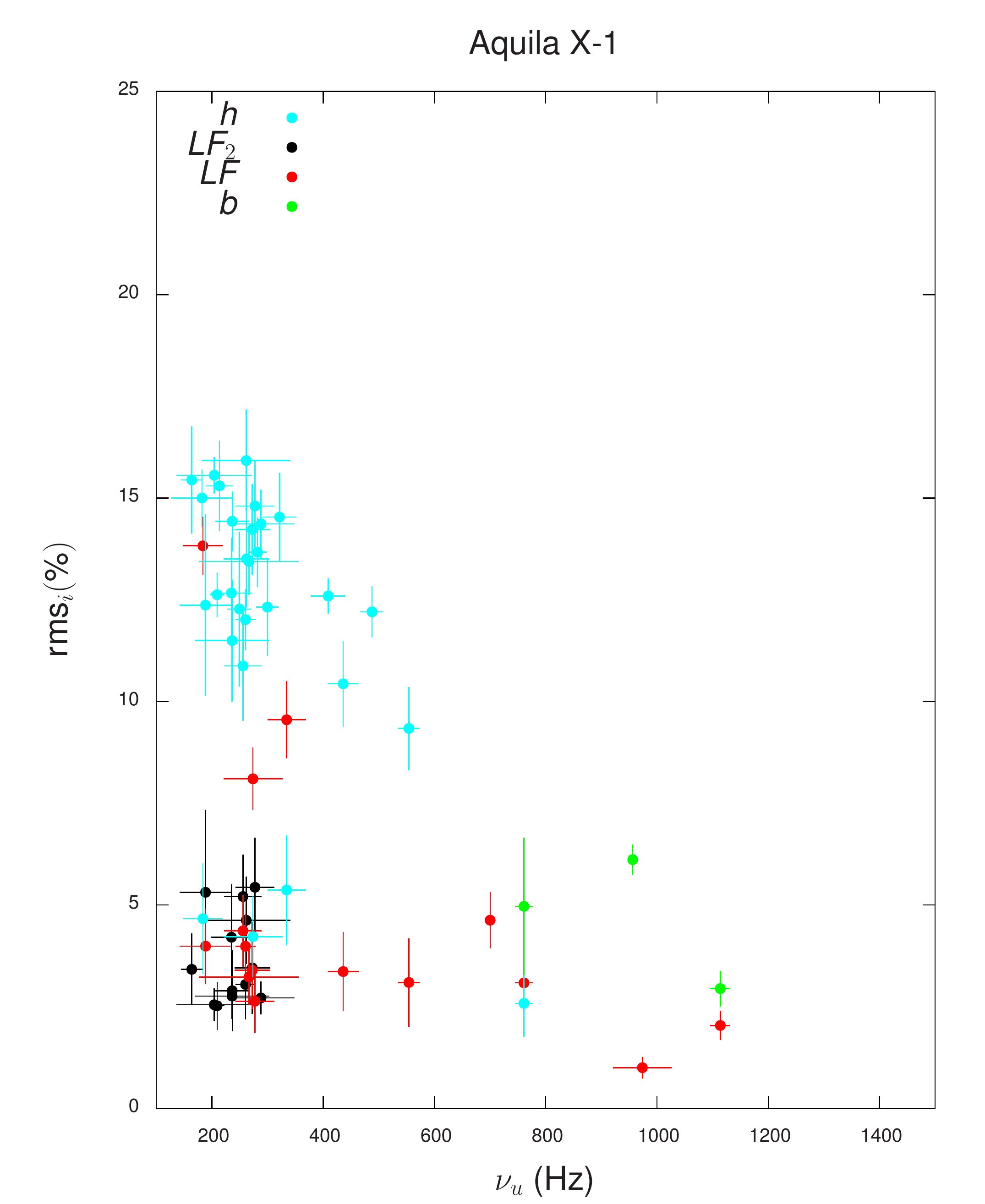}
    \caption{Same as Figure \ref{fig:rms_1728}; but for Aquila X-1. }
    \label{fig:rms_Aql}
\end{figure}

\begin{figure}
	\includegraphics[width=\columnwidth]{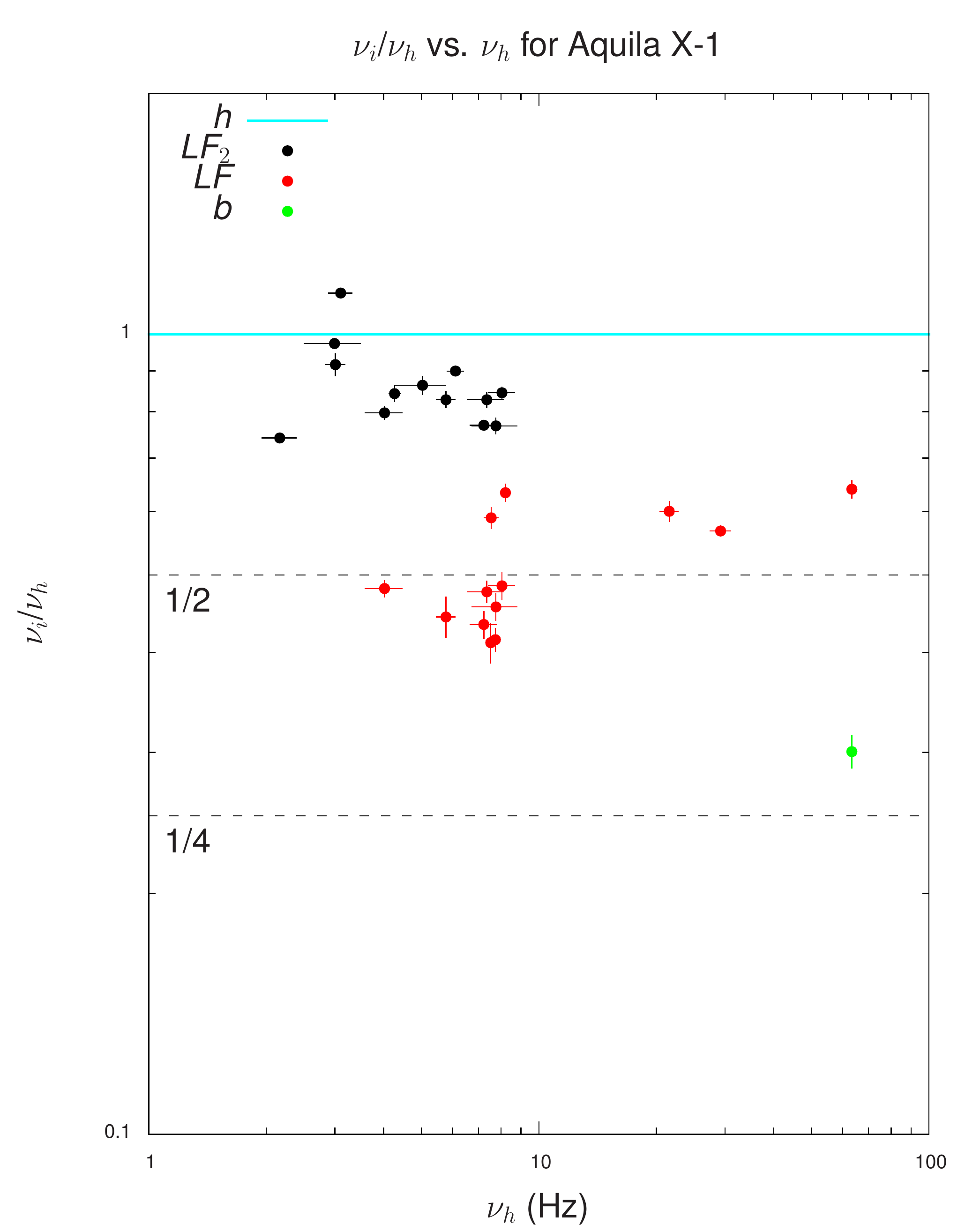}
    \caption{Same as Figure \ref{fig:LF_H}; for Aquila X-1. }
    \label{fig:LF_H_Aql}
\end{figure}

The power spectra of Aquila X-1 closely resemble those of the other sources in our sample. For the identification of L$_h$, L$_{LF}$ and L$_{LF_2}$ we use Figures \ref{fig:rms_Aql} and \ref{fig:LF_H_Aql}.
The upper kHz QPO at low frequencies is very broad, which explains the structure of the frequency-frequency plot below $\nu_u$=300 Hz, see Figure \ref{fig:freq_Aql}.

In 40049-01-02-02 at $\nu_u\sim$200 Hz, we fit an unusually broad L$_{LF}$ as compared to other power spectra with similar colours, and a narrow L$_{h}$ (see Figure \ref{fig:rms_Aql}).  We consider the identification of these features insecure and atypical for the source. We omit them from our frequency-frequency correlation fitting.

\subsection{SAX\ J1750.8-2900}
In SAX\ J1750.8-2900  signal to noise is low, so feature identifications are less secure. 
In 60035-01-02-01, near $\nu_u$=620 Hz, L$_h$ and L$_{LF}$ can only be fitted when fixing their widths (see online version of Table \ref{tab:pars}) and are then detected at 3$\sigma$ each. 

\begin{figure}
	\includegraphics[width=\columnwidth]{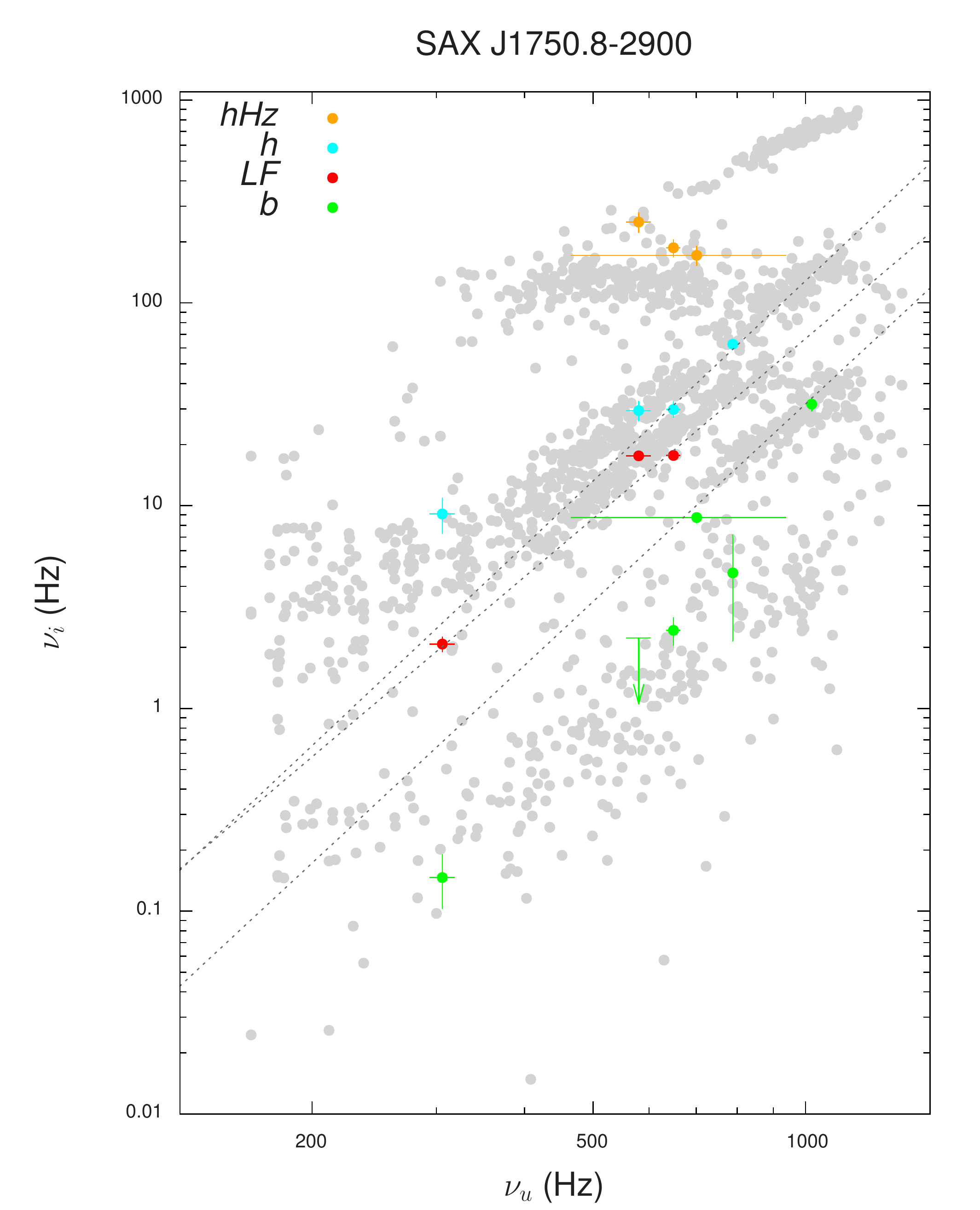}
    \caption{Same as Figure \ref{fig:freq_1728} but for SAX\ J1750.8-2900. }
    \label{fig:SAX}
\end{figure}

L$_h$ and L$_{LF}$ in 93432-01-03-04 with $\nu_u$ around 200 Hz are detected at >3$\sigma$ However, $\nu_h$ is offset from the $\nu_h$-$\nu_u$ correlations in other sources, see Figure \ref{fig:SAX}. Probably, due to low signal to noise, components between L$_u$ and L$_h$ remain undetected and are fitted with a broader L$_h$, shifting the centroid frequency to a higher value. We use this $\nu_h$ in our $\nu_h$-$\nu_u$ fit, because we cannot compare to similar observations of this source to confirm the possible broadening of L$_h$.

\subsection{4U\ 1915--05}
\begin{figure}
	\includegraphics[width=\columnwidth]{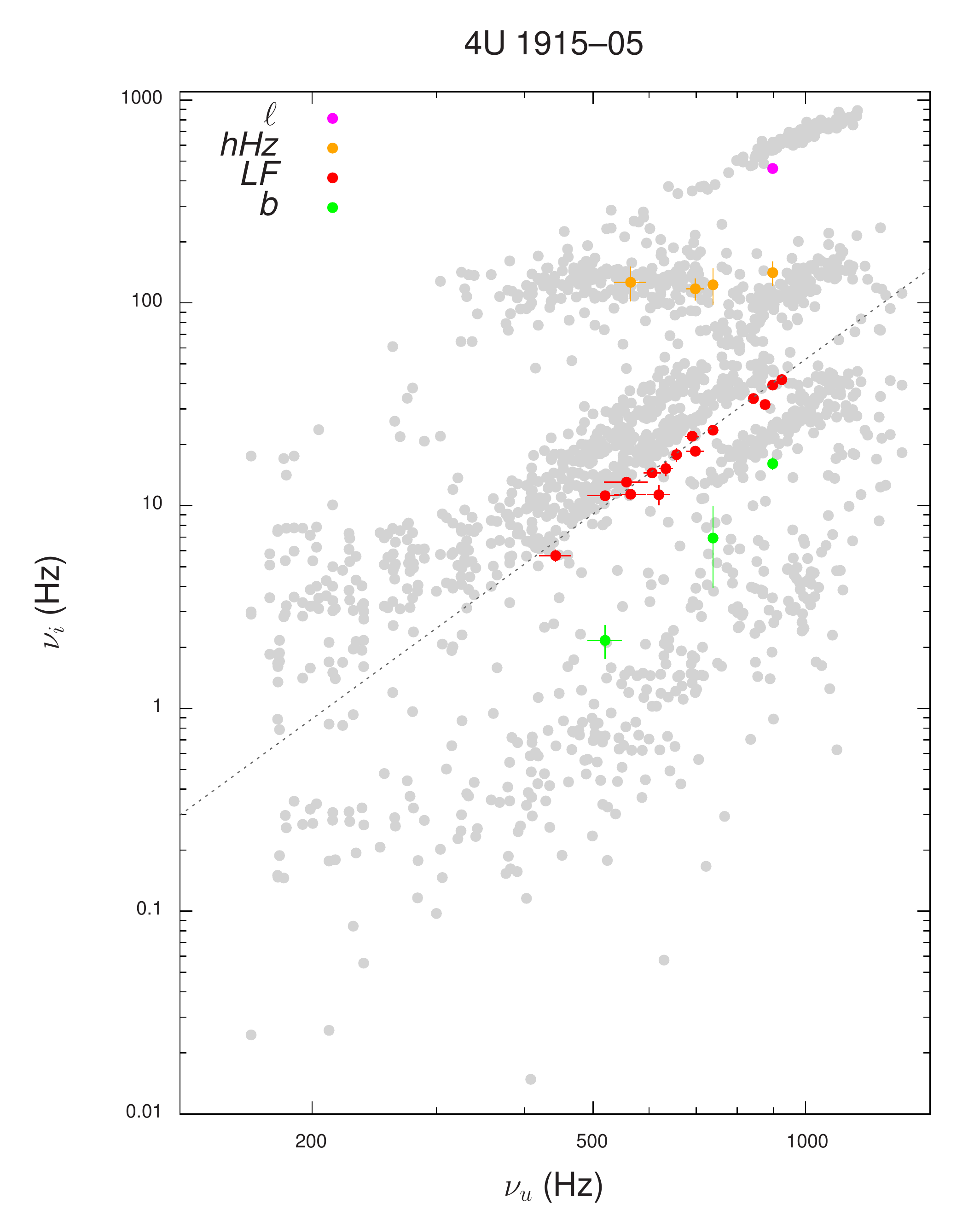}
    \caption{Same as Figure \ref{fig:freq_1728} but for 4U\ 1915--05. }
    \label{fig:1915}
\end{figure}

\begin{figure}
	\includegraphics[width=\columnwidth]{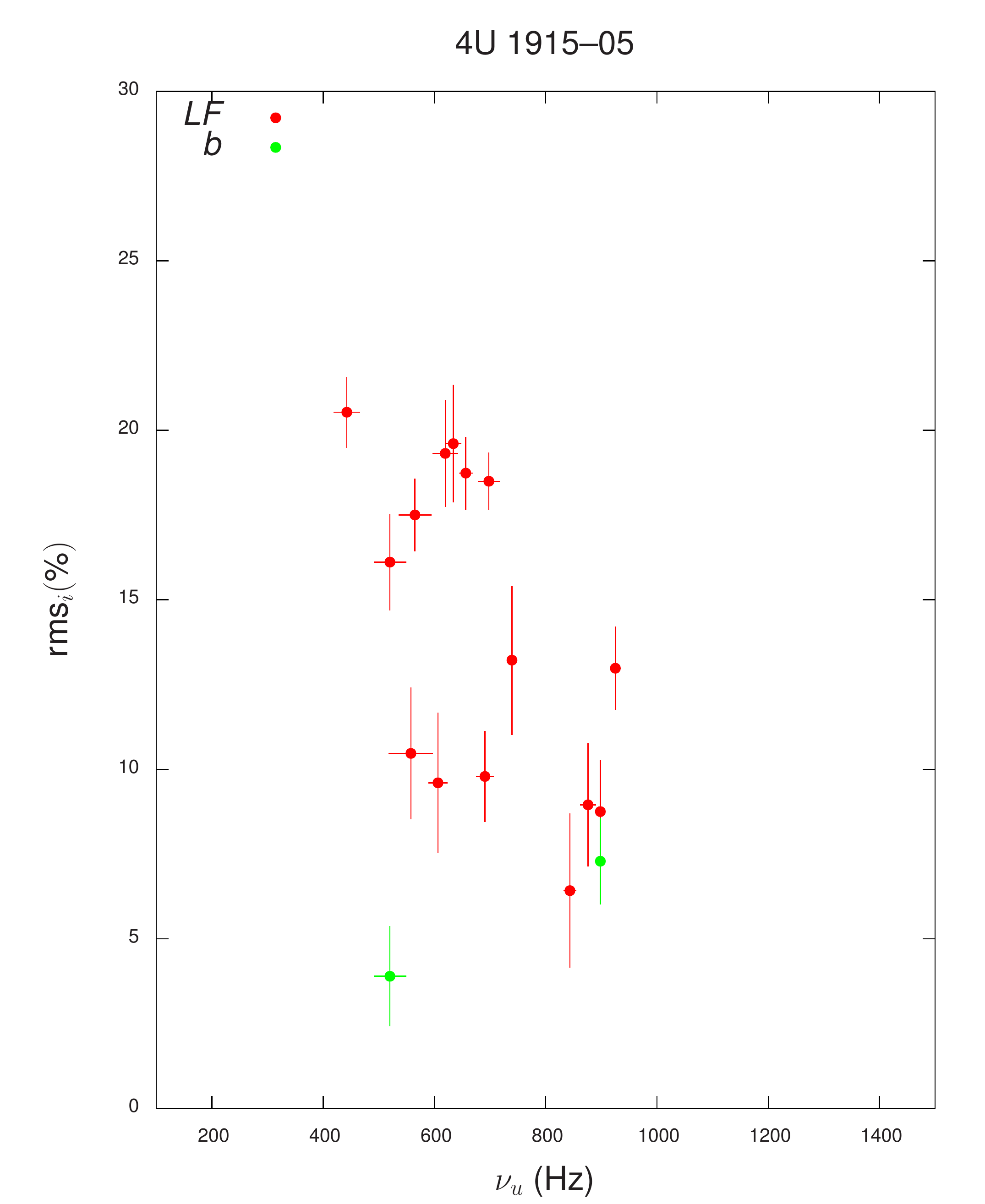}
    \caption{Same as Figure \ref{fig:rms_1728} but for 4U\ 1915--05. }
    \label{fig:19152}
\end{figure}

In this source we fit a power spectral feature that we can follow over a range of 600 Hz in $\nu_u$ at a frequency that systematically falls between that of L$_b$ and L$_{LF}$ as observed in other sources, see Figure \ref{fig:1915}. It could be a blend of  L$_b$ and L$_{LF}$ we fit there. As would be expected from a blend, the fractional rms-$\nu_u$ relation behaves similarly to L$_{h}$ (which has much higher rms, compare Figure \ref{fig:19152} to Figure \ref{fig:rms_1728}) in other sources (like 4U\ 1728--34). We designate this QPO as L$_{LF}$ in Table \ref{tab:pars} and (by plotting colour) in the figures.



\subsection{KS\ 1731--260}

 \begin{figure}
	\includegraphics[width=\columnwidth]{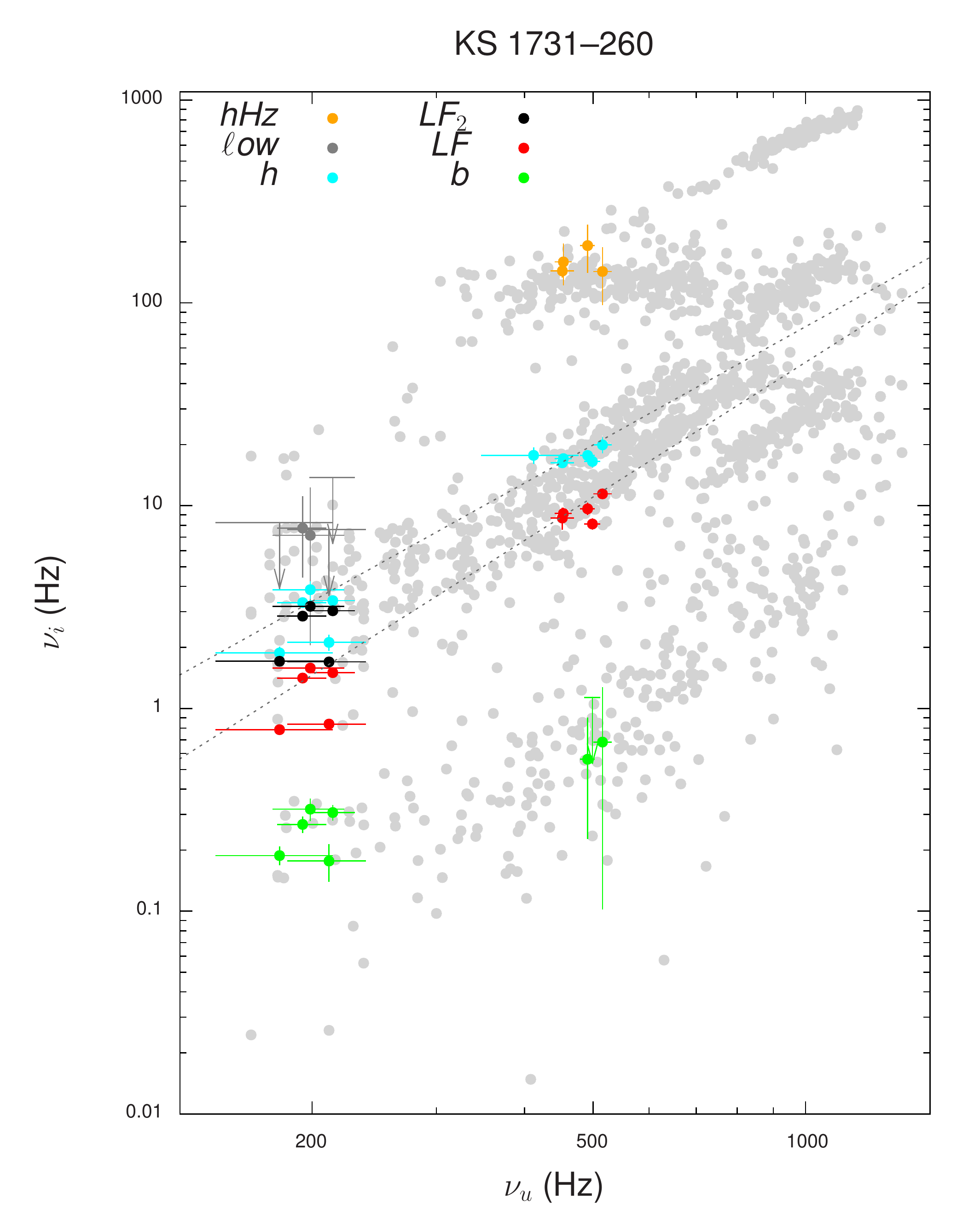}
    \caption{Same as Figure \ref{fig:freq_1728} but for KS\ 1731--260. }
    \label{fig:KS}
\end{figure}
\begin{figure}
	\includegraphics[width=\columnwidth]{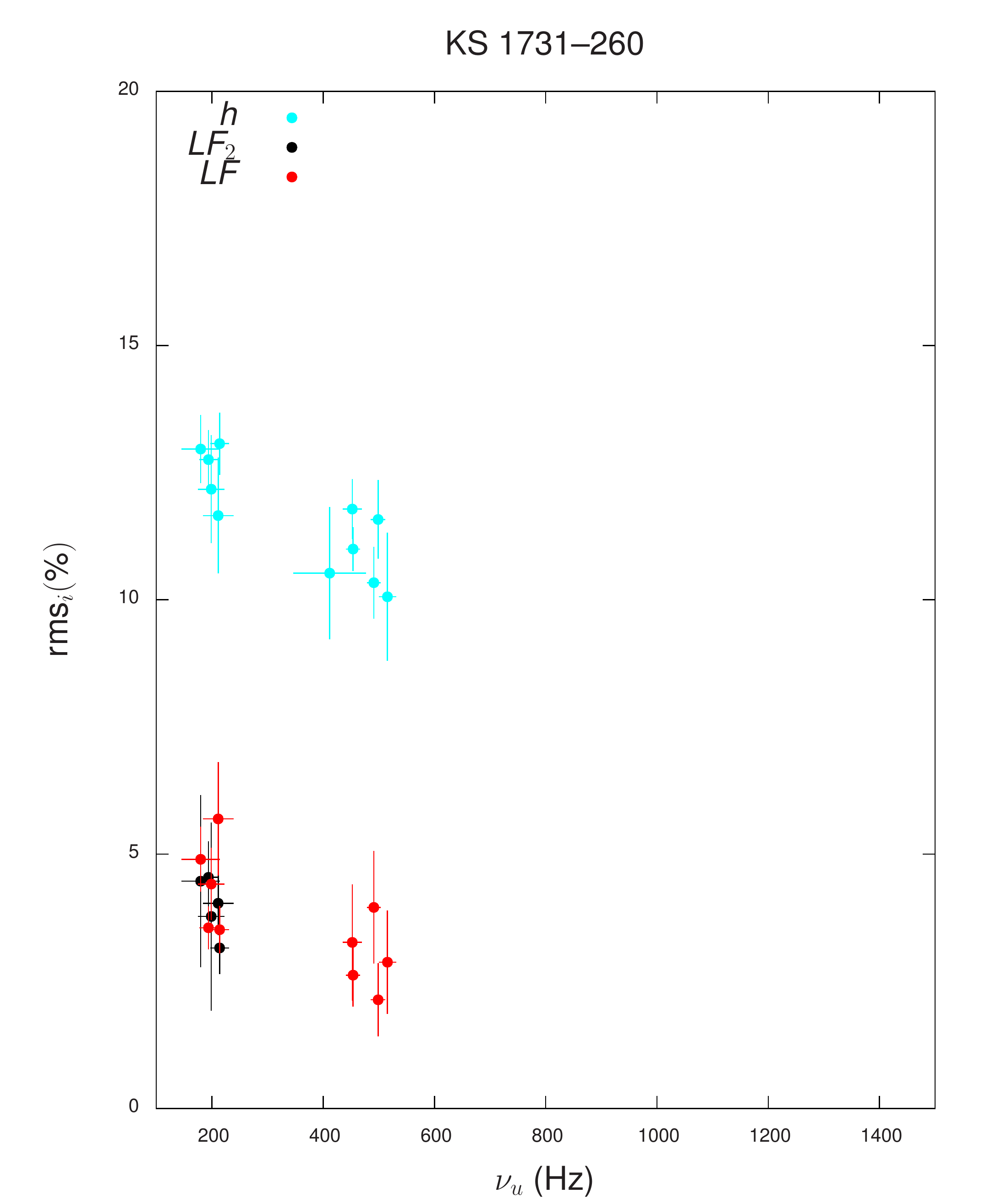}
    \caption{Same as Figure \ref{fig:rms_1728} but for KS\ 1731--260. }
    \label{fig:KS2}
\end{figure}

In this source  two distinct types of power spectrum are seen where $\nu_u$ is near 200 Hz (see Figure \ref{fig:KS}):  one where $\nu_{LF}\sim$0.8 Hz, $\nu_{LF_2}\sim$1.6 Hz and $\nu_h\sim$2.5 Hz, the other where $\nu_{LF}\sim$1.5 Hz, $\nu_{LF_2}\sim$3 Hz, and  $\nu_{h}\sim$4 Hz. In terms of rms, the features in both types are similar, see Figure \ref{fig:KS2}. To what extent this is related to the flattening of the correlations seen in other sources in this range is unclear.

\subsection{IGR\ J17191--2821}
\begin{figure}
	\includegraphics[width=0.93\columnwidth]{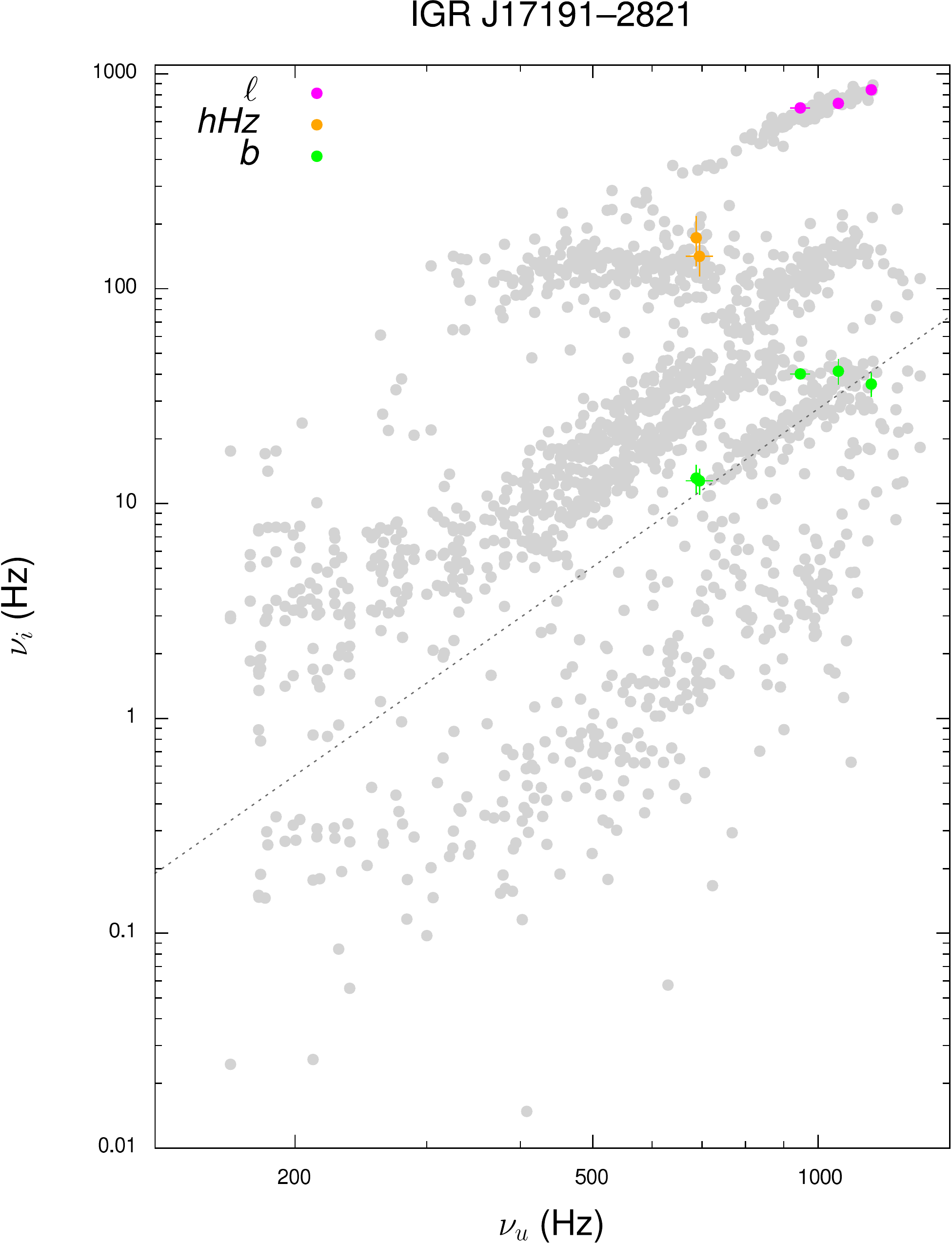}
    \caption{Same as Figure \ref{fig:freq_1728} but for  IGR\ J17191--2821. }
    \label{fig:IGRJ}
\end{figure}

Similarly to SAX\ J1750.8-2900, in IGR\ J17191--2821 we can fit L$_u$ accompanied by features at low frequency that are hard to identify due to low signal to noise. Blending could have affected these frequencies.
  We designate the low frequency component as L$_{b}$ in Table \ref{tab:pars} and Figure \ref{fig:IGRJ} (by plotting colour).

\bsp	
\label{lastpage}
\end{document}